\documentclass[%
  aps,
  superscriptaddress,
  longbibliography,
  12pt,
  onecolumn,
  a4paper,
  showpacs,
  showkeys,
  amsfonts, amssymb, amsmath]{revtex4-2}

\usepackage[utf8]{inputenc}
\usepackage[caption=false]{subfig}
\usepackage{hyperref} 

\hypersetup{pdfpagemode={UseOutlines},
bookmarksopen=true,
bookmarksopenlevel=0,
hypertexnames=false,
colorlinks=true, 
citecolor=blue, 
linkcolor=red, 
urlcolor=black, 
pdfstartview={FitV},
unicode,
breaklinks=true,
}
\usepackage{graphicx,amssymb,amsmath}
\usepackage{grffile}
\usepackage{color}
\usepackage[dvipsnames]{xcolor} 
\usepackage{siunitx}
\usepackage[normalem]{ulem}

\usepackage{float}

\usepackage{booktabs}

\usepackage{enumitem}
\setlist{nosep}

\usepackage{xspace}

\newcommand{\p}{\partial}

\newcommand{\kvec}{\boldsymbol{k}}
\newcommand{\dt}{\mathop{dt}}

\newcommand{\deltavec}{\boldsymbol{\delta}}

\newcommand{\kmax}{k_{\max}}
\newcommand{\knl}{k_{\mathrm{nl}}}

\newcommand{\bnabla}{\mathbf{\nabla}}
\newcommand{\uu}{\mathbf{u}}
\newcommand{\etamin}{\tilde\eta}

\newcommand{\coo}{\ensuremath{\mathrm{CO_2}}\xspace}
\newcommand{\cpp}{C\texttt{++}\xspace}

\newcommand{\modpath}[1]{\texttt{\path{#1}}}
\newcommand{\hrefmodpath}[2]{\href{#1}{\texttt{\path{#2}}}}

\newcommand{\fluidsimsolver}[1]{\hrefmodpath{https://foss.heptapod.net/fluiddyn/fluidsim/-/blob/branch/default/fluidsim/solvers/#1}{fluidsim.solvers.#1}}

\DeclareMathOperator{\sign}{sign}

\newlength{\figwidth}
\begin{document}

\title{Aliasing and phase shifting in pseudo-spectral simulations of the incompressible
Navier-Stokes equations}

\author{Clovis Lambert}
\author{Jason Reneuve}
\author{Pierre Augier}

\email[]{pierre.augier@univ-grenoble-alpes.fr}

\affiliation{Laboratoire des Ecoulements G\'eophysiques et Industriels, Universit\'e
Grenoble Alpes, CNRS, Grenoble-INP,  F-38000 Grenoble, France}

\begin{abstract}

Pseudo-spectral methods are widely used for direct numerical simulations of turbulence,
but the standard 2/3 truncation rule for dealiasing is computationally
expensive—accounting for up to 80\% of the total cost in three dimensions.
Phase-shifting methods provide a more efficient alternative by canceling aliasing
errors by combining nonlinear terms evaluated on shifted grids, allowing the same
physical resolution to be achieved on a coarser numerical grid. Despite their use in
high-resolution turbulence codes, these methods remain poorly documented in the
literature and no open-source implementation exists.

This paper presents a comprehensive analysis of phase-shifting dealiasing for
pseudo-spectral simulations of the incompressible Navier-Stokes equations. We derive
the aliasing mechanism from quadratic nonlinearities in discrete Fourier space and
explain how phase shifting cancels aliasing contributions exactly or approximately
depending on the time-stepping scheme. We describe and compare several
algorithms—including the exact and approximate RK2 phase-shifting schemes of
\citet{PattersonOrszag1971} and \citet{Rogallo1981}, and an extension to forced
flows—and discuss their interaction with different truncation geometries in three
dimensions.

All algorithms are implemented in the open-source framework Fluidsim, providing the
first publicly available implementation of phase-shifting dealiasing for
pseudo-spectral Navier-Stokes solvers. We evaluate the methods on two test cases: the
transition to turbulence of Taylor-Green vortices and forced homogeneous isotropic
turbulence at $Re_\lambda \approx 200$. Phase-shifting methods achieve speedups of up
to a factor of 3 compared to RK4 with 2/3 truncation at the same maximum wavenumber,
with small accuracy loss.

\end{abstract}

\maketitle

\section{Introduction}

Pseudo-spectral methods are widely used for fundamental studies of turbulence because
of their high accuracy and efficiency. The time evolution is computed in spectral
space, while nonlinear terms are computed in real space, resulting in a large part of
the computation involving spectral transforms. The quasi-linear complexity $N\log{N}$
of the Fast Fourier Transform algorithm and the very good efficiency of open-source
implementations make these methods attractive for high quality simulations.

A critical challenge in pseudo-spectral methods is the treatment of aliasing errors.
With $N$ grid points regularly spaced over a line of size $L$, the highest represented
wavenumber is the Nyquist frequency $k_N = (2\pi / L) N/2$. The nonlinear terms
computed from a solution without energy for $k > k_N$ can contain energy at wavenumbers
larger than $k_N$. Represented on a discretized basis, a signal which should be at a
wavenumber $k_{nl}$ larger than $k_N$ is ``aliased'' and actually appears at $k_{nl} -
2k_N$. This artifact of the discretization can introduce significant errors for
nonlinear equations.

The standard solution is to apply a 2/3 truncation rule. In each space direction in
spectral space, the modes with wavenumbers equal to or larger than $(2/3)k_N$ are
forced to zero, limiting the maximum wavenumber with energy to $\kmax = (2/3)k_N$. A
very similar alternative is the 3/2 extension method, which expands the fields on a
larger grid before the spectral transforms to compute the nonlinear terms. These two
methods are both exact and straightforward to implement without introducing numerical
parameters. However, they are computationally expensive. In three dimensions with a
squared truncation, only 30\% of the modes in Fourier space are retained. For
applications requiring correctness of the dynamics at the smallest scales, a 2/3
spherical truncation is preferable \cite{PattersonOrszag1971}, but retains less than
16\% ($\simeq (4/3) \pi ((2/3) \times 0.5)^3$) of the computed grid points. In
practice, approximately $80\%$ of the computational power is spent on dealiasing (see
section \ref{section-aliasing}). Given that turbulence studies typically require very
high resolutions, often pushing the limits of what is numerically feasible, improving
the efficiency of dealiasing is crucial for enabling more ambitious simulations within
available computational resources and limiting our \coo{} emission.

Phase-shifting methods offer a more efficient alternative. For example, combining phase
shifting with truncation allows resolving modes up to $\kmax = (2\sqrt{2}/3) k_N \simeq
0.94 k_N$ instead of $\kmax = (2/3) k_N$ for the same grid size, representing a
significant improvement in effective resolution. Despite this advantage, phase-shifting
methods remain poorly documented in the literature. Studies employing these methods
typically provide only brief mentions. For instance, \citet{Delache2014} describe their
approach as using ``the phase-shifting method \cite[]{Canuto2007} at a resolution of
$2048^3$, yielding $\kmax=965(=\sqrt{2}N/3)$". Similarly, \cite{Iyer2019} state that
``a combination of phase shifting and truncation is used to reduce aliasing errors,
with the highest resolved wavenumber $\kmax = \sqrt{2}N/3$" with only a citation to
\cite{Rogallo1981}. More recently, \cite{YEUNG2025} mention ``phase shifting and
truncation at wavenumbers beyond $\kmax = \sqrt{2}N/3$ in Fourier space
\cite[]{Canuto1988}". These citations reference comprehensive textbooks
\cite{Canuto1988, Canuto2007} whose sections on phase shifting are brief and
insufficient for implementation, or older publications \cite{Rogallo1981,
PattersonOrszag1971} that lack detailed numerical methodology and evaluation on
turbulent flows.

A related problem is the scarcity of open-source implementations. Many pseudo-spectral
codes are developed in-house or remain closed-source. While several open-source codes
exist—including \href{https://dedalus-project.org/}{Dedalus} \cite{burns2020dedalus},
\href{https://github.com/pmininni/GHOST}{GHOST} \cite{mininni2011hybrid},
\href{https://github.com/spectralDNS}{SpectralDNS} \cite{mortensen2016high},
\href{https://turtle.pages.mpcdf.de/turtle/index.html}{TurTLE}
\cite{lalescu2022particle}, and \href{https://github.com/schumakov/hit3d}{HIT3D}
\cite{chumakov2008priori}—none employ phase-shifting methods. This combination of
incomplete documentation and absent open-source implementations creates a significant
barrier to reproducibility and wider adoption of these efficient methods.

This study addresses these gaps by providing a comprehensive treatment of
phase-shifting dealiasing methods. We detail the numerical methods used fo this work
(section \ref{section-num-methods}), explain the physical origin of aliasing in
turbulent flows (section \ref{section-aliasing}), describe and compare different
phase-shifting algorithms (section \ref{section-phase-shifting}), and evaluate their
numerical performance and simulation quality on decaying flows (section
\ref{section-decaying}) and forced incompressible homogeneous isotropic turbulence
(section \ref{section-forced}). We provide in Fluidsim the first open-source
implementation of these algorithms , with clear guidelines for parameter selection. Our
goal is to make efficient pseudo-spectral simulations with phase-shifting readily
accessible to the turbulence research community.

\section{Numerical methods and overview}

\label{section-num-methods}

This study combines theoretical presentation with numerical validation on realistic
turbulent flows. All numerical results are obtained with the computational fluid
dynamics framework \href{https://fluidsim.readthedocs.io}{Fluidsim} \cite{fluidsim}
(version \href{https://pypi.org/project/fluidsim/0.9.0/}{0.9.0}) through time
advancement of nonlinear equations. All data used in this article are available in this
\href{https://doi.org/10.5281/zenodo.19554006}{dataset} \cite{Dataset}. We consider
three test cases of increasing complexity to illustrate the effects of aliasing and the
performance of different dealiasing methods:

\begin{itemize}

\item In sections \ref{section-aliasing} and \ref{section-phase-shifting}, we first
consider the simplest nonlinear equation in one dimension $\p_t S(x, t) = - \sign(S)
S^2$, using the solver \fluidsimsolver{nl1d}. This academic case allows clear
visualization of aliasing effects.

\item In sections \ref{section-aliasing} to \ref{section-decaying}, we examine the
transition to turbulence of Taylor–Green vortices \cite{Taylor-Green}, a canonical test
case for three-dimensional turbulence codes.

\item In section \ref{section-forced}, we study forced incompressible homogeneous
isotropic turbulence (HIT), a demanding and realistic test case.

\end{itemize}

For the two three-dimensional cases, the Fluidsim solver \fluidsimsolver{ns3d} is used,
which solves the incompressible Navier-Stokes equations

\begin{subequations} \label{eq:N-S-incompr}
\begin{alignat}{2}
\bnabla\cdot\uu & = 0, \label{eq:N-S-incompr-cont}\\
\frac{\partial \uu}{\partial t} + \uu\cdot\bnabla\uu & = -\bnabla p + \nu \bnabla^2 \uu, \label{eq:N-S-incompr-qdm}
\end{alignat}
\end{subequations}

where $\mathbf{u}$ is the velocity field, $p$ is the rescaled kinematic pressure, and
$\nu$ is the kinematic viscosity. For the forced simulations presented in section
\ref{section-forced}, a random time-correlated forcing term is added to equation
(\ref{eq:N-S-incompr-qdm}) to inject energy at large scales at a constant rate equal to
unity.

All three models are carried out with periodic boundary conditions. All simulations
based on \fluidsimsolver{ns3d} are carried out in a cubic domain of size $L_x = L_y =
L_z = 2\pi$ with $N^3$ grid points. We present two types of simulations: relatively
long runs of a few overturn times for physical analysis, and very short runs of a few
time steps for performance profiling using the CProfile tool. For long simulations, the
time step is computed adaptively based on a CFL condition, with the CFL coefficient
equal to 0.4 for second-order Runge-Kutta schemes and 1.0 for the fourth-order
Runge-Kutta scheme.

Both solvers employed in this study are pseudo-spectral codes based on Fast Fourier
Transforms (FFTs). For \fluidsimsolver{ns3d}, FFTs are performed via the Python library
Fluidfft \cite{fluidfft}, which delegates the actual FFT computation to optimized
libraries such as FFTW, FFTW\_MPI, PFFT, or P3DFFT. For this study, we use the Fluidfft
plugin \modpath{fluidfft-fftwmpi}, based on the library \modpath{libfftw_mpi}, which is
sufficient for the moderately-sized simulations on a few nodes presented here. The
dealiasing methods and the FFT library are orthogonal components in Fluidsim's
architecture, allowing advanced phase-shifting dealiasing methods to be combined with
any FFT library supporting high levels of MPI parallelism.

As discussed in section \ref{section-phase-shifting}, dealiasing methods based on
phase-shifting require modified time-stepping schemes. All schemes described and used
in this study are implemented in Python-Numpy code in the module
\hrefmodpath{https://foss.heptapod.net/fluiddyn/fluidsim/-/blob/branch/default/fluidsim/base/time_stepping/pseudo_spect.py}{fluidsim.base.time_stepping.pseudo_spect}
\footnote{The documentation for this module is available in the Fluidsim documentation
at
\url{https://fluidsim.readthedocs.io/en/latest/generated/fluidsim.base.time_stepping.pseudo_spect.html}}.
The numerical kernels (functions used during time stepping) utilize the meta-compiler
Transonic and the Python-Numpy compiler Pythran. These tools provide excellent code
readability while achieving performance comparable to \cpp. Moreover, efficient pure
\cpp code can be generated by transpiling these Python implementations, enabling their
use in native-language projects without Python runtime dependencies.

The Fluidsim package is fully implemented in Python-Numpy but achieves good overall
performance through the use of Transonic and Pythran for performance-critical sections.
For Fourier-based solvers, we will show that FFT operations account for a very large
fraction of the total computational cost (typically 80\%, see Section
\ref{section-decaying}), while the remaining time is spent in compiled and optimized
functions. Additional strengths of Fluidsim include its modular architecture,
comprehensive output capabilities (saving and plotting of multiple physical
quantities), robustness through extensive testing, ease of installation, and
user-friendly interface for standard fluid mechanics simulation tasks.

We present only direct numerical simulations (DNS) with standard molecular viscosity.
For DNS of turbulent flows, all scales of motion down to the smallest dissipative
scale—the Kolmogorov length scale $\eta$—must be resolved. The product $\kmax\eta$ is
commonly used to assess resolution quality, where $\kmax$ is the maximum resolved
wavenumber and $\eta$ the minimal Kolmogorov length scale. In practice, many studies
choose the grid resolution and Reynolds number such that $\kmax\eta\gtrsim 1$. The
exact requirement depends on the quantities of interest; larger values ($\kmax\eta= 2$,
4, or even 8) are necessary when the precise dynamics in the far dissipative range are
important.

The parameter $\kmax\eta$ is particularly relevant when studying aliasing and
dealiasing methods, as aliasing errors become more severe for smaller values of
$\kmax\eta$ (see Section~\ref{sec:kmaxetamin:alias:error}). Therefore, we examine
different values of $\kmax\eta$ ranging from 0.5 to 3 in our Taylor–Green simulations
to assess dealiasing performance under various resolution conditions. For the forced
HIT simulations, we consider a standard value of $\kmax\eta\simeq 2$, representative of
well-resolved turbulence.

\section{Description and examples of aliasing}


\label{section-aliasing}

In this section, we explain how the Fast Fourier Transform (FFT) method, based on
Discrete Fourier Transform (DFT) computation, generates aliasing errors. We illustrate
aliasing with examples and demonstrate its consequences on the spectral and physical
fields of simulations in one and three dimensions.

\subsection{Aliasing description}

Aliasing errors are introduced by the FFT method due to the discretization of signals,
the periodicity of the domain, and the orthogonality rule of the Fourier basis. If a
mode with a wavenumber greater than the Nyquist wavenumber $k_N = (2\pi/L)N/2$ is
created in the equations, it will alias to a wavenumber within $\mathcal{K} = [\delta
k-k_N, k_N]$ (with $\delta k = 2\pi/L$ the wavenumber spacing), as depicted in the
Supplementary material \cite{supp} or in books \cite[p.32-41]{Canuto1988} and
\cite[p.434-442]{Canuto2007} for a detailed mathematical explanation of aliasing errors
generated for a single field.

Here, we present the generation of aliasing errors due to nonlinear terms in the
Navier-Stokes equations, i.e.\ quadratic terms of the form $u(x)v(x)$. These terms
create modes with wavenumbers higher than the span of the wavenumber range
\cite{PattersonOrszag1971,Rogallo1981}. For simplicity, we focus on the one-dimensional
case, but the generalization to the three-dimensional case is straightforward. In a
numerical discretization using $N$ equally spaced grid points at positions $x \in [0,
L]$, the discrete Fourier transform represents only the wavenumbers $k \in
\mathcal{K}$. Consider the field $u(x)$. Its discrete Fourier transform is given by:

\begin{equation}
\hat{u}(k_n) = \frac{1}{L} \int_{0}^{L} u(x) e^{-i k_n x}dx,
\label{eq:u_fft_1d}
\end{equation}
and the corresponding inverse Fourier transform is given by:
\begin{equation}
u(x) = \sum_{k_n \in \mathcal{K}} \hat{u}(k_n) e^{i k_n x},
\label{eq:u_expansion_1d}
\end{equation}
where $\hat{u}(k_n)$ is the discrete Fourier coefficient of $u$ at wavenumber $k_n$.
Note that an integral notation is used in Eq.~\eqref{eq:u_fft_1d} for compactness; this
integral is understood as a finite sum over grid points, $\int_0^L (\cdot)\,dx \equiv
\delta x \sum_{j=0}^{N-1} (\cdot)|_{x_j}$, with $\delta x = L/N$. The nonlinear product
of field $u$ with another field $v$ can be written as:

\begin{equation}
w(x) = u(x) v(x)
     = \sum_{k_n \in \mathcal{K}} \sum_{k_m \in \mathcal{K}}
       \hat{u}(k_n) \hat{v}(k_m) e^{i(k_n + k_m)x}.
\label{eq:w_product_convo}
\end{equation}
In practice, these convolution sums, which require $(N^d)^2$ operations in $d$
dimensions, are not computed. Instead, the pseudo-spectral method computes two inverse
FFTs (IFFTs) to obtain $u$ and $v$, then computes the product in physical space $w(x) =
u(x)v(x)$, and finally performs one FFT on $w$ to return to spectral space. This
process requires three Fourier transforms, resulting in an order of $3N^d \log N^d$
operations.

However, this is mathematically equivalent to calculating the convolution sums, which
is necessary to understand the origin of aliasing. The corresponding discrete Fourier
coefficients of the field $w$ are obtained from the discrete transform of the product:

\begin{equation}
\hat{w}(k) = \frac{1}{L} \int_{0}^{L} \sum_{k_n \in \mathcal{K}} \sum_{k_m \in \mathcal{K}}
\hat{u}(k_n) \hat{v}(k_m) e^{i(k_n + k_m - k)x}dx = \sum_{k_n \in \mathcal{K}} \sum_{k_m \in \mathcal{K}}
\hat{u}(k_n) \hat{v}(k_m) \frac{1}{L} \int_{0}^{L} e^{i(k_n + k_m - k)x}dx.
\label{eq:discrete_transform_w}
\end{equation}
Using $2k_N = (2\pi/L)N$ and the orthogonality property of the Fourier basis
\cite{Canuto1988,Canuto2007}, we obtain:

\begin{equation}
\frac{1}{L} \int_{0}^{L} e^{i(k_n + k_m - k)x}dx =
\begin{cases}
1, & \text{if } k_n + k_m = k + q2k_N, \quad q \in \mathbb{Z},\\[4pt]
0, & \text{otherwise}
\end{cases}
\label{eq:orthogonality}
\end{equation}
or equivalently,

\begin{equation}
\frac{1}{L} \int_{0}^{L} e^{i(k_n + k_m - k)x}dx = \delta[k_n + k_m - k \bmod 2k_N].
\label{eq:Kronecker}
\end{equation}
To verify this: when $k_n + k_m - k = q(2\pi/L)N$ for some $q \in \mathbb{Z}$,
substituting the grid positions $x_j = j\,\delta x$ with $\delta x = L/N$ gives
$e^{i(k_n + k_m - k)x_j} = e^{i 2\pi q j} = 1$ for all $j$, so the discrete sum equals
$L$ as expected. Eq.~\eqref{eq:discrete_transform_w} therefore becomes:

\begin{equation}
\hat{w}(k)
= \sum_{k_n \in \mathcal{K}} \sum_{k_m \in \mathcal{K}}
  \hat{u}(k_n) \hat{v}(k_m) \,
  \delta[k_n + k_m - k \bmod 2k_N].
\label{eq:rogallo_convolution}
\end{equation}
Note that since $k_n \in \mathcal{K}$ and $k_m \in \mathcal{K}$, then $k_n + k_m \in
2\mathcal{K}$, meaning that it cannot be greater than $2k_N$ or less than $-2k_N$. This
implies that $q = \{0, \pm 1\}$ for Eq.~\eqref{eq:orthogonality}.

Finally, writing the double sum in Eq.~\eqref{eq:rogallo_convolution} as a single sum
over the combined index $\knl = k_n + k_m$, and using the fact that in
Eq.~\eqref{eq:orthogonality} the non-aliased part corresponds to $q = 0$ and the
aliased part to $q = \pm 1$, the Fourier coefficient of the nonlinear product can be
rewritten by separating the aliased and non-aliased contributions as:
\begin{equation}
\hat{w}(k) = \underbrace{\sum_{\substack{k_n + k_m = k, \\ k \in \mathcal{K}}} \hat{u}(k_n) \hat{v}(k_m)}_{\hat{w}_{\mathrm{clean}}(k)}
           + \underbrace{\sum_{\substack{k_n + k_m = k \pm 2k_N, \\ k \in \mathcal{K}}} \hat{u}(k_n) \hat{v}(k_m)}_{\hat{w}_{\mathrm{alias}}(k)},
\label{eq:rogallo_separated}
\end{equation}

\begin{figure}[H]
\centerline{\includegraphics[width=1.8\figwidth]{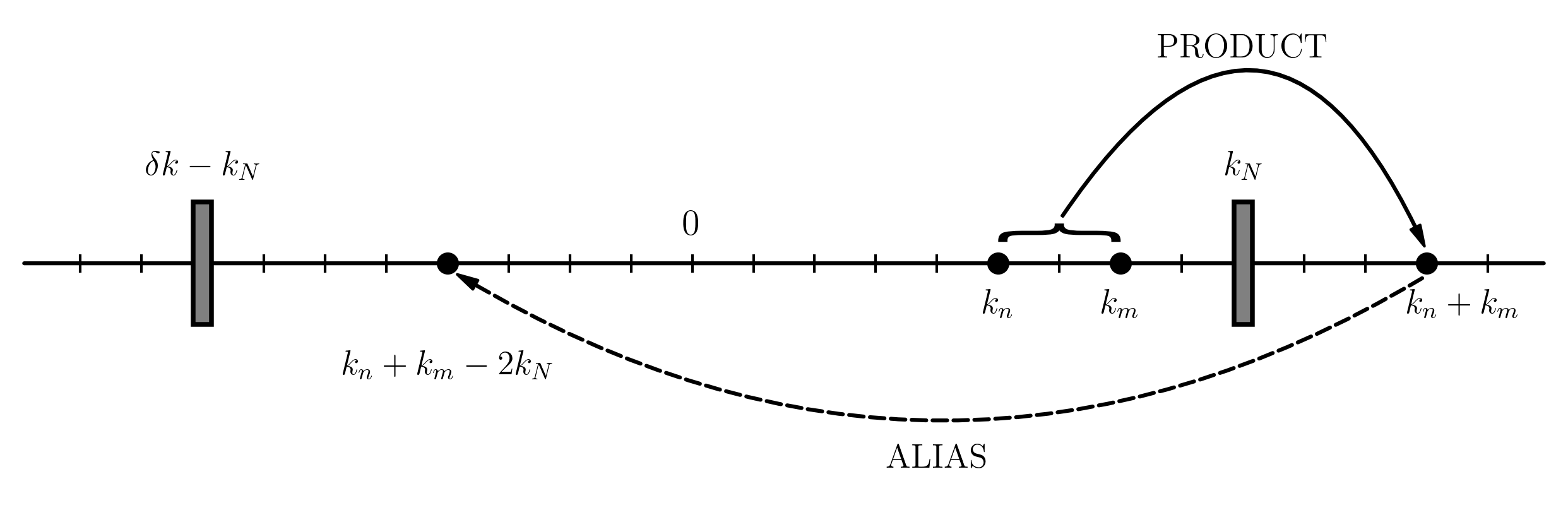}}
\caption{Illustration of aliasing generated by nonlinear terms in one-dimensional
spectral space, adapted from \cite{Rogallo1981}, with $2k_N = (2\pi / L) N$ and $k_m +
k_n = \knl$. Grey patches represents the domain limits. \label{fig:illu_alias_1d}}
\end{figure}

Eq.~\eqref{eq:rogallo_convolution} and Eq.~\eqref{eq:rogallo_separated} correspond to a
circular convolution in discrete Fourier space and make the aliasing mechanism
explicit: when sampled at $N$ points, any nonlinear contribution $\knl = k_n + k_m$
that lies outside the resolved range $\mathcal{K}$ is indistinguishable from the mode
within the resolved range at a distance of a multiple of $2k_N$ from $\knl$. In other
words, the mode $\knl$ is folded back into the resolved range $\mathcal{K}$ through the
modulo operation at $k_{\mathrm{alias}} = \knl - 2k_N$ when $k_n + k_m > 2k_N$ (see
Fig.~\ref{fig:illu_alias_1d}) and at $k_{\mathrm{alias}} = \knl + 2k_N$ when $k_n + k_m
< -2k_N$. This aliasing therefore adds energy from the nonlinear mode $\knl$ to the
$k_{\mathrm{alias}}$ mode, which already contains energy, creating purely numerical
noise.

\subsection{Standard dealiasing with truncation}

\begin{figure}[]
\centerline{
  \includegraphics[width=0.9\figwidth]{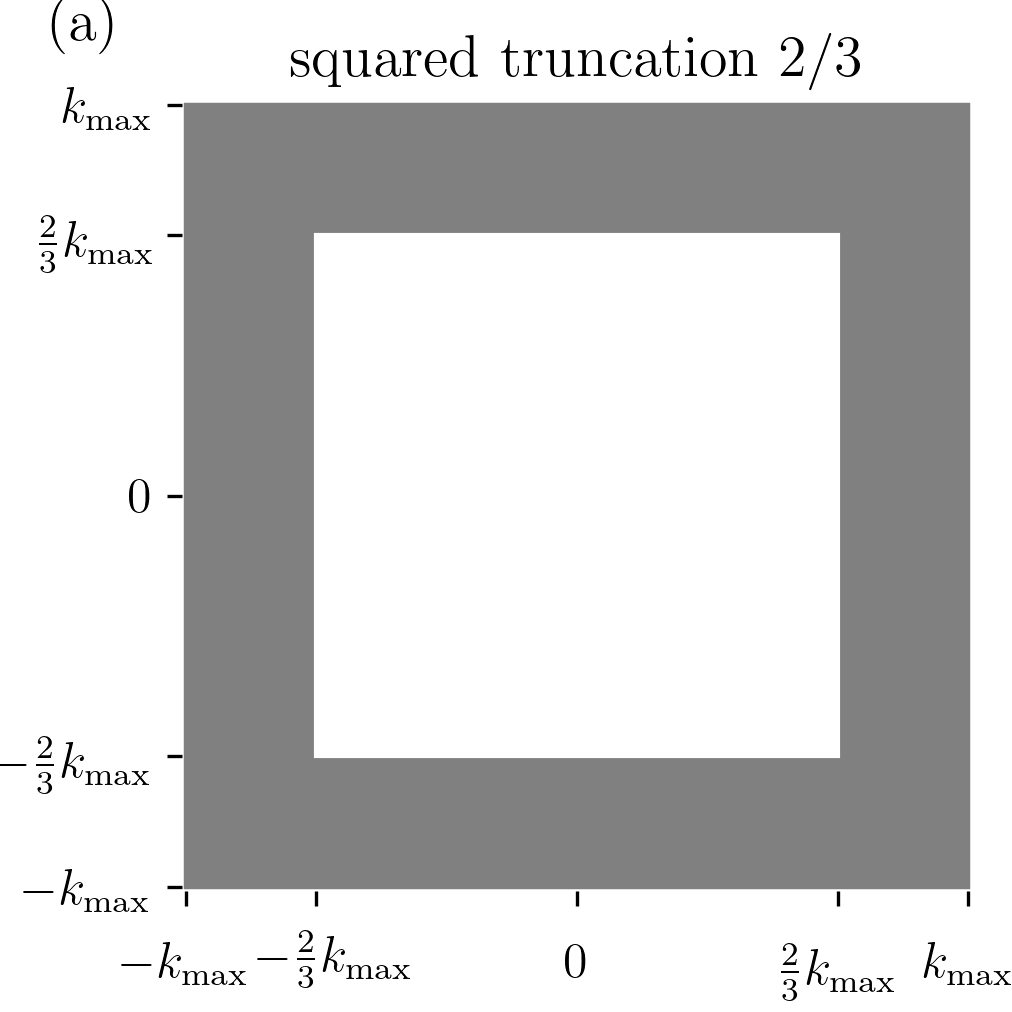}
  \includegraphics[width=0.9\figwidth]{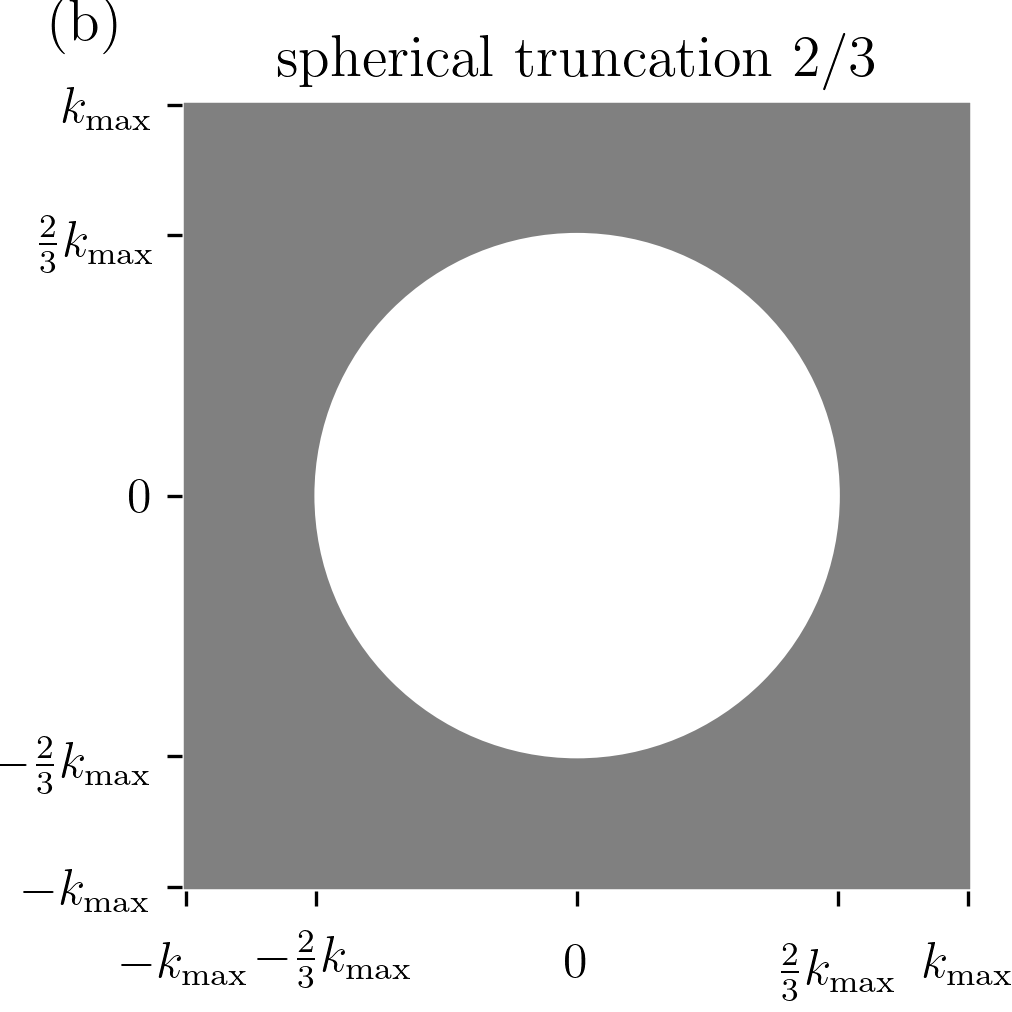}
}
\caption{Common truncations used without phase shifting. Modes are set to zero in the
gray regions. In Fluidsim, (a) is referred to as ``cubic'' and (b) as ``spherical''.
\label{fig:truncation:common}}
\end{figure}

The standard approach to dealiasing in pseudo-spectral methods is to set to zero all
Fourier modes whose wavenumber exceeds a threshold $\kmax$, chosen so that any
nonlinear interaction between retained modes cannot produce an aliased wavenumber
within the resolved range. Two common choices are illustrated in
Fig.~\ref{fig:truncation:common}. The cubic truncation,
Fig.~\ref{fig:truncation:common}(a), zeroes all modes for which any single wavenumber
component satisfies $|k_i| \geq (2/3)k_N$; it retains $30\%$ of all modes (in 3D). The
spherical truncation, Fig.~\ref{fig:truncation:common}(b), instead zeroes all modes
with $|\kvec| \geq (2/3)k_N$; it is isotropic at small scales
\cite{PattersonOrszag1971} and retains only $16\%$ of the modes (in 3D), but provides a
more uniform spatial resolution in all directions.

Throughout this study, we use the spherical truncation exclusively, parameterized by a
truncation coefficient $C_t$ defined such that $\kmax = C_t k_N$. The standard
dealiasing choice corresponds to $C_t = 2/3$, and we will see that phase-shifting
methods allow $C_t$ to be increased beyond this value while still controlling aliasing
errors.

To illustrate the aliasing mechanisms concretely, we first consider a simple
one-dimensional nonlinear model before turning to three-dimensional turbulence.

\subsection{Example: quadratic model in 1D}\label{sec:sign-function}

To illustrate the consequences of aliasing errors, we first consider one of the
simplest nonlinear models:
\begin{equation}
\p_t S(x, t) = - \sign(S) S^2,
\label{eq:model-1d}
\end{equation}
implemented in the package \fluidsimsolver{nl1d}. For simplicity, the length of the
domain is taken as $2\pi$ so that wavenumbers are integers. The initial condition $S(x,
t=0)$ is chosen such that $S(x,t=0)>0$ for all $x\in[0,2\pi]$. Since $S$ is positive,
$\text{sign}(S) = 1$, meaning that the 1D model from Eq.~\eqref{eq:model-1d} becomes
simply:
\begin{equation}
\p_t S(x, t) = -S^2,
\label{eq:model-1d-2}
\end{equation}
and the exact solution is:
\begin{equation}
S_\text{exact}(x, t) = \frac{S(x, t=0)}{1 + S(x, t=0) t}.
\label{eq:1d_exact}
\end{equation}
We see that the solution remains positive so that only Eq.~\eqref{eq:model-1d-2} needs
to be considered. The second derivative is $\p_{tt} S(x, t) = 2S^3$ and, after one RK2
time step, when retaining terms of order $dt^2$ minimum, the state becomes $S(x, t=dt)
= S(x, t=0) - dtS(x, t=0)^2 + dt^2S(x, t=0)^3$.

In order for aliases to be clearly identified after advancing in time, we consider an
initial condition $\hat{S}(k,t=0)$ with a simple frequency content:
\begin{equation}
S(x,t=0) = 1 + 0.7 \cos(k_0 x),
\label{eq:1d_initial}
\end{equation}
where $k_0 = 10$. Since $S(x, t=0)^2$ and $S(x, t=0)^3$ contain terms in $\cos(2k_0x)$
and $\cos(3k_0x)$, the nonlinearity creates two new frequencies $k_{nl1} = 2k_0 = 20$
and $k_{nl2} = 3k_0 = 30$. The second harmonic is generated due to the interaction
between the harmonic $2k_0$ and the initial peak $k_0$. Remarkably, in contrast to the
Euler scheme, RK2 schemes take into account this interaction in one time step.

\begin{figure}[]
\includegraphics[width=2\figwidth]{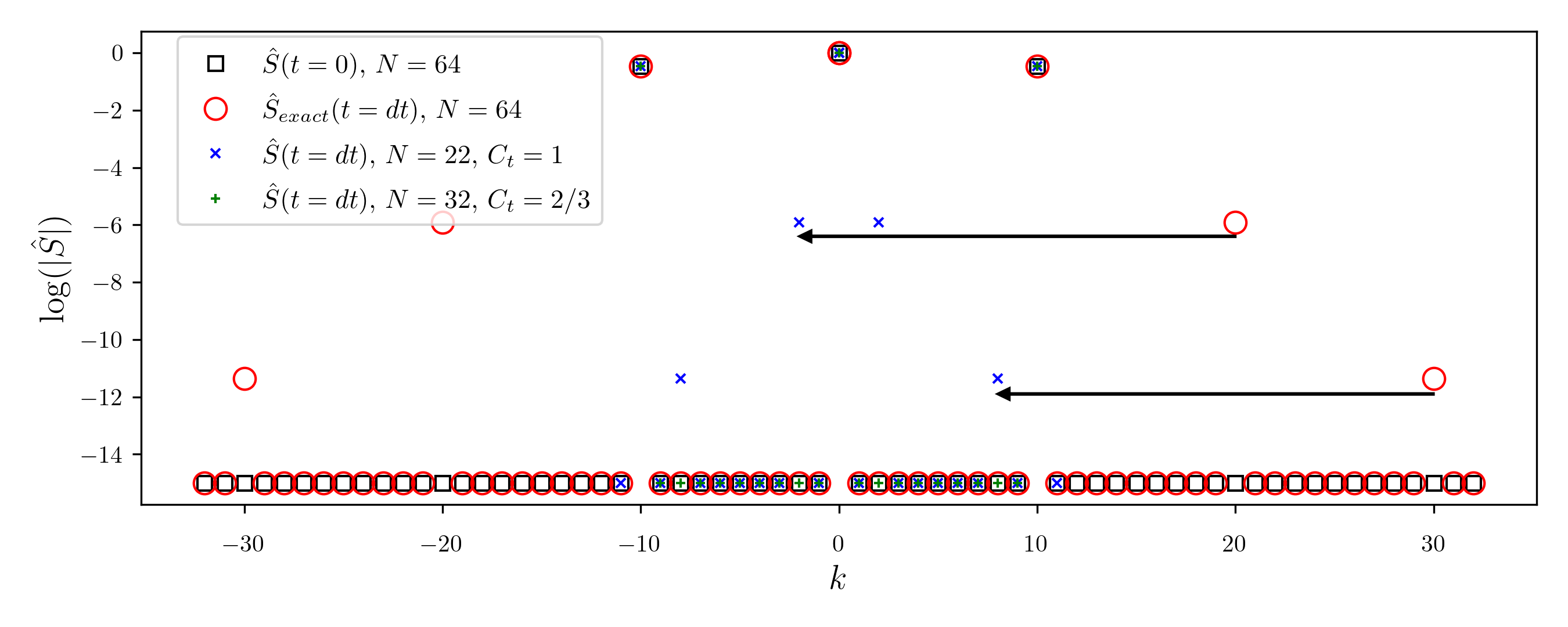}
\caption{Numerical and exact solutions of Eq.~\eqref{eq:model-1d} in spectral space
after one RK2 scheme time step. The arrows indicate how unresolved modes are aliased
(see Fig.~\ref{fig:illu_alias_1d}). \label{fig:1d-alias}}
\end{figure}

We then consider different numerical setups with different numbers of grid points $N$,
corresponding to different values of $k_N$. Fig.~\ref{fig:1d-alias} represents
$\log(|\hat S|)$ for $t = 0$ (squares) and after one time step at $t = dt$. For the
exact solution (circles), there is signal at the nonlinear wavenumbers $k_{nl1}$ and
$k_{nl2}$. We now consider what happens after one RK2 step for an under-resolved case
with $N = 22$, without truncation ($C_t = 1$). Since $k_N = 11$, both nonlinear
wavenumbers are not resolved and are aliased at $k_{\mathrm{alias,1}} = \pm (k_{nl1} -
2 k_N) = \pm 2$ and $k_{\mathrm{alias,2}} = \pm (k_{nl2} - 2 k_N) = \pm 8$. Finally, we
consider a case dealiased with standard 2/3 truncation with about the same $\kmax$.
Thus, the number of points is increased to $N = 32$ and the truncation coefficient
decreased to $2/3$, which gives a maximum wavenumber of $\kmax = 10$ but with $k_N =
16$. Thus, $k_{\mathrm{alias,1}} = \pm 12$, which is truncated so that all aliases are
removed.

For real-valued fields, $\hat{S}(-k) = \hat{S}^*(k)$, so only positive wavenumbers need
to be stored. In this case, a nonlinear mode that would alias into the negative range
instead folds onto its mirror wavenumber: $k_{\mathrm{alias}} = 2k_N - \knl$ rather
than $\knl - 2k_N$, and the alias appears in the stored positive-wavenumber spectrum.

\subsection{3D flow: Taylor-Green vortices}\label{sec:T-G_vortices}

We now turn to the study of a real flow: the transition to turbulence of horizontal
Taylor-Green vortices. The Taylor–Green vortex is an unsteady and exact solution of the
incompressible Navier–Stokes equations derived by \citet{Taylor-Green}. It represents a
simple yet fundamental configuration for studying vortex dynamics in periodic domains.
For sufficiently large Reynolds number, the flow is unstable and transitions to
turbulence, leading to energy dissipation and the decay of the flow, which eventually
vanishes \cite{Taylor-Green,Debonis2013}. The corresponding solution, without its
temporally decaying component and with zero vertical velocity (i.e., horizontal
Taylor–Green vortices), is used as the initial condition in the Taylor–Green
simulations presented in this paper. The horizontal components of the initial velocity
are:

\begin{subequations} \label{eq:T-G-init}
\begin{alignat}{2}
v_x & = \sin(2\pi x / L_x) \cos(2\pi y / L_y) \cos(2\pi z / L_z), \label{eq:T-G-vx}\\
v_y & = -\cos(2\pi x / L_x) \sin(2\pi y / L_y) \cos(2\pi z / L_z). \label{eq:T-G-vy}
\end{alignat}
\end{subequations}

\begin{table}[]
\centering \begin{tabular}{| c | c | c | c | c | c | c | c |}
\toprule
$N$ & $Re$ & scheme & $C_t$ & $k_{\max} \tilde\eta$ & number of proc. & CPU.h speedup & error (\%) \\
\midrule
192 & 1600 & RK4 & 2/3 & 0.75 & 32 & 1.0 & 0.0 \\
128 & 1600 & RK4 & 1 & 0.75 & 32 & 5.0 & 25.43 \\
\bottomrule
\end{tabular}
 \caption{List of the
Taylor-Green simulations carried out for sub-section \ref{sec:T-G_vortices}. Error is
defined by Eq.~\eqref{eq:quality} as a percentage relative to the reference simulation
fully dealiased with truncation (first row).}\label{tab:192}
\end{table}

\begin{figure}[]
\centerline{\includegraphics[width=\figwidth]{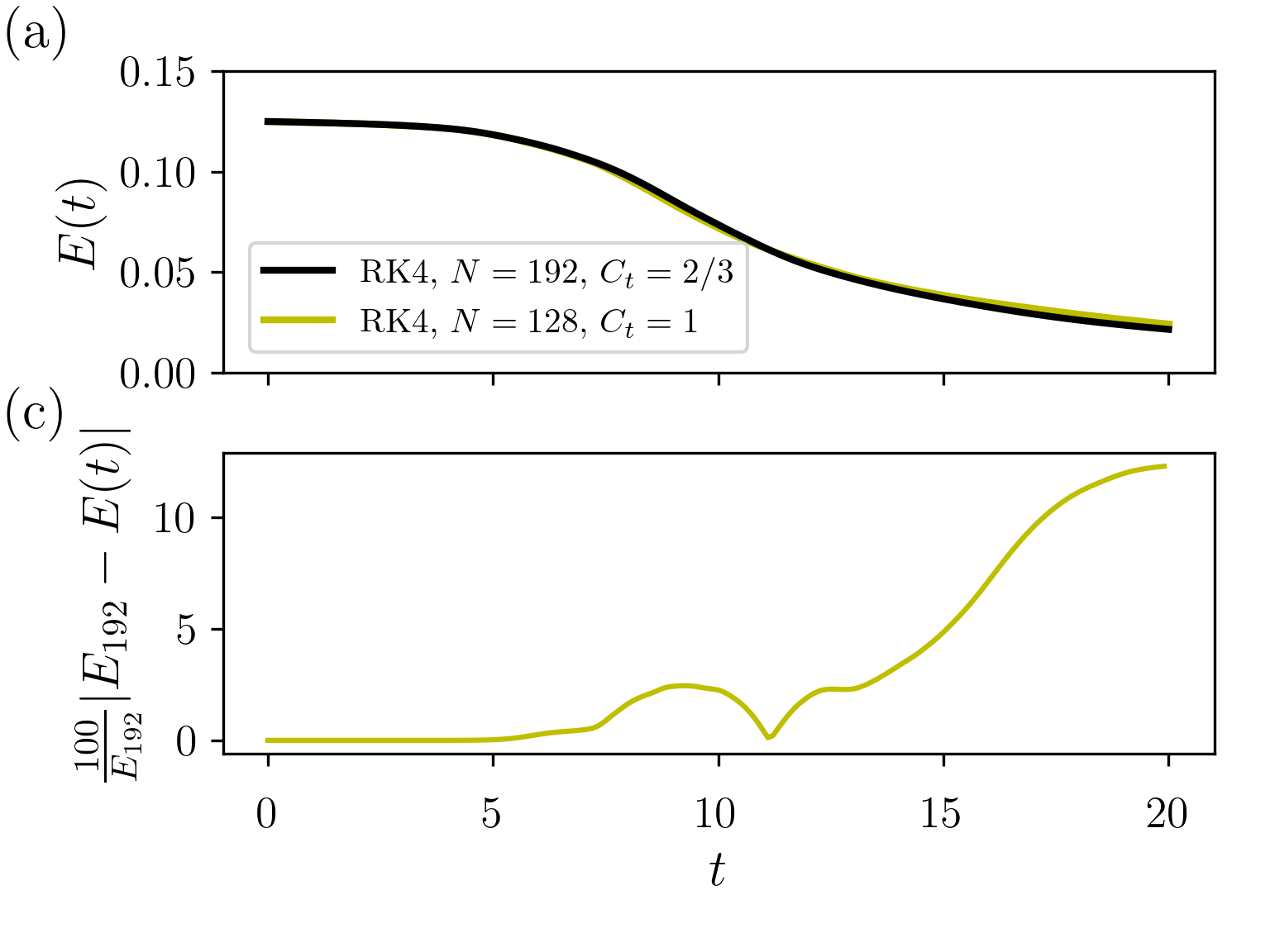}
  \includegraphics[width=\figwidth]{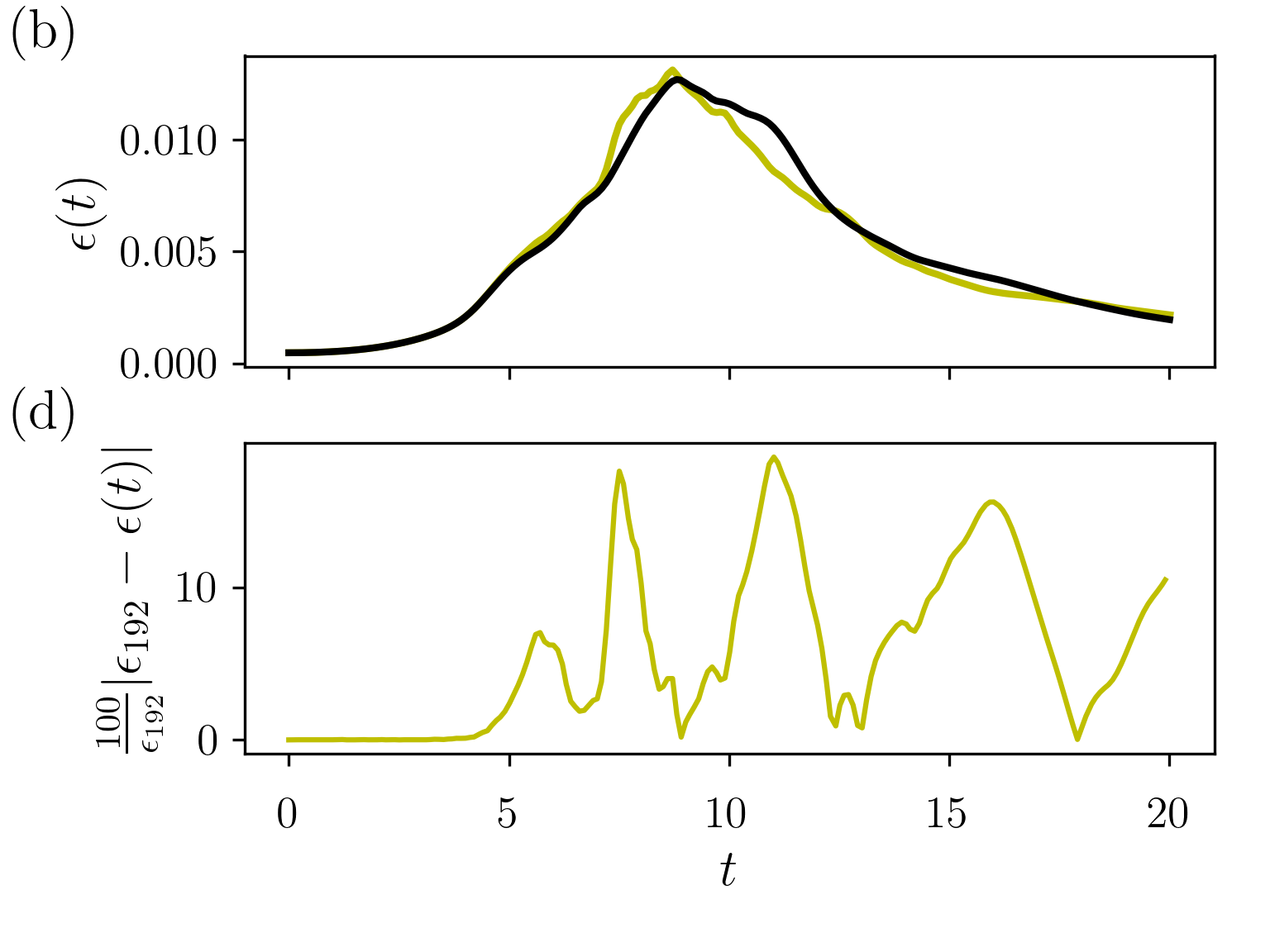} }
\caption{Temporal evolution of (a) total kinetic energy $E(t)$ (b) energy dissipation,
(c) energy relative error, and (d) energy dissipation relative error for dealiased
($C_t=2/3$, black curves) and aliased ($C_t=1$, light curves) simulations at $\kmax
\etamin \simeq 0.7$ (see Table~\ref{tab:192}). \label{fig:spatialmeans:192} }
\end{figure}

\begin{figure}[]
\centerline{
  \includegraphics[width=\figwidth]{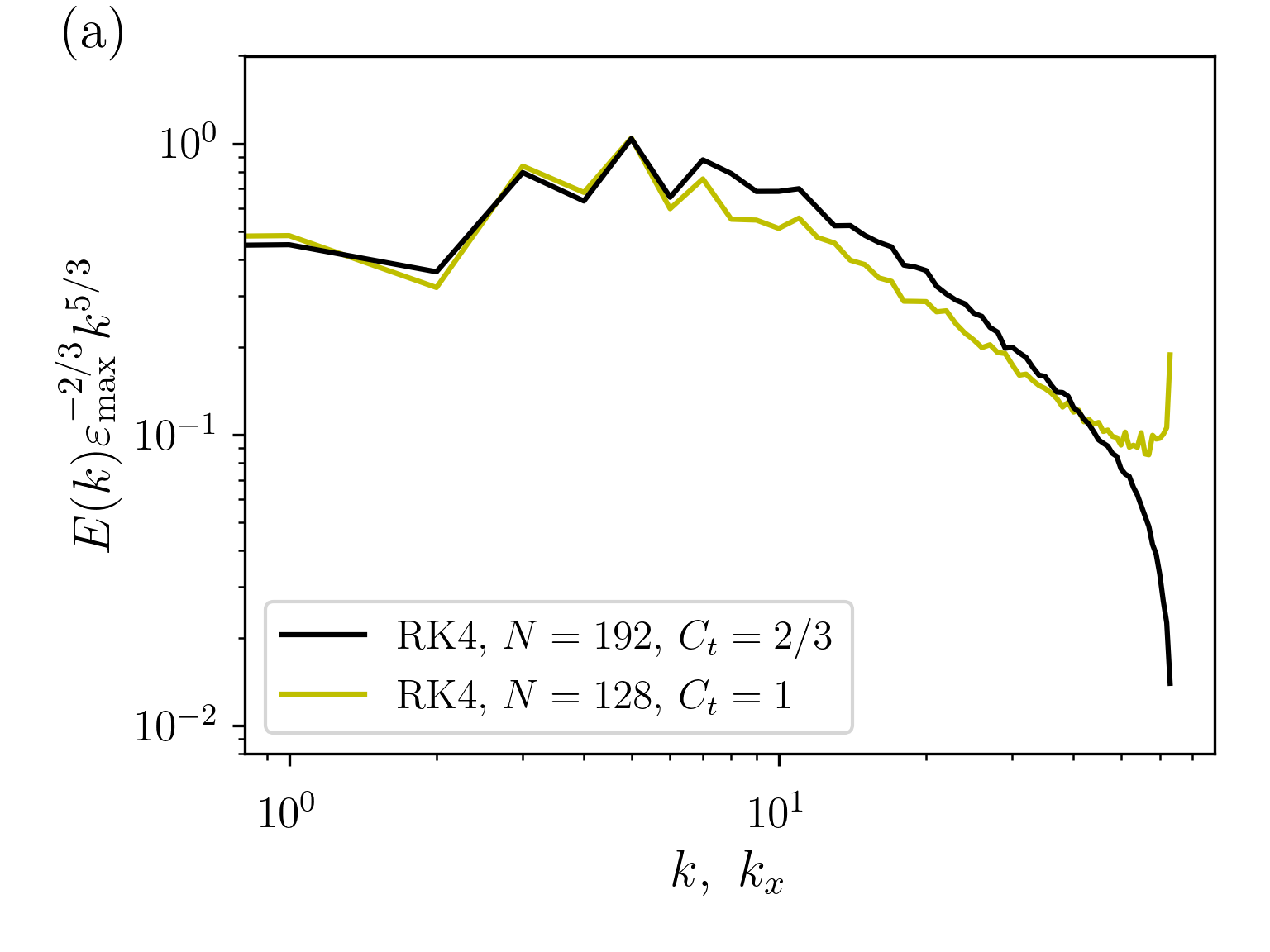}
  \includegraphics[width=\figwidth]{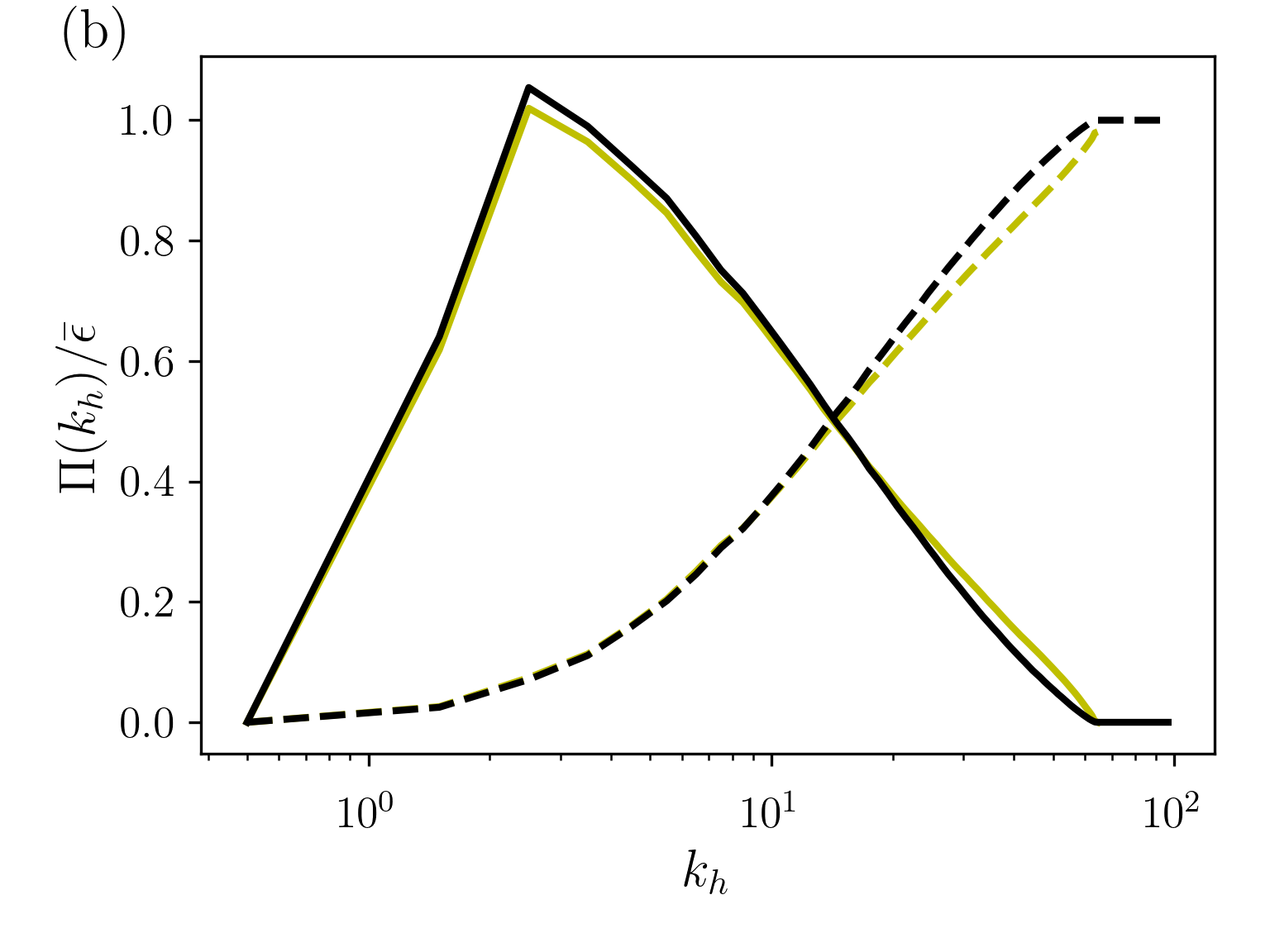}
}
\caption{Spectral statistics averaged between $t=9$ and $t=14$ for dealiased (black
curves) and aliased (light curves) simulations for $\kmax \etamin \simeq 0.7$ (see
Table~\ref{tab:192}). (a) Compensated 1D energy spectra versus wavenumber modulus
$k_x$. (b) Nonlinear energy flux (solid lines) and cumulated dissipation energy (dashed
lines). \label{fig:spectra:192}}
\end{figure}

\begin{figure}[]
\centerline{
  \includegraphics[width=1.15\figwidth]{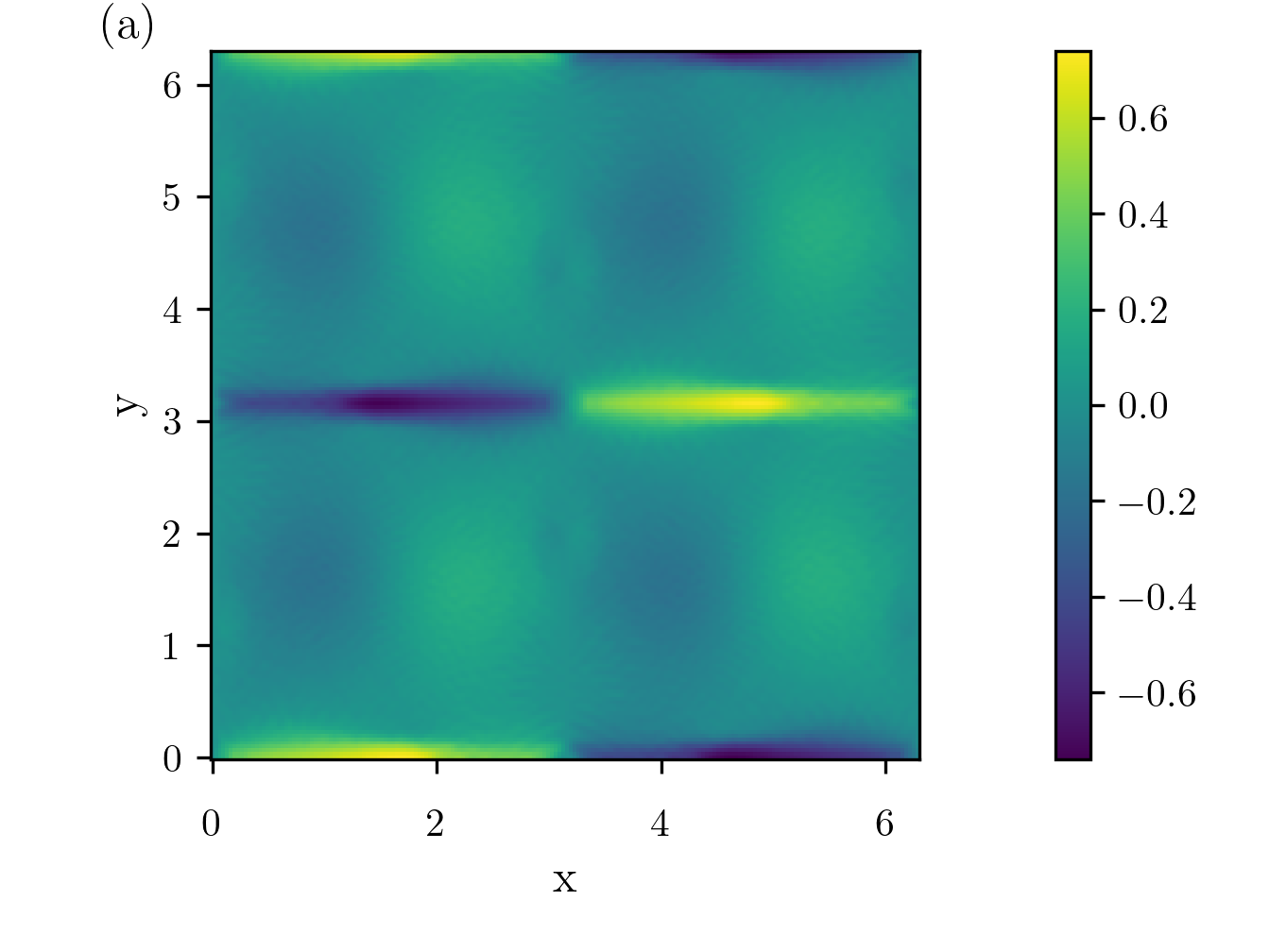}
  \includegraphics[width=1.15\figwidth]{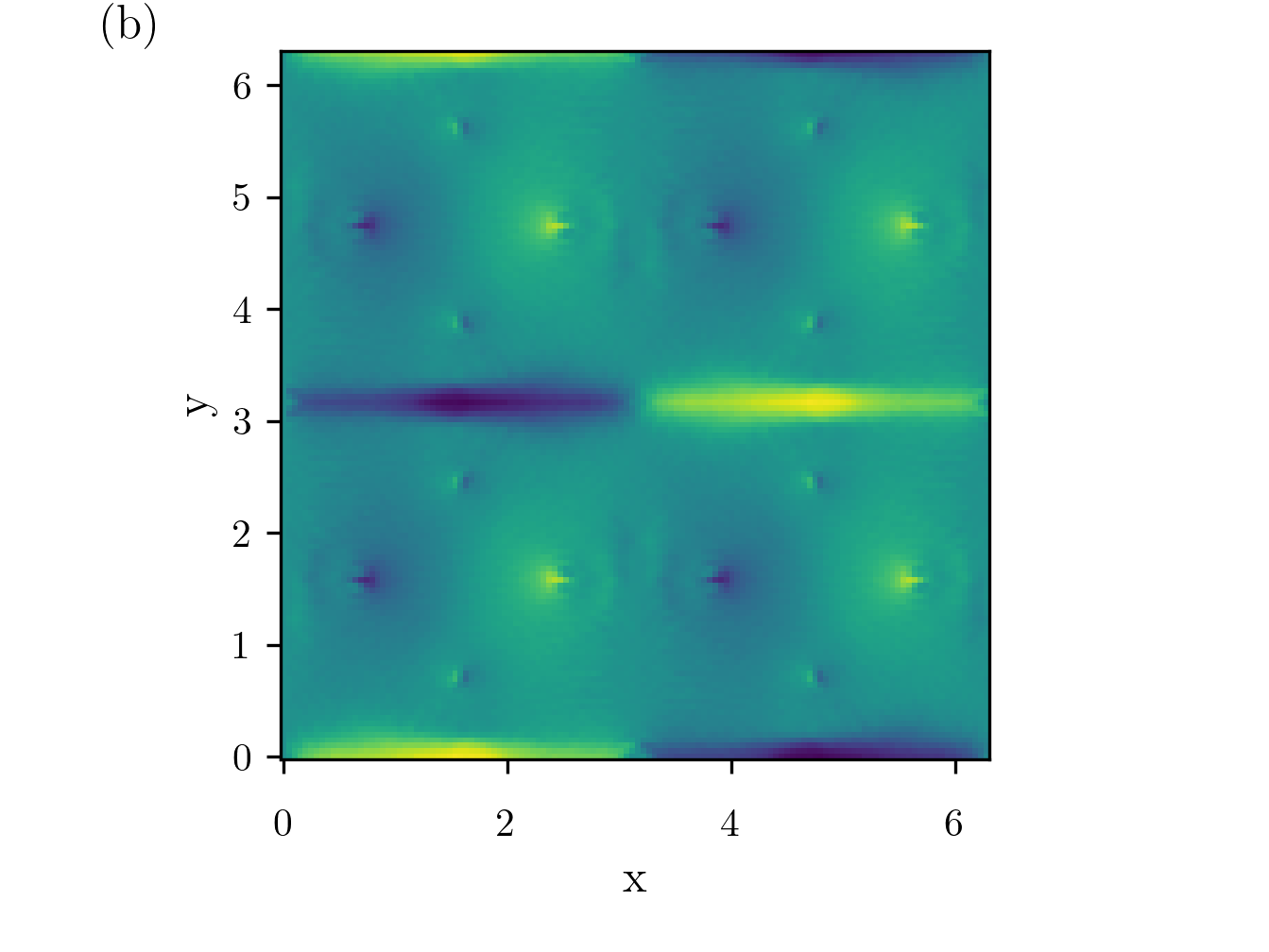}
}
\caption{Cross-section of the velocity field component $v_x$ at $t=9$ (close to the
peak of dissipation) for the simulations (a) without aliasing and (b) with aliases (see
Table~\ref{tab:192}).}\label{fig:phys-fields:192}
\end{figure}

We first consider two simulations (using the RK4 scheme) with the same $\kmax\etamin$
(i) fully dealiased with 2/3 truncation ($N = 192$) and (ii) with aliasing errors ($N =
128$). This yields a relatively small value of $\kmax\etamin=0.75$ so that dealiasing
errors are large. These simulations are presented in Table~\ref{tab:192} and in
Figs.~\ref{fig:spatialmeans:192}, \ref{fig:spectra:192} and \ref{fig:phys-fields:192}.

Fig.~\ref{fig:spatialmeans:192} shows the temporal evolution of the total energy and
energy dissipation for both resolutions, as well as the corresponding errors in the
aliased case relative to the dealiased case. The figure shows differences, and thus
errors in the aliased case, particularly at the transition to turbulence, i.e., when
nonlinear terms induce the energy cascade to smaller scales. The minimum (in time)
Kolmogorov scale is computed from the maximum (in time) mean energy dissipation
$\varepsilon_{\max}$ as $\etamin = (\nu^3 /\varepsilon_{\max})^{(1/4)}$. Note that the
time-average dissipation is approximately 0.6 times smaller than $\varepsilon_{\max}$
so that the average Kolmogorov length scale would be 1.14 times larger than $\etamin$.

Fig.~\ref{fig:spectra:192} shows the corresponding (a) energy spectra and (b) nonlinear
energy fluxes and cumulative dissipation in both cases, averaged over $9<t<14$,
i.e.\ when turbulence is strongest. The cumulative dissipation and the nonlinear
energy flux are the running integral over wavenumbers of the local spectral dissipation
rate and the nonlinear transfert, respectively. The former is zero at small
wavenumbers, rises through the dissipative range, and reaches the total mean
dissipation rate at the largest wavenumbers.

We first examine the reference simulation without aliasing (black curves). The
compensated spectra in Fig.~\ref{fig:spectra:192}(a) are roughly constant for $4 < k <
10$, but no clear power-law scaling or inertial range is apparent. This is consistent
with the moderate Reynolds number $Re = 1600$ typical of Taylor-Green transition
studies, at which scale separation between the energy-containing and dissipative scales
is limited. Accordingly, the nonlinear flux in Fig.~\ref{fig:spectra:192}(b) shows no
plateau, and its decrease mirrors the rise of the cumulative dissipation, confirming
the absence of a clear separation between the scales excited during the Taylor-Green
instability and the dissipative range.

Note that the wavenumber range differs between the two panels. In
Fig.~\ref{fig:spectra:192}(a) the spectra extend to $\kmax = 64$, while
Fig.~\ref{fig:spectra:192}(b) extends to $k_N = \kmax / C_t = 96$. The range $64 < k <
96$ therefore corresponds to the truncated modes: the flux and cumulative dissipation
are constant there, making the truncation directly visible.

Turning to the aliased simulation (light curves), the differences from the reference
are pronounced, especially in the inertial range and at the smallest scales near
$\kmax$. These spectral differences are consistent with the errors observed in the
total energy and dissipation rate, and explain why aliasing errors grow most strongly
at the transition to turbulence, when the inertial range expands and the nonlinear
interactions intensify.

A cross-section of one component of the velocity field for $t=9$, i.e.\ just before
the transition to turbulence, is shown in Fig.~\ref{fig:phys-fields:192}. We see for
the aliased simulation (Fig.~\ref{fig:phys-fields:192}(b)) regularly spaced small dots
corresponding to numerical errors.

To quantify the error between a given simulation and a reference simulation without
aliasing, we define an error index based on the 1D energy spectra, i.e.\ the
spectra as functions of $k_x$, $k_y$, and $k_z$. Taking $E_{\mathrm{ref}}$ as the
reference energy spectrum, we define the one-dimensional continuous error as:
\begin{subequations} \label{eq:quality}
\begin{alignat}{2}
\mathrm{Error}_x = \frac{1}{(t_{end} - t_{start})\int d\log(k_x)}
  \int_{t_{start}}^{t_{end}}dt\int d\log(k_x) \frac{100 |E(k_x,t)-E_{ref}(k_x,t)|}{E_{ref}(k_x,t)}.
\end{alignat}
\end{subequations}
The corresponding discrete form is expressed in Appendix~\ref{anx:annex_0}. The error
index reported in the tables throughout this article is the mean of the three
quantities from Eq.~\eqref{eq:quality} over all spatial directions. The time
$t_{\mathrm{start}}$ is taken before the energy dissipation peak at $9$ overturn times,
and $t_{\mathrm{end}} = t_{\mathrm{start}} + 5$ overturn times, after the peak. The
integration over $\log(k_x)$ rather than $k_x$ ensures that each decade of wavenumbers
contributes equally, avoiding undue weight on the largest scales. The results are
presented in Table~\ref{tab:192}, with a significant error of $25.43\%$, illustrating
the necessity of dealiasing methods to remove these errors.

Table~\ref{tab:192} also displays speedup values, defined throughout this article as
the total simulation time of the reference simulation divided by the total simulation
time of the given simulation. The aliased simulation at $N = 128$ achieves a speedup of
$5$ relative to the dealiased one at $N = 192$ for the same $\kmax$. With a computation
time expected to scale as $N^4$ (this will be discussed later), the computational
speedup is expected to be $(192/128)^4 \approx 5$ which is consistent with our results.
This means that approximately 80\% of the computational cost of the dealiased
simulation is due to the larger grid size required for dealiasing, illustrating the
value of developing dealiasing methods that reduce the domain size for the same
$\kmax$, such as the phase-shifting method.

\section{Description of phase shifting methods}


\label{section-phase-shifting}

In this section, we describe how phase-shifting methods work. These methods modify the
time-stepping algorithm by evaluating some quantities on a phase-shifted grid, so that
aliasing contributions cancel when the results are combined. Our goal is to provide a
full explanation of the phase-shifting algorithms used in pseudo-spectral codes solving
the Navier-Stokes equations. Because these algorithms involve interactions with
different truncation geometries, multi-stage time-stepping schemes, and randomization
of the phase shift, we first build intuition in the simpler setting of problems
depending on a single variable.

\subsection{Phase shifting for one-dimensional problems}

\subsubsection{General description}

We consider equations that can be written in spectral space as:
\begin{equation}
\p_t S = \sigma S + \mathcal{N}(S),
\label{eq:base-eq}
\end{equation}
where $S$ represents the state of the system (potentially containing several
variables), $\sigma(\kvec)$ is a real or complex coefficient depending on the
wavenumber, and $\mathcal{N}(S)$ is an operator with at most second-order
nonlinearities. The incompressible Navier-Stokes equations \eqref{eq:N-S-incompr} fall
into this form, with $\sigma(\kvec) = -\nu k^2$ and $\mathcal{N}(S)$ corresponding to
the projection of the advective term $-\uu\cdot\bnabla\uu$ onto the divergence-free
subspace in Fourier space.

We consider explicit time-stepping schemes for which the linear term $\sigma S$ is
integrated exactly and which are built on stages based on approximations that can be
written as
\begin{equation}
\p_t \log(S) = \sigma + \frac{\mathcal{N}(S)}{S} \simeq \sigma + \frac{\mathcal{N}_a}{S_a},
\end{equation}
where $S_a$ is the approximate state and $\mathcal{N}_a$ the approximate nonlinear
term. Integrating over a sub-step duration $\tau$ from $t_0$ to $t_0 + \tau$ gives:
\begin{equation}
\frac{S(t_0+\tau)}{S_0 e^{\sigma \tau}} = 1 + \frac{\mathcal{N}_a}{S_a} \tau,
\end{equation}
or, retaining only first-order terms in $\tau$:
\begin{equation}
S(t_0+\tau) = (S_0 + \mathcal{N}_a \tau) e^{\sigma \tau}.
\end{equation}
The time schemes considered in this study differ by their number of stages (which
determines their order) and by the choice of $\tau$, $S_a$ and $\mathcal{N}_a$ at each
stage. For example, the simple forward Euler scheme corresponds to one stage with $\tau
= \dt$, $S_a = S_0 \equiv S(t_0)$, and $\mathcal{N}_a = \mathcal{N}_0 \equiv
\mathcal{N}(S_0)$. The standard Runge-Kutta 2 (RK2) scheme corresponds to two stages:
for the first stage, $\tau = \dt/2$, $S_a = S_0$, and $\mathcal{N}_a = \mathcal{N}_0$;
for the second stage, $\tau = \dt$, $S_a = S_1$, and $\mathcal{N}_a = \mathcal{N}_1 =
\mathcal{N}(S_1)$, where $S_1$ is the result of the first stage. All schemes
implemented in Fluidsim are presented in detail in the
\href{https://fluidsim.readthedocs.io/en/latest/generated/fluidsim.base.time_stepping.pseudo_spect.html}{pseudo-spectral
time stepping documentation}
\footnote{\url{https://fluidsim.readthedocs.io/en/latest/generated/fluidsim.base.time_stepping.pseudo_spect.html}}

\citet{PattersonOrszag1971} describe a method to remove aliases by modifying the
time-stepping scheme. We define a phase-shifting operator $\mathcal{P}_{\Delta}$ that
shifts a field $X$ by a distance $\Delta$ in one direction, i.e.\ it represents a
translation of the grid on which $X$ is evaluated. In spectral space, this operator
reads:
\begin{equation}
\mathcal{P}_{\Delta}(X) = e^{ik\Delta}X,
\label{eq:shift_operator}
\end{equation}
where $k$ is the wavenumber. The corresponding inverse operator is
$\mathcal{P}^{-1}_{\Delta}(X) = e^{-ik\Delta}X$. For the forward Euler scheme, the
phase-shifting method consists in replacing the nonlinear term with the average:
\begin{equation}
\mathcal{N}_a = \tfrac{1}{2} \left(\mathcal{N}_0 + \widetilde{\mathcal{N}_0}\right),
\label{eq:sum_ps_non_ps}
\end{equation}
where $\widetilde{\mathcal{N}_0}$ is the shifted-and-deshifted nonlinear term defined
by:
\begin{equation}
\widetilde{\mathcal{N}_0} \equiv \mathcal{P}^{-1}_{\Delta}\!\left[\mathcal{N}\!\left(\mathcal{P}_{\Delta}(S_0)\right)\right].
\end{equation}

To show that Eq.~\eqref{eq:sum_ps_non_ps} removes the aliasing error, we return to the
nonlinear product. Applying the shift operator to the Fourier coefficients and the
inverse shift operator to the sums in Eq.~\eqref{eq:rogallo_separated} gives
\cite{Rogallo1981}:

\begin{equation}
\widetilde{w_k}
= e^{-ik\Delta}\sum_{\substack{k_n + k_m = k, \\ k \in \mathcal{K}}} \hat{u}(k_n) \hat{v}(k_m) e^{i(k_n + k_m)\Delta}
+ e^{-ik\Delta}\sum_{\substack{k_n + k_m = k \pm 2k_N, \\ k \in \mathcal{K}}} \hat{u}(k_n) \hat{v}(k_m) e^{i(k_n + k_m)\Delta},
\label{eq:rogallo_ps_1}
\end{equation}
which simplifies to:
\begin{equation}
\widetilde{w_k}
= \sum_{\substack{k_n + k_m = k, \\ k \in \mathcal{K}}} \hat{u}(k_n) \hat{v}(k_m)
+ e^{\pm i 2k_N \Delta}\sum_{\substack{k_n + k_m = k \pm 2k_N, \\ k \in \mathcal{K}}} \hat{u}(k_n) \hat{v}(k_m),
\label{eq:rogallo_ps_2}
\end{equation}
or in compact form:
\begin{equation}
\widetilde{w_k}
= \hat{w}_{\mathrm{clean}}(k)  + e^{\pm i 2k_N \Delta}\hat{w}_{\mathrm{alias}}(k).
\label{eq:rogallo_ps_3}
\end{equation}
The clean, non-aliased term is unchanged while the aliased term acquires the phase
factor $e^{\pm i 2k_N \Delta}$. \citet{PattersonOrszag1971} propose choosing $\Delta =
\delta x/2 = L/(2N) = \pi/(2k_N)$ (with $\delta x = L/N$ the grid spacing), i.e.\ a
shift of half a grid cell. The phase factor then becomes $e^{\pm i\pi} = -1$. Averaging
$\hat{w}(k)$ and $\widetilde{w_k}$ as in Eq.~\eqref{eq:sum_ps_non_ps} causes the
aliased contributions to cancel exactly, leaving only the clean term:
\begin{equation}
\frac{\hat{w}(k) + \widetilde{w_k}}{2}
= \frac{2\hat{w}_{\mathrm{clean}}(k)  + \hat{w}_{\mathrm{alias}}(k) -\hat{w}_{\mathrm{alias}}(k)}{2} = \hat{w}_{\mathrm{clean}}(k).
\label{eq:rogallo_ps_4}
\end{equation}

This strategy extends naturally to higher-order schemes such as Adams-Bashforth or
Runge-Kutta. For the standard RK2, the first stage uses $\tau = \dt/2$, $S_a = S_0$,
and $\mathcal{N}_a = \tfrac{1}{2}(\mathcal{N}_0 + \widetilde{\mathcal{N}_0})$, while
the second stage uses $\tau = \dt$, $S_a = S_1$, and $\mathcal{N}_a =
\tfrac{1}{2}(\mathcal{N}_1 + \widetilde{\mathcal{N}_1})$. We refer to this scheme as
``RK2 phase-shift exact'' since all aliases are (in 1D) cancelled exactly. However, the
cost of a simulation is approximately doubled at the same resolution, since four
evaluations of the nonlinear term are required per time step instead of two.

In contrast, \citet{Rogallo1981} exploits the structure of the RK2 scheme with the
trapezoidal rule (Heun's method) to dealias the nonlinear term used in the final update
without any additional evaluation of the nonlinear term. This scheme, which we call
``RK2 phase-shift approx'', removes the aliases approximately, with a residual aliasing
error scaling as $\dt^2$ \cite{Rogallo1981}. Setting aside the complications specific
to three dimensions (discussed below), the scheme consists of a first stage with $\tau
= \dt$, $S_a = S_0$, and $\mathcal{N}_a = \mathcal{N}_0$ (aliased), and a second stage
with $\tau = \dt$, $S_a = S_0$, and:

\begin{equation}
\left. \frac{\mathcal{N}}{S}\right\vert_a = \frac{\tfrac{1}{2}(\mathcal{N}_0 +
\widetilde{\mathcal{N}_1})}{S_0 e^{\sigma \dt /2}}.
\label{eq:Rogallo1981}
\end{equation}

\subsubsection{Application to the quadratic 1D model}

\begin{figure}[]
\centerline{
\includegraphics[width=\figwidth]{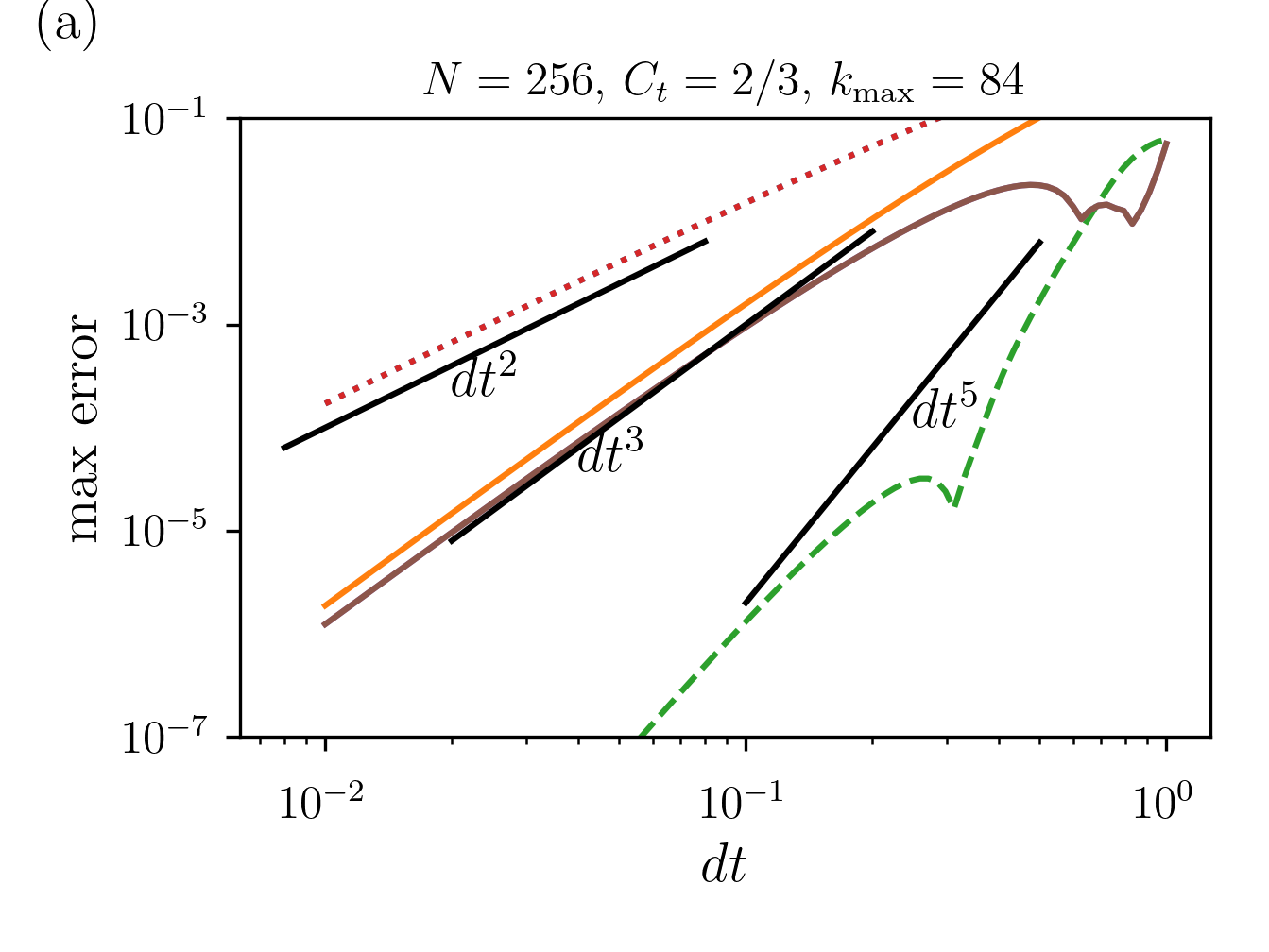}
\includegraphics[width=\figwidth]{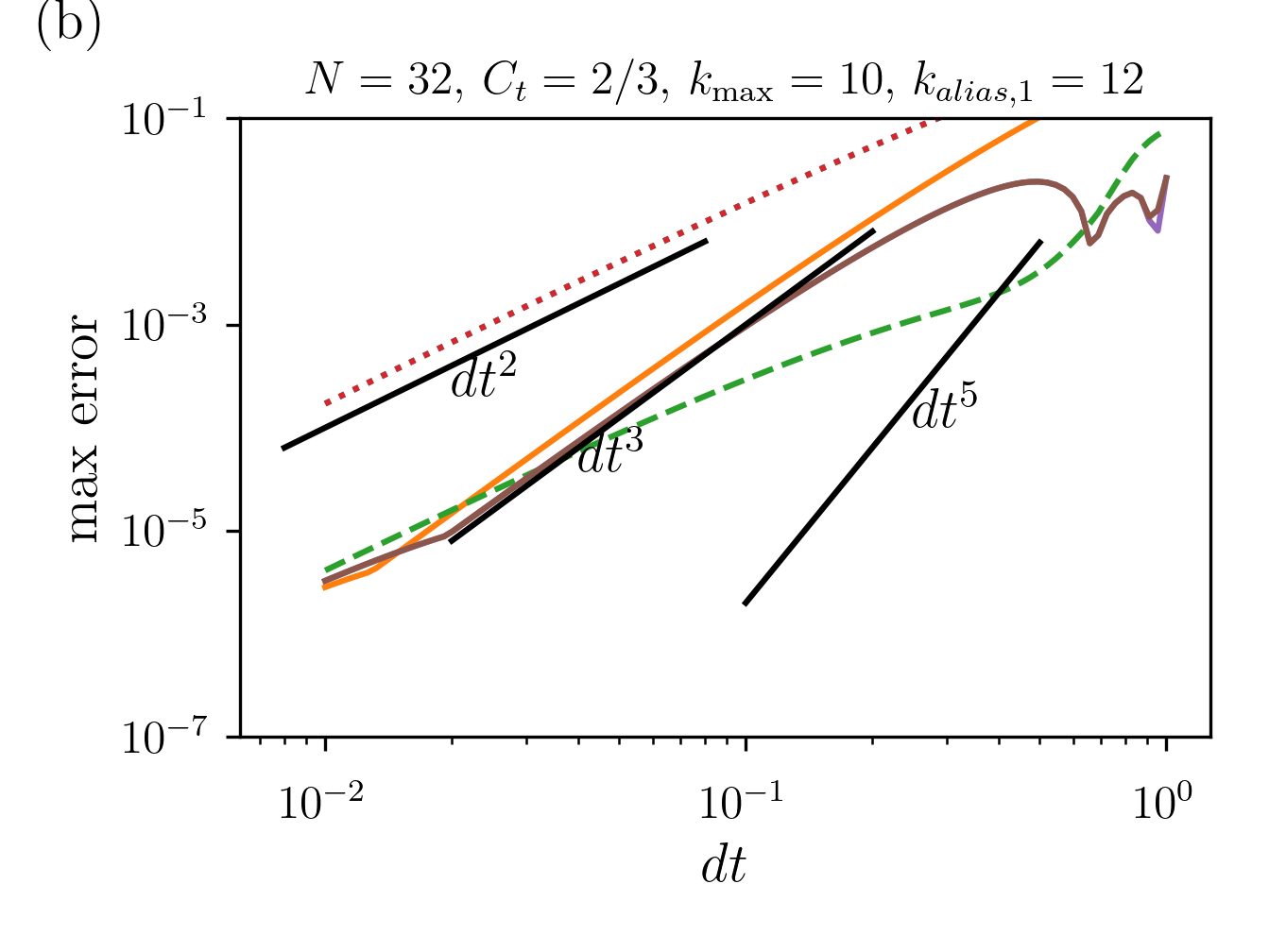}
}
\centerline{
\includegraphics[width=1.765625\figwidth]{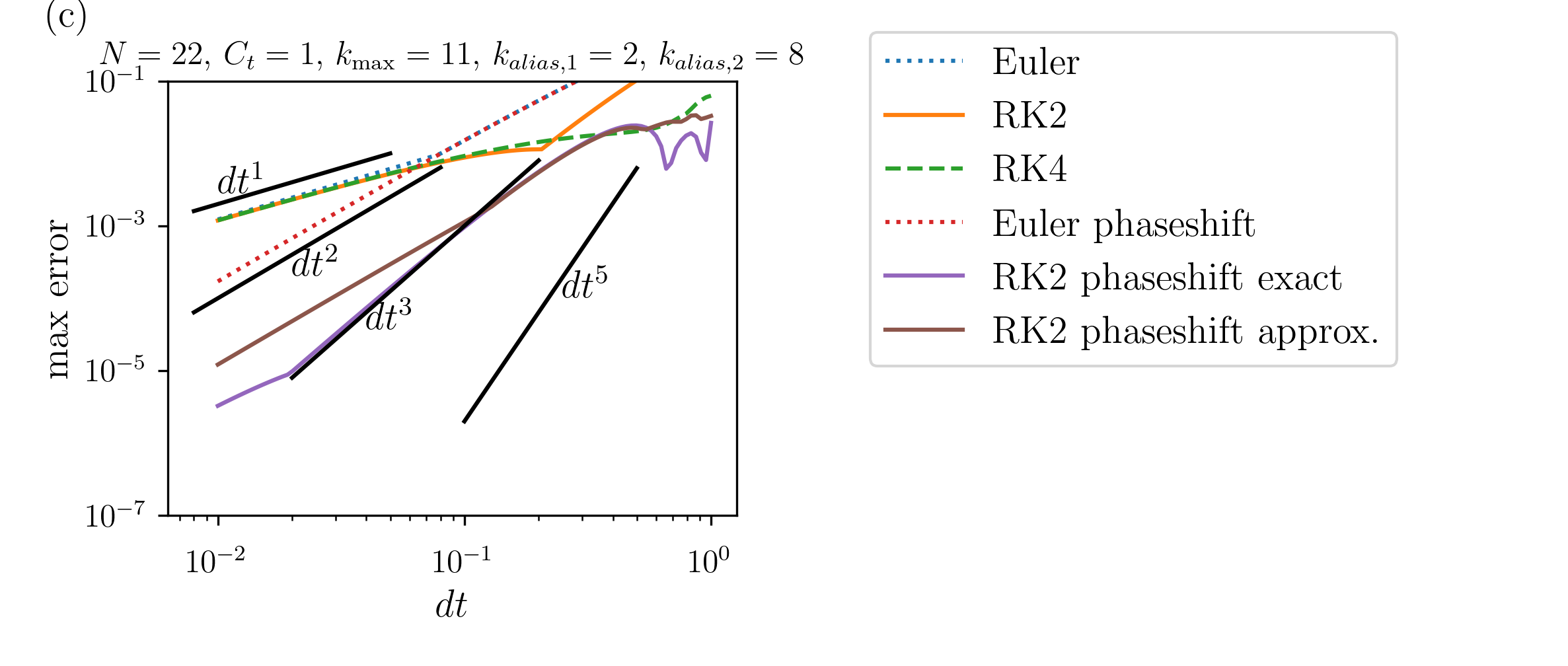}
}
\caption{Maximum error ($\max_k|S(k, dt) - S_\text{exact}(k, dt)|$) for one time step
versus time increment $dt$ for all time schemes implemented, at $N = 256$ and $C_t =
2/3$ (a), $N = 32$ and $C_t = 2/3$ (b) and $N = 22$ and $C_t = 1$ (c). Note that for
(a) and (b) the curves of phase-shifting base schemes overlap. With $k_0 = 10$, one
time step creates one nonlinear mode $k_{nl,1} = 20$ for all schemes and a second one
at $k_{nl,2} = 30$ for multi-step schemes. \label{fig:error1d}}
\end{figure}

We first use the one-dimensional (1D) model defined by Eq.~\eqref{eq:model-1d-2} to
verify the behavior of the different algorithms and our implementations.
Fig.~\ref{fig:error1d} displays the maximum of $|S(k, dt) - S_\text{exact}(k, dt)|$ as
a function of the time step $dt$ for standard schemes (Euler, RK2, and RK4) and for the
two RK2 phase-shifting schemes described above. We consider three numerical cases: (i)
a well-resolved case without aliases ($N=256$, $C_t=2/3$, Fig.~\ref{fig:error1d}(a));
(ii) an under-resolved case with aliases removed by 2/3 truncation ($N=32$, $C_t=2/3$,
Fig.~\ref{fig:error1d}(b)); and (iii) an under-resolved case without truncation
($N=22$, $C_t=1$, Fig.~\ref{fig:error1d}(c)). The latter two cases were studied in
subsection~\ref{sec:sign-function} and Fig.~\ref{fig:1d-alias}, where aliases were
shown to be produced for the last case. Since we study the error as a function of the
time increment $dt$, it is useful to recall standard scaling results: the error for one
time step scales as $dt^2$, $dt^3$, and $dt^5$ for Euler, RK2, and RK4 schemes,
respectively. The ``RK2 phase-shift approx'' scheme, for which the first stage of the
RK2 scheme does not use phase shifting, retains a residual aliasing error scaling as
$dt^2$ \cite{Rogallo1981}.

In case (i), Fig.~\ref{fig:error1d}(a), the simulation is perfectly resolved with no
aliasing and negligible signal at the largest wavenumbers. For small time steps, all
schemes follow their expected scaling laws: $\dt^2$ for Euler, $\dt^3$ for RK2, $\dt^5$
for RK4 and phase-shifting schemes curves overlap and follow RK2 scaling at $\dt^3$.

Case (ii) is shown in Fig.~\ref{fig:error1d}(b) for $N = 32$ with 2/3 truncation, so
that $k_N = 16$ and $\kmax = 10$ (as in Fig.~\ref{fig:1d-alias}). Here, $k = k_0 = 10$
is the largest non-truncated mode and all nonlinear interactions are unresolved.
Consequently, the error for RK4 (dashed line) no longer follows the $\dt^5$ scaling.
The self-interaction of mode $k_0$ would force the unresolved mode at $k=20$, which
would alias to $k=12$, but this mode is truncated so the alias is suppressed. The mode
at $3k_0$, arising from the interaction between $k_0$ and the transiently generated
$2k_0$ mode at intermediate sub-steps (in RK2 and higher-order schemes), is also absent
here.

Case (iii), in Fig.~\ref{fig:error1d}(c), has approximately the same $\kmax$ but uses
$N = 22$ without truncation, so aliases are present. For the three standard schemes
without phase shifting, the aliasing error manifests as a plateau at small $\dt$
scaling as $\dt^1$. ``Euler phase-shift'' as well as ``RK2 phase-shift exact'' schemes
show no visible difference from the alias-free case (i), confirming exact alias
cancellation. For the ``RK2 phase-shift approx'' scheme, the aliasing error is largely
suppressed compared to the standard RK2, but a small residual error scaling as $\dt^2$
remains, as expected \cite{Rogallo1981}.

\subsection{Extension to the multidimensional case}

We now generalize the aliasing error analysis and the phase-shifting method to multiple
dimensions, focusing on the three-dimensional case. Eq.~\eqref{eq:rogallo_separated}
extends straightforwardly by replacing 1D wavenumbers with 3D wavevectors: the
convolution sums become three-dimensional, and aliasing occurs whenever any component
of the nonlinear wavevector $\kvec_{\mathrm{nl}} = \kvec_n + \kvec_m$ exceeds $k_N$ in
absolute value. The three-dimensional shift operator is:
\begin{equation}
\mathcal{P}_{\mathbf{\Delta}}(X) = e^{i \kvec \cdot \mathbf{\Delta}}X,
\label{eq:shift_operator_3d}
\end{equation}
where $\mathbf{\Delta}$ is the shift vector whose components each take values $0$ or
$\delta x/2$. The phase-shifted nonlinear term, generalizing
Eq.~\eqref{eq:rogallo_ps_3}, acquires a factor $e^{\pm i 2k_N \Delta_i}$ for each
out-of-range direction $i$.

\begin{figure}[]
\centerline{
  \includegraphics[width=1.0\figwidth]{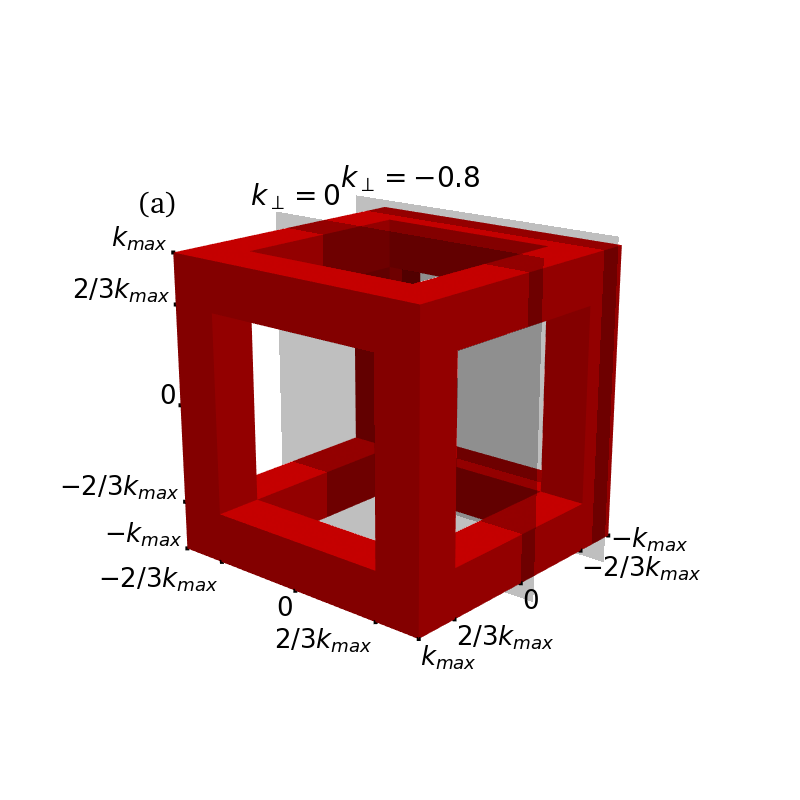}
}
\centerline{
  \includegraphics[width=0.9\figwidth]{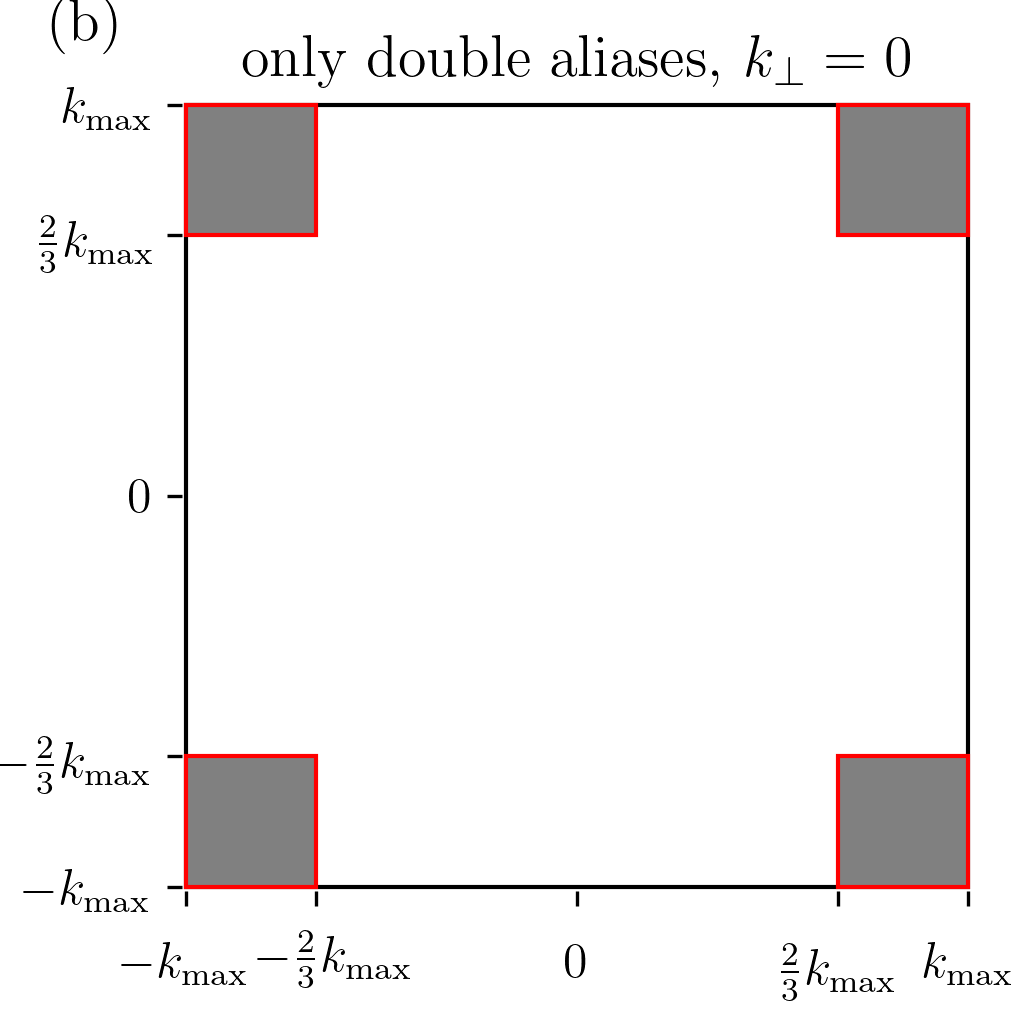}
  \includegraphics[width=0.9\figwidth]{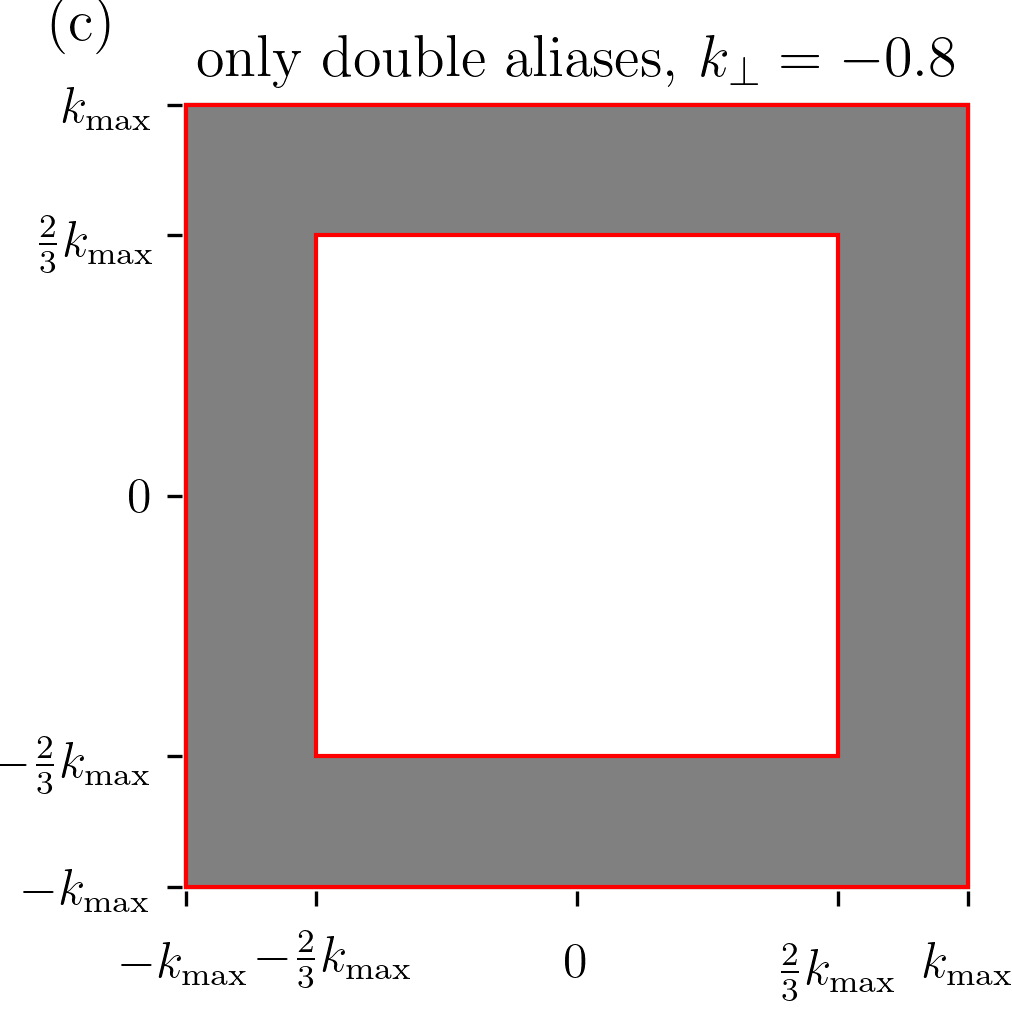}
}
\caption{(a) 3D visualisation of regions removed only by double aliases truncation, as
in \citet{Rogallo1981}. Panels (b) and (c) show 2D cuts (represented in grey on (a))
respectively at $-(2/3) \kmax < k_\perp = 0 < (2/3) \kmax$ and $\kmax < k_\perp = -0.8
< -(2/3) \kmax$, respectively. Modes are set to zero in the gray regions and red
contours represents regions associated with double aliases.
\label{fig:truncation:double}}
\end{figure}

A fully exact removal of all alias types would require evaluating the nonlinear terms
on all $2^d$ possible shift combinations, i.e.\ eight evaluations per time step in
3D. \citet{PattersonOrszag1971} observe that using only the isotropic shift ($\Delta_x
= \Delta_y = \Delta_z = \delta x/2$) already removes the dominant alias contributions
at twice the cost of an unshifted evaluation, as can be seen by classifying aliases
according to how many components of $\kvec_{\mathrm{nl}}$ are out of range. A
\emph{single alias} arises when exactly one component satisfies $|k_{\mathrm{nl},i}| >
k_N$; a \emph{double alias} when two components do; and a \emph{triple alias} when all
three do. The isotropic shift, which sets each factor to $e^{\pm i\pi} = -1$, therefore
cancels single aliases and triple aliases, but leaves double aliases unchanged. The
regions in Fourier space associated with the remaining double aliases are shown in
Fig.~\ref{fig:truncation:double}. Further details on the generalization of aliasing and
phase shifting to 3D are given in the supplementary material \cite{supp}. In the
following, this scheme using isotropic shift exactly at each sub-steps will be referred
to as ``RK2 phase-shift exact''.

\begin{figure}[]
\centerline{
  \includegraphics[width=0.9\figwidth]{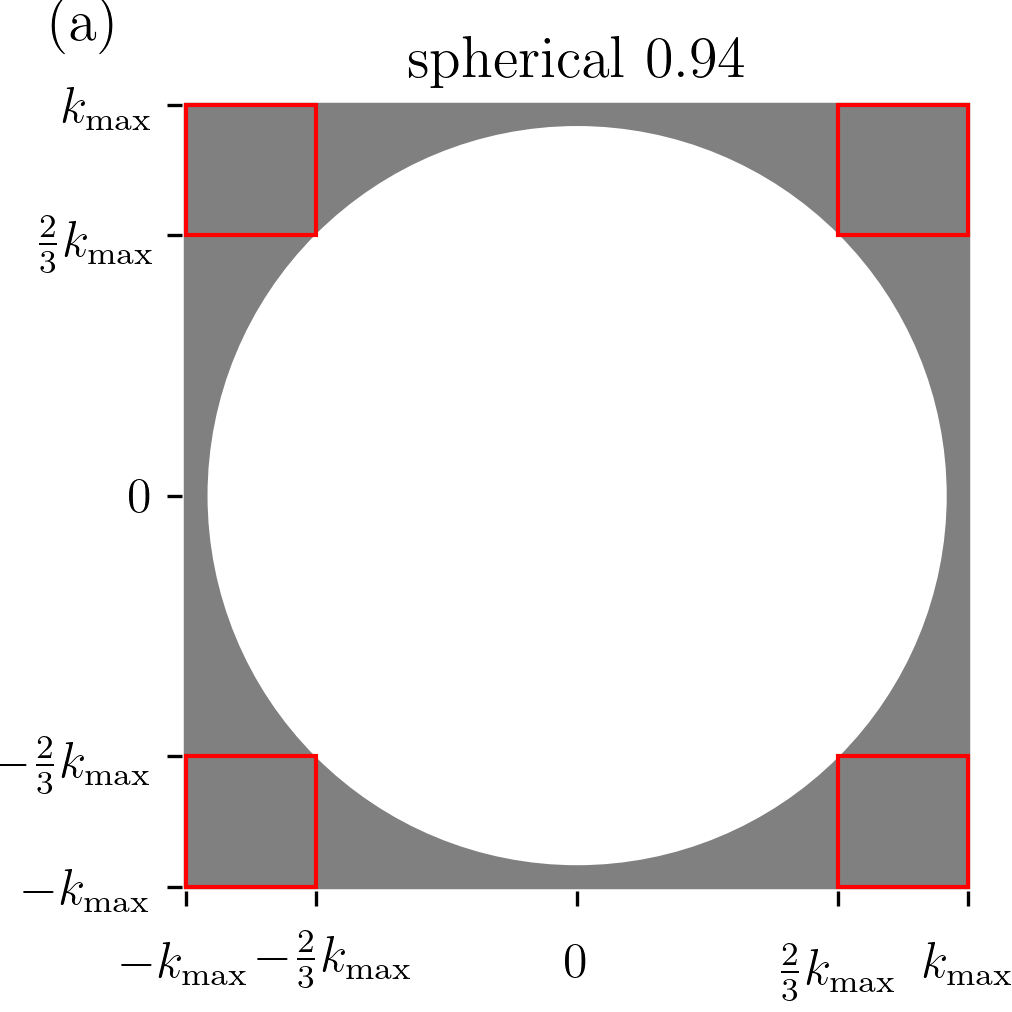}
  \includegraphics[width=0.9\figwidth]{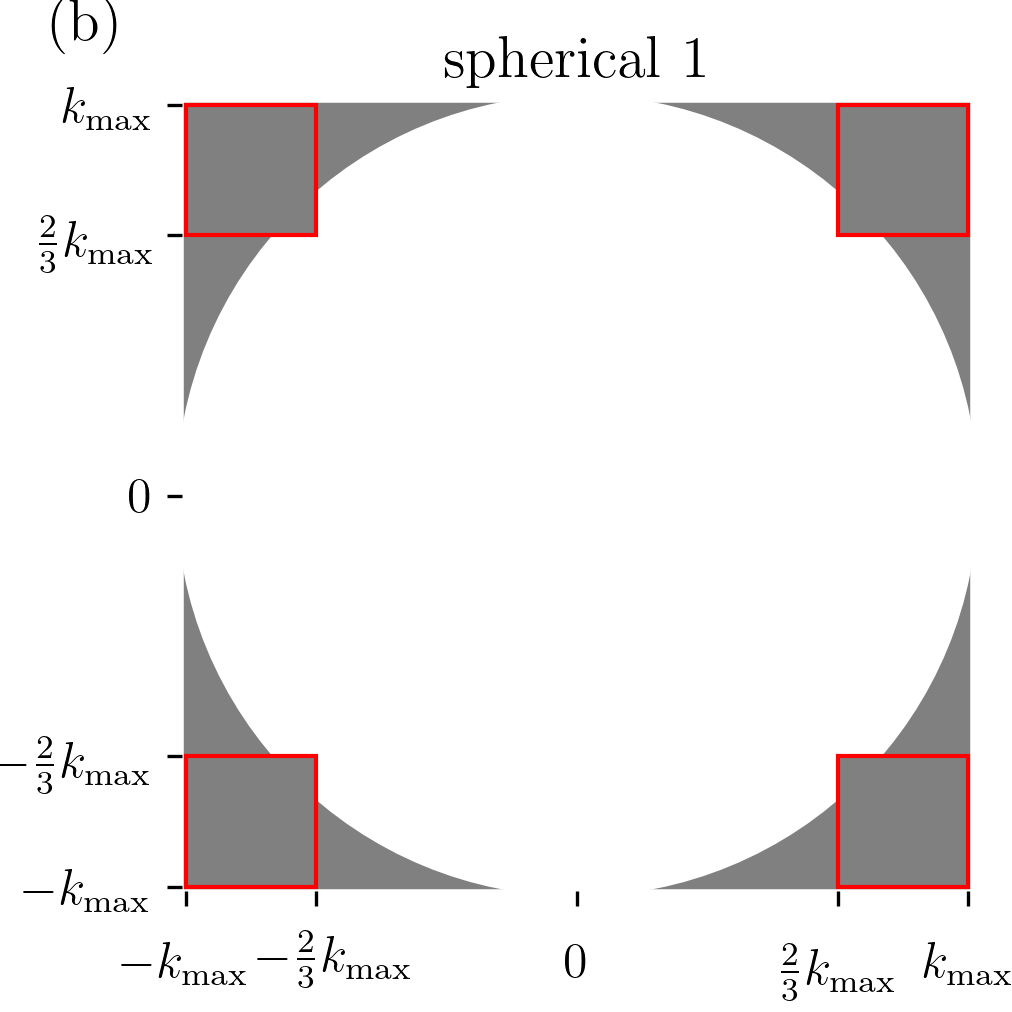}
}
\caption{Same as Fig.~\ref{fig:truncation:common} but with spherical truncations with
double aliases removing. Red contours indicate regions with double aliases. In
Fluidsim, both (a) and (b) are referred to as ``no\_multiple\_aliases''. Panel (a)
corresponds to the method of \citet{PattersonOrszag1971}.
\label{fig:truncation:spherical:double:aliases}}
\end{figure}

\citet{PattersonOrszag1971} also argued that a spherical truncation is preferable to a
cubic truncation for a more faithful representation of the small scales, despite
retaining fewer modes overall. They proposed to remove the remaining double aliases by
applying a spherical truncation at wavenumbers larger than $2\sqrt{2}/3\, k_N \simeq
0.94\, k_N$. As shown in Fig.~\ref{fig:truncation:spherical:double:aliases}(a), this
truncation naturally eliminates all double aliases.

The scheme of \citet{Rogallo1981}, already presented in Eq.~\eqref{eq:Rogallo1981}, is
an evolution of ``RK2 phase-shift approx'' scheme (in 3D with isotropic shift) and
differs from the ``RK2 phase-shift exact'' scheme in two respects in three dimensions.
First, for symmetry, the two nonlinear term evaluations use two distinct shift vectors:
$\mathbf{\Delta} = \deltavec_\alpha$ for the first stage ($\mathcal{N}_a =
\widetilde{\mathcal{N}_{0\alpha}}$) and $\mathbf{\Delta} = \deltavec_\beta$ for the
second stage, with components chosen such that $|\delta_{\alpha i} - \delta_{\beta i}|
= 0.5\,\delta x_i$ for each direction $i$. The second-stage approximation then becomes:

\begin{equation}
\left. \frac{\mathcal{N}}{S}\right\vert_a =
 \frac{\tfrac{1}{2}\!\left(\widetilde{\mathcal{N}_{0\alpha}} + \widetilde{\mathcal{N}_{1\beta}}\right)}{S_0 e^{\sigma \dt /2}}.
\end{equation}

Second, the components of $\deltavec_\alpha$ are drawn randomly at each time step from
$[0, \delta x]$ at uniform probability to avoid time-correlated errors. This scheme is
referred to as ``RK2 phase-shift random'' in the following. \citet{Rogallo1981} also
adopts a different truncation than \citet{PattersonOrszag1971}: only double aliases are
removed, so the computational domain in Fourier space is a cube from which thin
parallelepipeds touching its edges are excluded, as shown in
Fig.~\ref{fig:truncation:double}.

In the comparisons that follow, we evaluate the efficiency and accuracy of four
schemes: RK4, RK2, ``RK2 phase-shift exact'', and ``RK2 phase-shift random''. In the
paper we adopt two truncation strategies as intermediate between those of
\citet{PattersonOrszag1971} and \citet{Rogallo1981}. First, in this section, we only
use spherical truncation (as in Figs.~\ref{fig:truncation:common}(b) and
\ref{fig:truncation:spherical:double:aliases}(a)) parameterized by a truncation
coefficient $C_t$ such that $\kmax = C_t k_N$. At the end of Section
\ref{section-decaying}, we also run simulations for which we additionally truncate the
regions associated with double aliases that survive the spherical truncation when $C_t
> 0.94$, as shown in Fig.~\ref{fig:truncation:spherical:double:aliases}(b) for $C_t =
1$. Note that increasing $C_t$ from $0.94$ to $1$ corresponds to a factor of $1.2$
increase in the number of retained modes.

\section{Performance and accuracy for a decaying flow}


\label{section-decaying}

The primary motivation for dealiasing algorithms based on phase shifting is
computational efficiency: they allow larger and longer simulations for a given
allocation on a computing cluster, or equivalently the same simulation with a reduced
allocation. They can also achieve the same physical resolution on a coarser numerical
grid, which reduces memory requirements. Since most phase-shifting algorithms are
approximate, we must evaluate them on a specific flow and compare physical results with
and without phase shifting, and across different resolutions. As we will show, both
performance and accuracy depend on the choice of numerical parameters, and these two
objectives must be considered together when selecting a numerical setup.

In this section, we study simulations of the transition to turbulence from Taylor-Green
vortices. Performance is assessed through benchmarking and profiling, while accuracy is
evaluated using the spectral error metric defined in Section~\ref{sec:T-G_vortices}.

\subsection{Effect of resolution for simulations without phase shifting}

Before studying phase-shifting algorithms, we first characterize the effect of
resolution alone on simulations without phase shifting, so as to establish a baseline
for comparison.

\subsubsection{Effect of resolution for RK4 truncated simulations}

\label{sec:effet:res:truncated:simuls}

We consider simulations dealiased with 2/3 truncation at four grid sizes: $N = 128$,
$256$, $400$, $640$, and $768$ (Table~\ref{tab:23}). Since we compare resolutions,
these simulations are run in parallel with a reasonable number of cores for each grid
size.

\begin{table}[]
\centering \begin{tabular}{| c | c | c | c | c | c | c | c |}
\toprule
$N$ & $Re$ & scheme & $C_t$ & $k_{\max} \tilde\eta$ & number of proc. & CPU.h speedup & error (\%) \\
\midrule
768 & 1600 & RK4 & 2/3 & 3.0 & 32 & 1.0 & 0.0 \\
640 & 1600 & RK4 & 2/3 & 2.5 & 20 & 3.6 & 0.86 \\
400 & 1600 & RK4 & 2/3 & 1.56 & 10 & 10.5 & 3.06 \\
256 & 1600 & RK4 & 2/3 & 1.0 & 4 & 34.4 & 10.4 \\
128 & 1600 & RK4 & 2/3 & 0.5 & 2 & 518.9 & 24.11 \\
\bottomrule
\end{tabular}
 \caption{Simulations dealiased
with 2/3 truncation used in Section~\ref{sec:effet:res:truncated:simuls}. Error is
defined by Eq.~\eqref{eq:quality} as a percentage relative to the reference simulation
at $N=768$ (first row).}\label{tab:23}
\end{table}

\begin{figure}[]
\centerline{\includegraphics[width=2\figwidth]{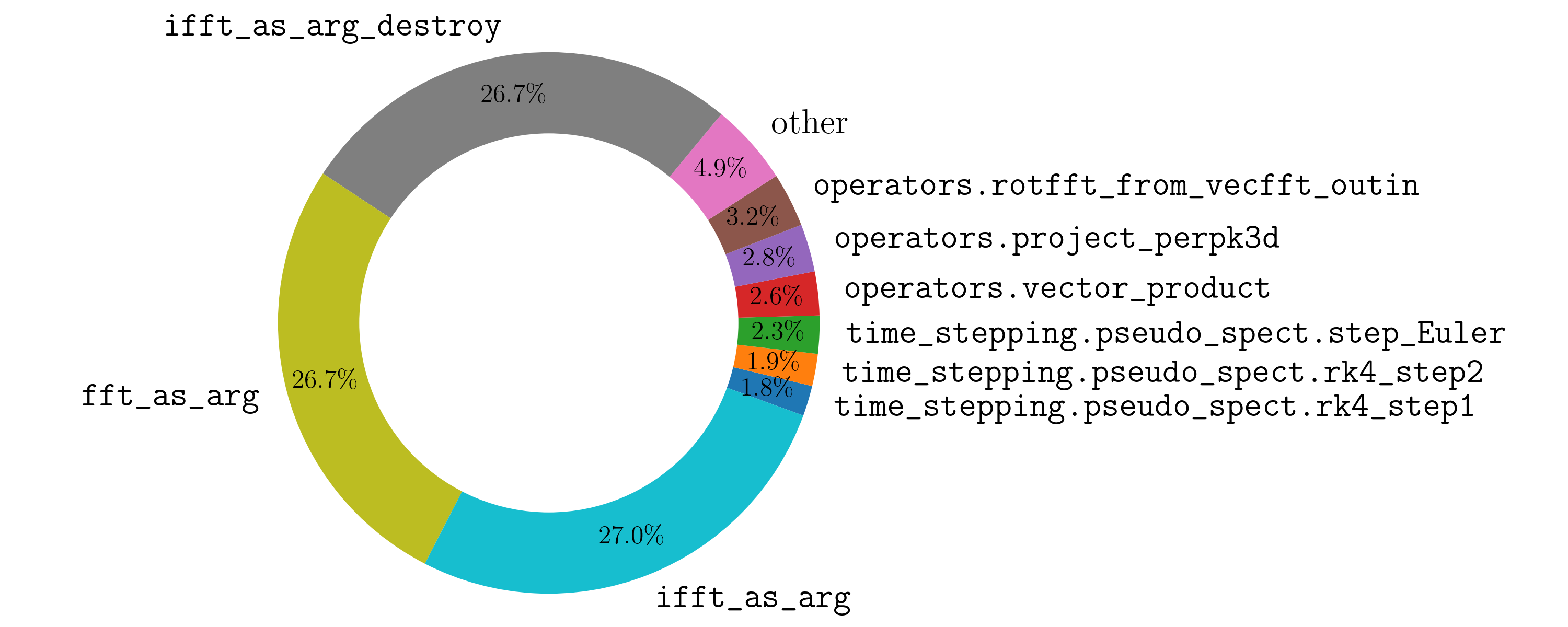}}
\caption{Distribution of computation time across different tasks for a simulation with
RK4 at $N = 768$ and $C_t = 2/3$. The profile is computed in parallel on $32$ cores and
over a short run of $0.01$ overturn times. \label{fig:profiling:pie_resol}}
\end{figure}

Fig.~\ref{fig:profiling:pie_resol} shows the distribution of computation time across
different functions for a standard 3D Navier-Stokes simulation with the RK4 scheme at
the largest grid size considered here. Three quarters of the time is spent in Fourier
transforms (FFT and IFFT). The remaining quarter is spent on linear and nonlinear
operations, truncation, and the Runge-Kutta stages.

\begin{figure}[]
\centerline{
  \includegraphics[width=1.1\figwidth]{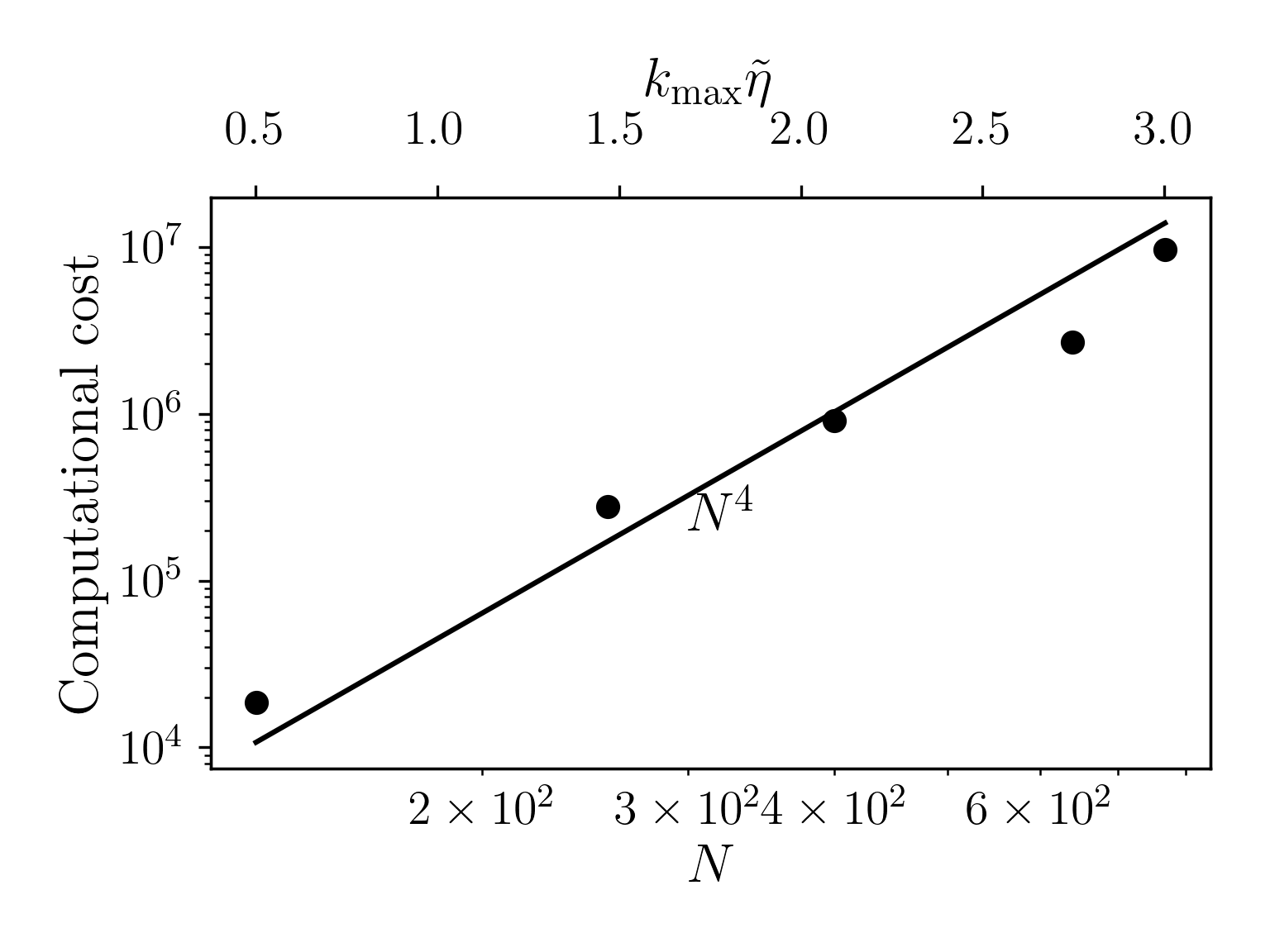}}
\caption{Computation time multiplied by the number of cores versus grid size $N$ for
the simulations in Table~\ref{tab:23} (RK4 scheme and 2/3 truncation). The solid line
shows the expected scaling $\propto N^4$. \label{fig:speedup-sizes}}
\end{figure}

To assess scalability, Fig.~\ref{fig:speedup-sizes} shows the computation time
normalized by the number of cores as a function of $N$. The expected scaling is $N^4$:
a factor $N^3$ from the three-dimensional grid and a factor $N$ from the time step,
which shrinks proportionally with the grid spacing under a CFL condition. The figure
shows some deviations from this ideal scaling, likely due to varying parallelization
efficiency across different grid sizes and core counts. Nevertheless, the overall trend
follows the $N^4$ slope, confirming that the expected scaling holds.

\begin{figure}[]
\centerline{
  \includegraphics[width=\figwidth]{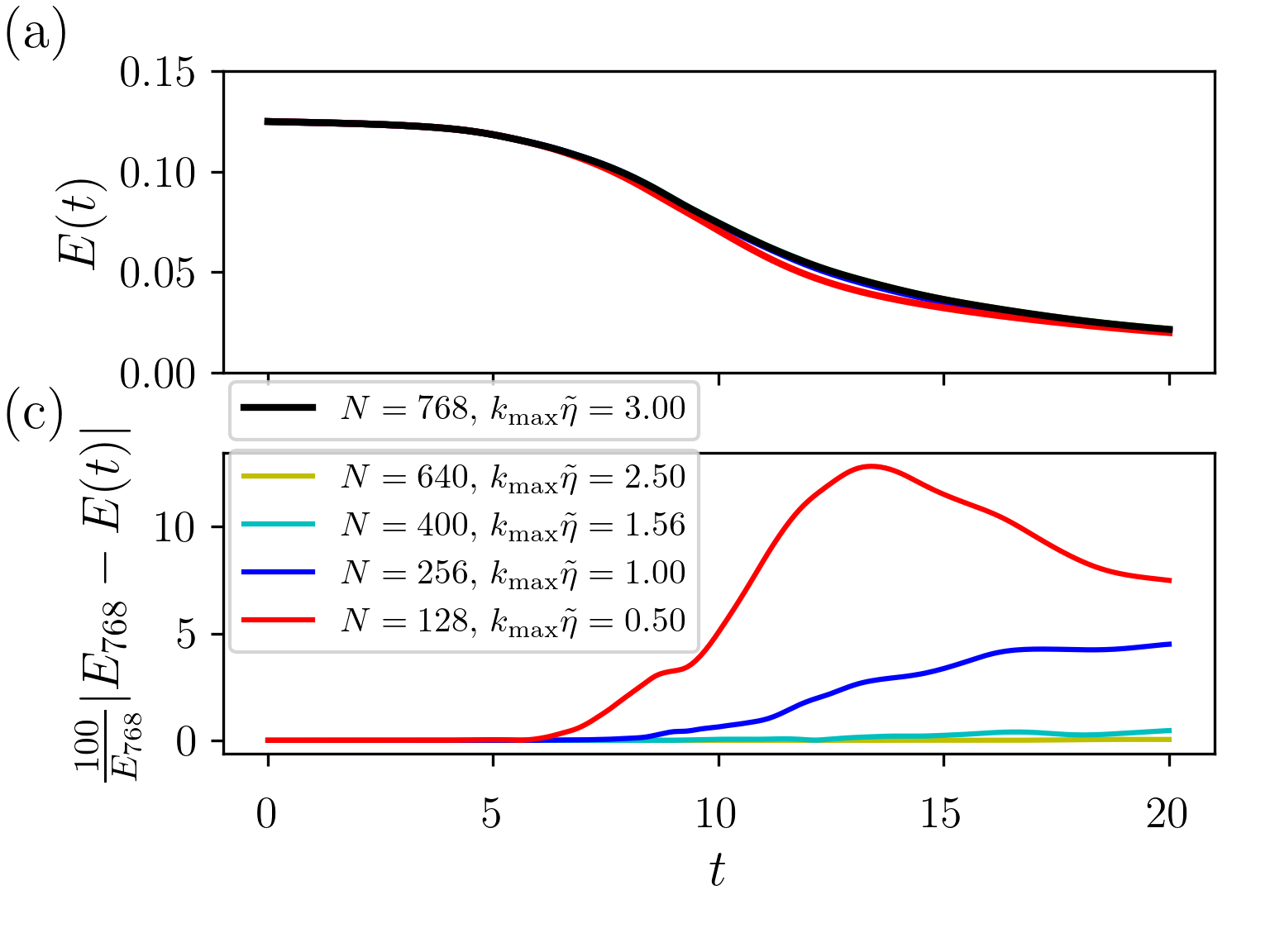}
  \includegraphics[width=\figwidth]{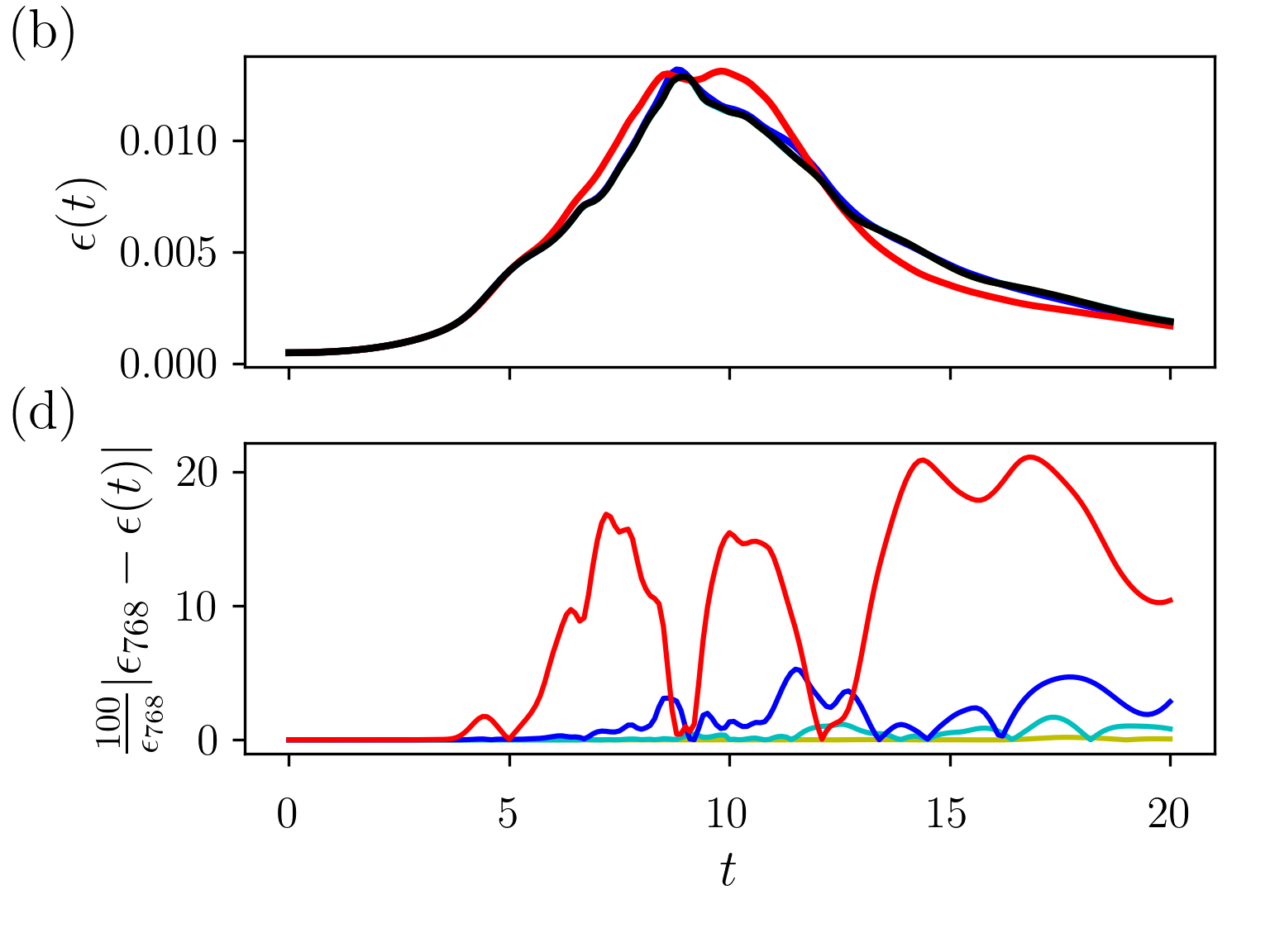}}
\caption{Same as Fig.~\ref{fig:spatialmeans:192} but for fully dealiased simulations
listed in Table~\ref{tab:23}. Note that for $N>128$, curves overlap on (a) and (b).
\label{fig:spatialmeans:dealiased}}
\end{figure}

Fig.~\ref{fig:spatialmeans:dealiased}(a) shows the temporal evolution of the kinetic
energy for each resolution, and Fig.~\ref{fig:spatialmeans:dealiased}(c) the
corresponding relative error in percent. Similarly,
Fig.~\ref{fig:spatialmeans:dealiased}(b) shows the energy dissipation rate and
Fig.~\ref{fig:spatialmeans:dealiased}(d) its relative error. All simulations follow the
same energy and dissipation evolution during the initial overturn times, with
negligible errors. As the flow approaches the peak of energy dissipation at the
transition to turbulence, differences begin to appear and grow with decreasing
resolution. Considering only the temporal evolution of the spatially averaged kinetic
energy (Fig.~\ref{fig:spatialmeans:dealiased}(c)), the average relative difference with
respect to the highest-resolution run is approximately $12\%$, about half the value
computed from the spectra with Eq.~\eqref{eq:quality} (reported in Table~\ref{tab:23}).
Also, very small errors are observed on the toral energy at $\kmax \etamin \geq 1.5$
($N \geq 400$). Both observations illustrate the need to compute finer error on energy
spectra.

\begin{figure}
\centerline{
    \includegraphics[width=\figwidth]{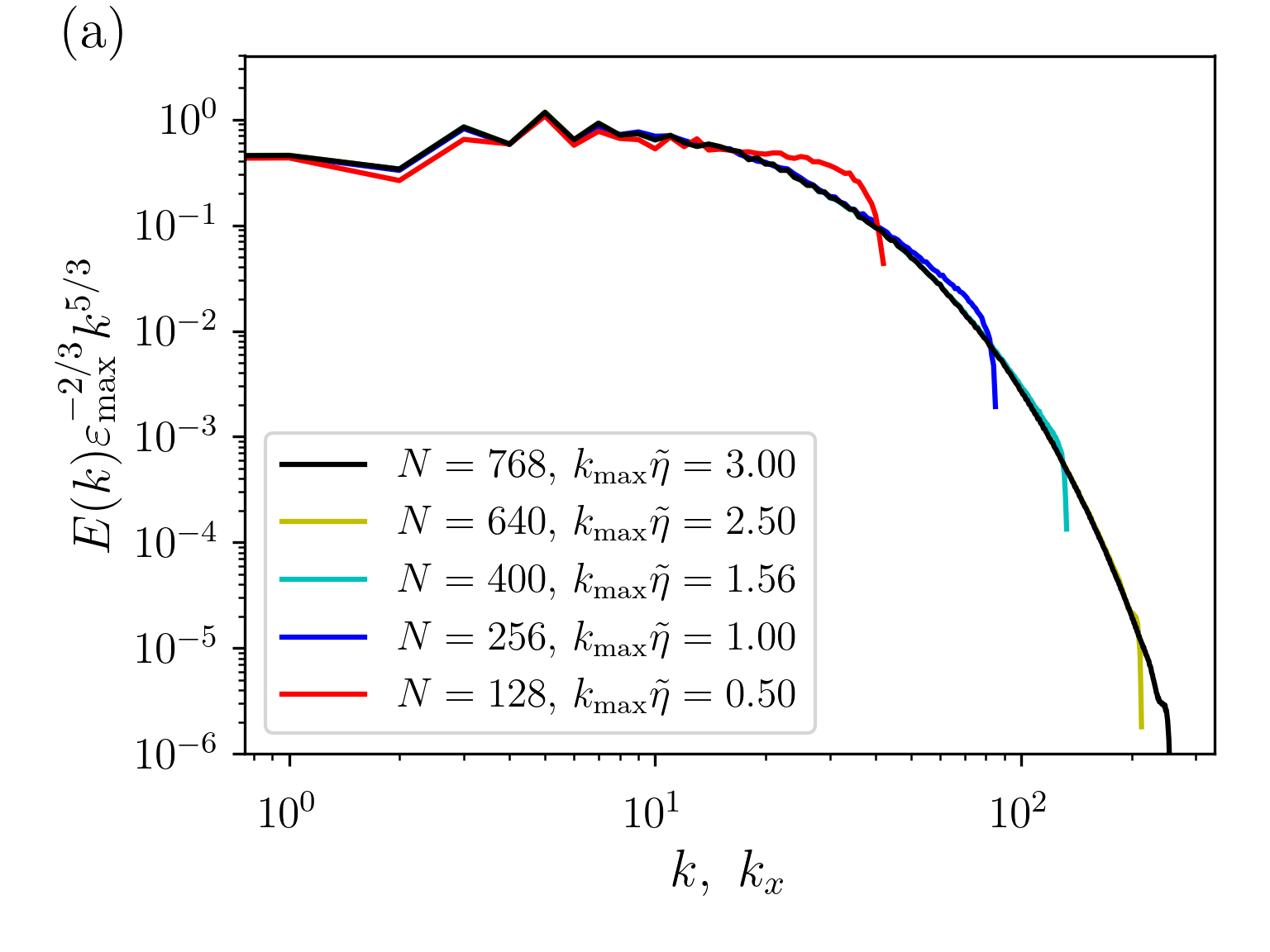}
    \includegraphics[width=\figwidth]{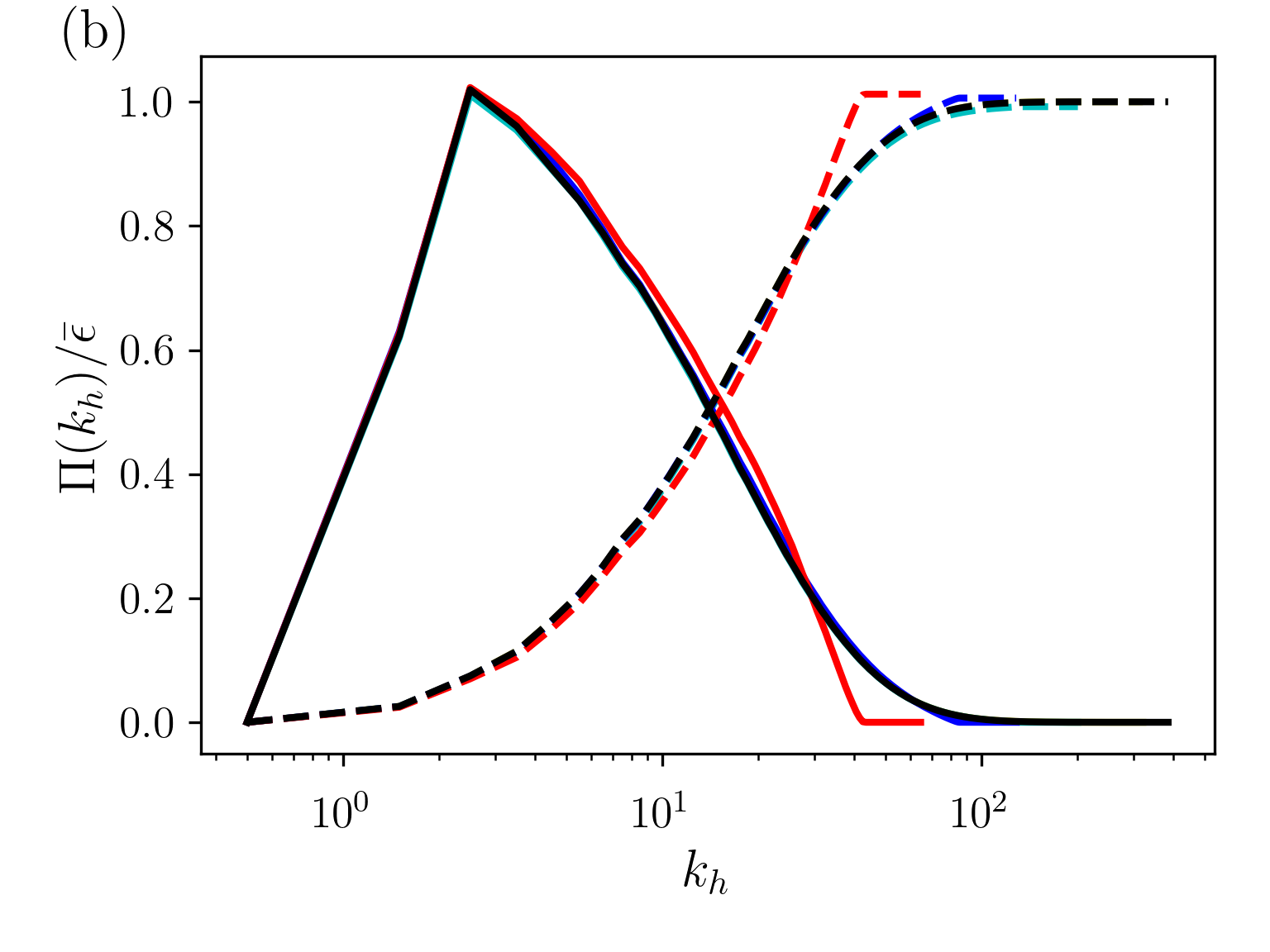}}
\caption{Same as Fig.~\ref{fig:spectra:192} but for fully dealiased simulations listed
in Table~\ref{tab:23}. Note that for $N \geq 128$, curves overlap for most of the
spectral domain. \label{fig:spectra:dealiased}}
\end{figure}

Fig.~\ref{fig:spectra:dealiased} shows the energy spectra, energy flux, and cumulated
dissipation for the various resolutions. Clear differences appear with decreasing
resolution in the energy spectra, fluxes, and cumulated dissipation, especially in the
inertial range. Table~\ref{tab:23} confirms that the error increases sharply as
resolution decreases. At low resolutions ($N \leq 256$), the inertial range is not
fully developed, so the energy spectra deviate substantially from the high-resolution
reference at $N = 768$, resulting in large errors.

\subsubsection{Effect of $\kmax\etamin$ on aliasing error}
\label{sec:kmaxetamin:alias:error}

Having quantified the effect of resolution on fully dealiased simulations, we now
examine how resolution affects aliasing errors. We already studied in
Section~\ref{sec:T-G_vortices} how aliasing errors affect a simulation at small
$\kmax\etamin = 0.75$. Here, we consider analogous simulations but for $\kmax\etamin >
3$ (see Table~\ref{tab:high_resol}).

\begin{table}[]
\centering \begin{tabular}{| c | c | c | c | c | c | c | c |}
\toprule
$N$ & $Re$ & scheme & $C_t$ & $k_{\max} \tilde\eta$ & number of proc. & CPU.h speedup & error (\%) \\
\midrule
800 & 1600 & RK4 & 2/3 & 3.13 & 50 & 1.0 & 0.0 \\
800 & 1600 & RK4 & 1 & 4.69 & 50 & 1.0 & 2.65 \\
640 & 1600 & RK4 & 1 & 3.75 & 64 & 2.6 & 2.71 \\
\bottomrule
\end{tabular}
 \caption{Taylor-Green
simulations with and without aliasing at high resolutions, used in
Section~\ref{sec:kmaxetamin:alias:error}. Error is defined by Eq.~\eqref{eq:quality} as
a percentage relative to the reference RK4 simulation at $N = 800$ with $C_t =
2/3$.}\label{tab:high_resol}
\end{table}

\begin{figure}
\centerline{
  \includegraphics[width=\figwidth]{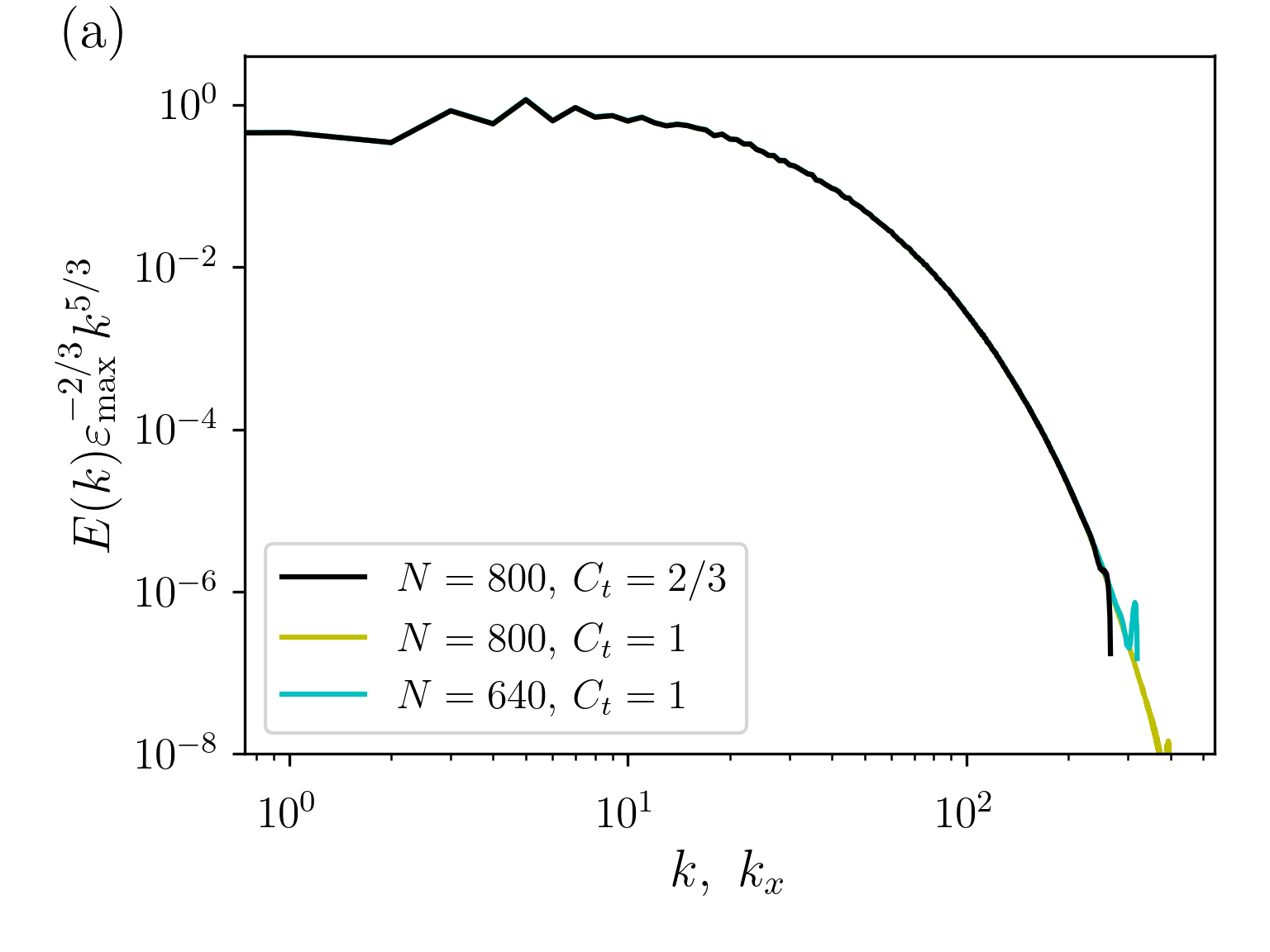}
  \includegraphics[width=\figwidth]{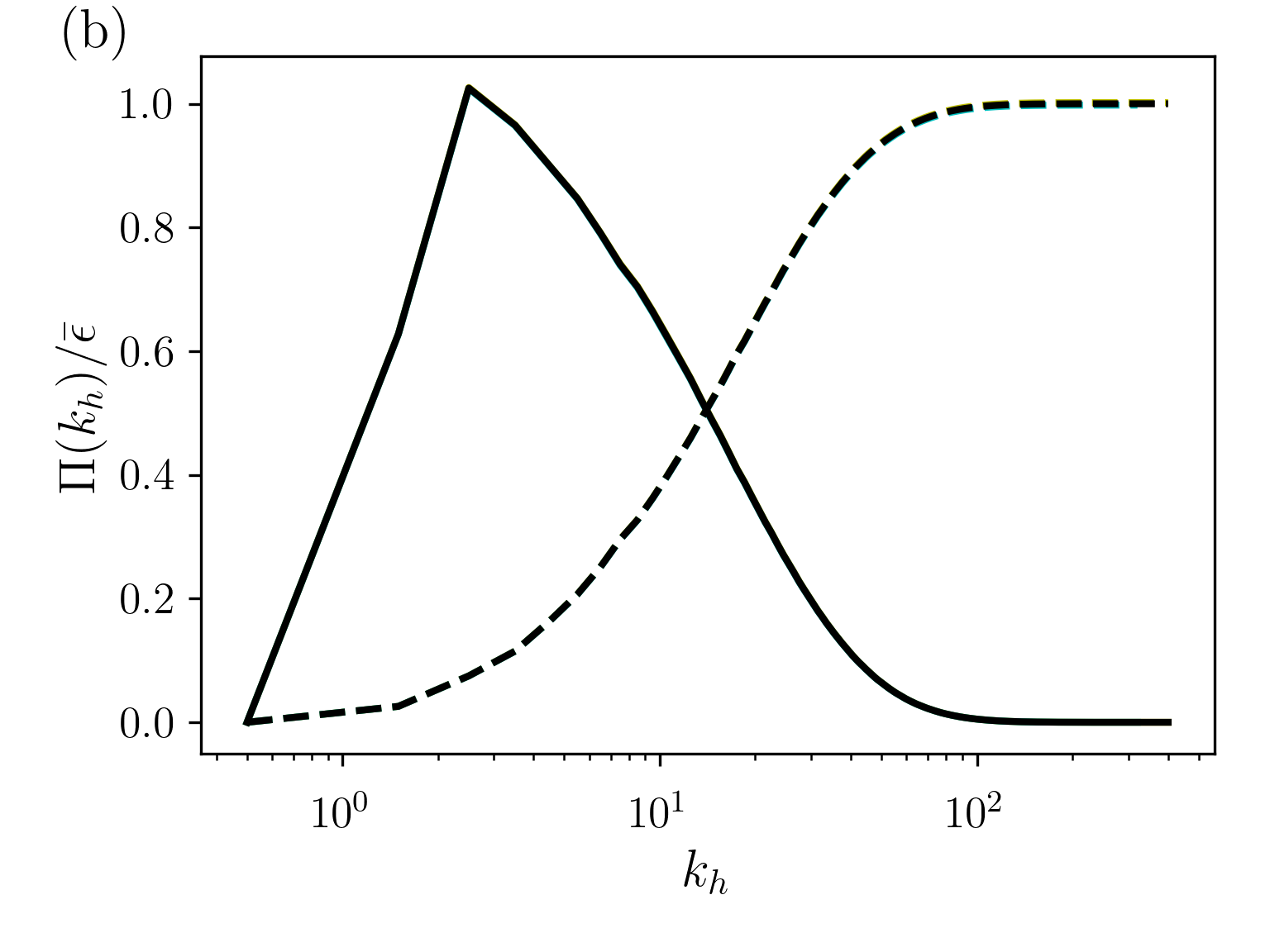} }
\caption{Same as Fig.~\ref{fig:spectra:192} but for RK4 simulations at various $\kmax
\etamin > 3$, corresponding to Table~\ref{tab:high_resol}. Note that curves overlap for
most of the spectral domain. \label{fig:spectra:RK4}}
\end{figure}

Fig.~\ref{fig:spectra:RK4} shows the energy spectra, energy fluxes, and cumulated
dissipation for these simulations with and without aliasing. The results are
particularly close across all three quantities, with discrepancies appearing only after
the peak of the nonlinear energy flux and at the smallest scales. Despite the slightly
different values of $\kmax$, we compute the corresponding error and speedup, reported
in Table~\ref{tab:high_resol}, using the truncated simulation as the reference.

\begin{figure}[]
\centerline{
  \includegraphics[width=1.1\figwidth]{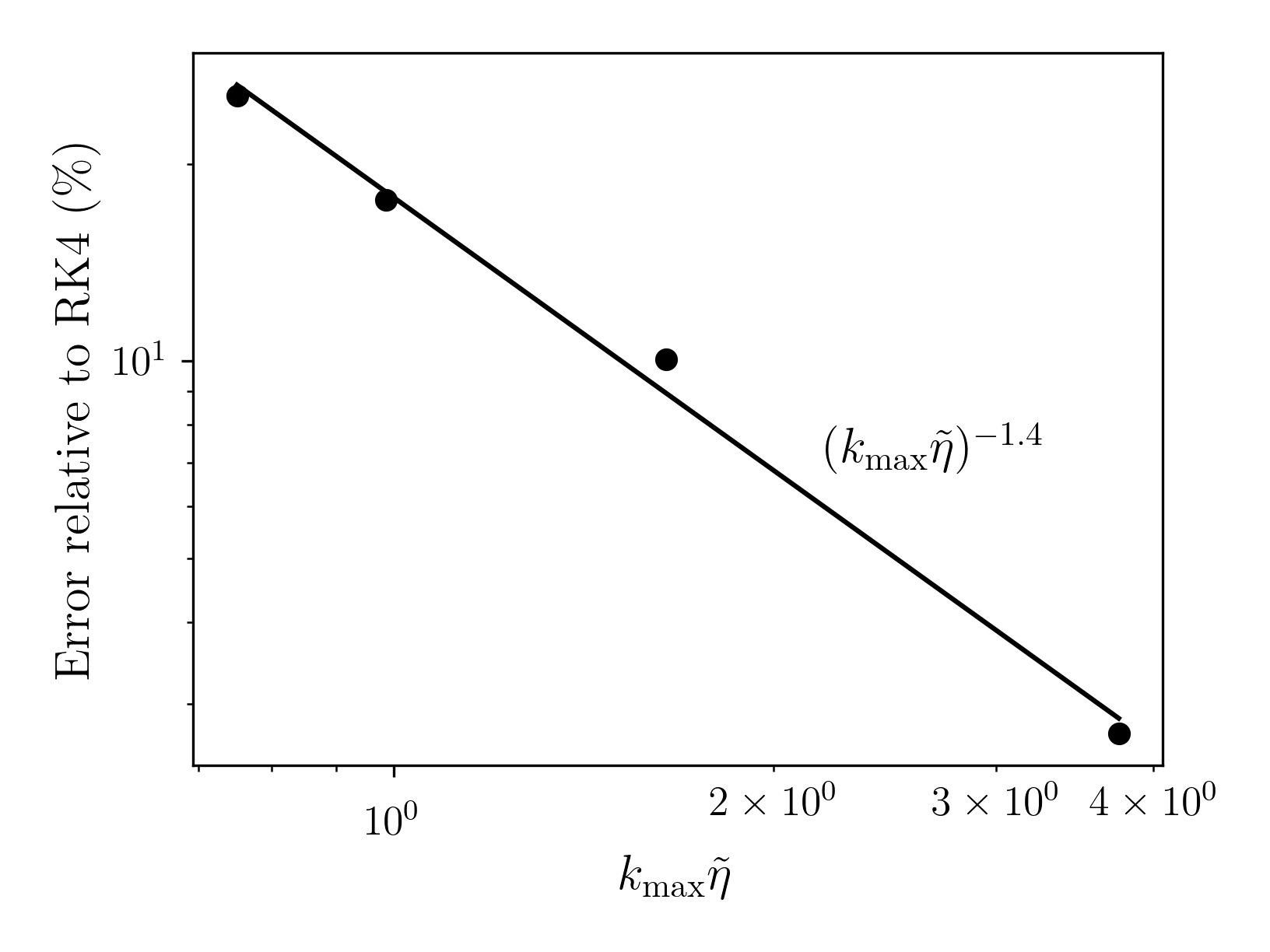}}
\caption{Spectral error in \% (see Eq.~\eqref{eq:quality}) between a dealiased
reference simulation (RK4, $C_t = 2/3$) and an aliased simulation ($C_t = 1$) with the
same $\kmax\etamin$, as a function of $\kmax\etamin$. The first and last points
correspond to the simulations in Tables~\ref{tab:192} and \ref{tab:high_resol},
respectively. \label{fig:error-sizes}}
\end{figure}

These results allow a direct comparison of aliasing error between well-resolved and
poorly resolved simulations. The poorly resolved simulations from
Section~\ref{sec:T-G_vortices} (with $N = 128$ and $\kmax\etamin = 0.7$) showed a much
larger spectral error (Table~\ref{tab:192}) than the aliased simulation at $N = 640$
(Table~\ref{tab:high_resol}), even accounting for the slight overestimation introduced
by the fact that $\kmax^{640} > \kmax^{800,\, C_t = 2/3}$. This confirms that aliasing
errors decrease with increasing $\kmax\etamin$. We verify this trend with
Fig.~\ref{fig:error-sizes}, which shows the error computed from 4 couples (for
different $\kmax\etamin$) of dealiased ($C_t = 2/3$) and aliased ($C_t = 1$)
simulations. We see that the error decreases as $(\kmax\etamin)^{-1.4}$ and become
negligible in the limit of large $\kmax\etamin$.

\subsection{Evaluation of methods based on phase shifting and truncation}

We now investigate how phase-shifting methods improve simulation efficiency and how it
affects accuracy.

\subsubsection{Profiles and speedup}

We profile the different algorithms implemented in Fluidsim on short Taylor-Green
simulations as described in Section~\ref{sec:T-G_vortices}, focusing on time-stepping
performance. To isolate the effect of the time-stepping algorithm, we always compare in
this subsection simulations run with the same number of cores.

\begin{figure}[]
\centerline{\includegraphics[width=\figwidth]{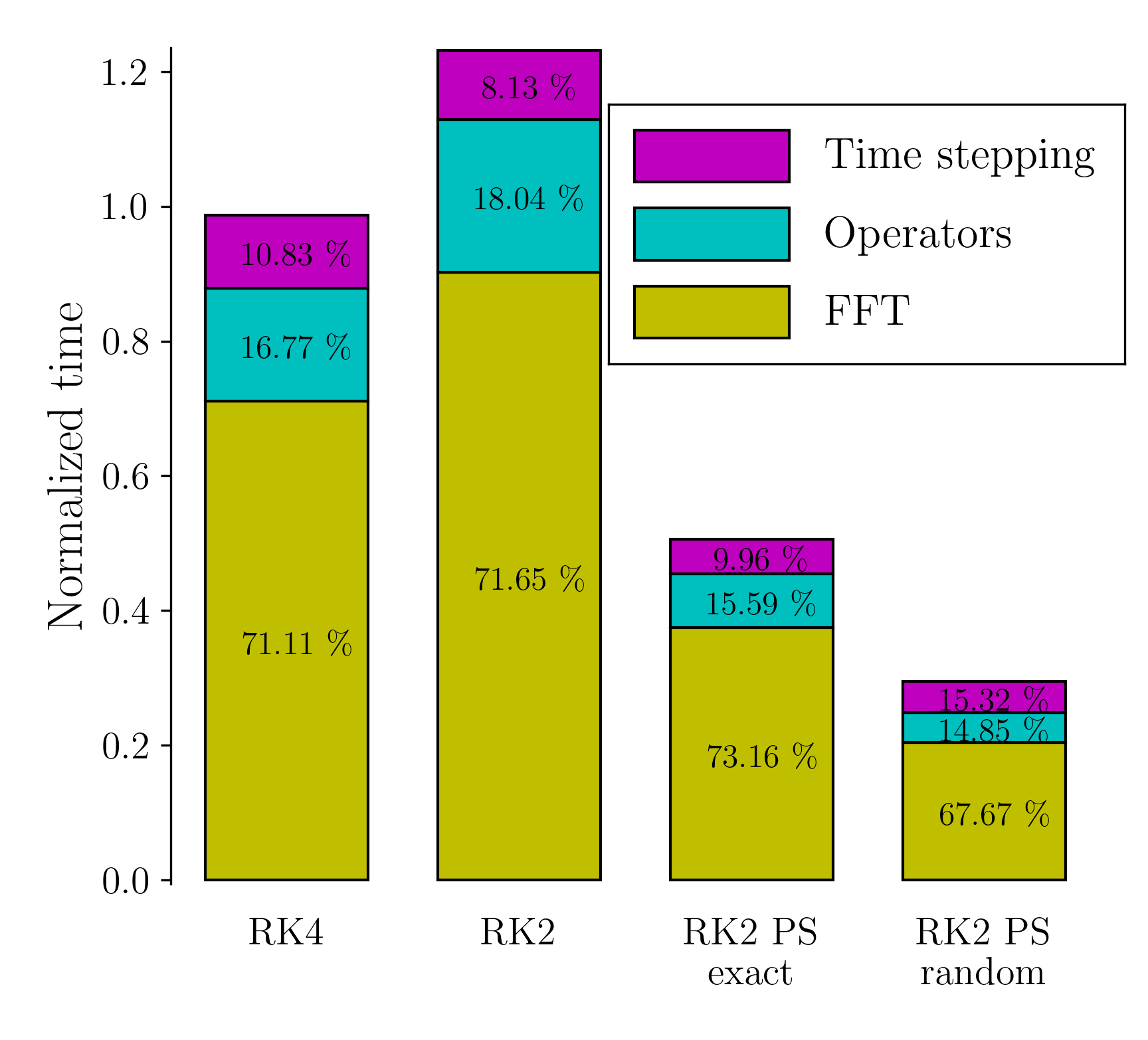}}
\caption{Distribution of computation time across different tasks (Fourier transforms,
operators, time stepping) for four parallel simulations run on 12 cores with different
time schemes up to the same physical time and with the same $\kmax$. For RK4 and RK2,
$N = 432$ and $C_t = 2/3$. For the two phase-shifting simulations, $N = 288$ and $C_t =
1$. Bar heights are proportional to elapsed time normalized by the RK4 elapsed time;
percentages are computed relative to the elapsed time of each method individually.
\label{fig:profiling}}
\end{figure}

Fig.~\ref{fig:profiling} compares the distribution of computation time across different
tasks for three simulations sharing the same maximum wavenumber $\kmax$: one using the
RK4 scheme (left), one using RK2 scheme (center left) and two using RK2-based
phase-shifting methods, exact (center right) and approximate (right). Because the RK4
scheme allows a larger CFL number than RK2, the RK2 and phase-shifting simulations
require more time steps to reach the same physical time. As a consequence, the
simulation time for RK2 simulation is greater than for RK4 simulation, it is therefore
appropriate to compare phase-shifting simulations to 2/3 truncation simulations with
the RK4 scheme rather than the RK2 scheme. For phase-shifting simulations, the
significantly smaller number of grid points reduces the computational cost, allowing
these simulations to complete in less than half the time required for the RK4
simulation, which counterbalances the impact of the increased CFL constraint. The
figure also shows that the fraction of time spent in the time-stepping stage is larger
for ``RK2 phase-shift random'' method than for RK4.

\begin{figure}[]
\centerline{\includegraphics[width=1.56875\figwidth]{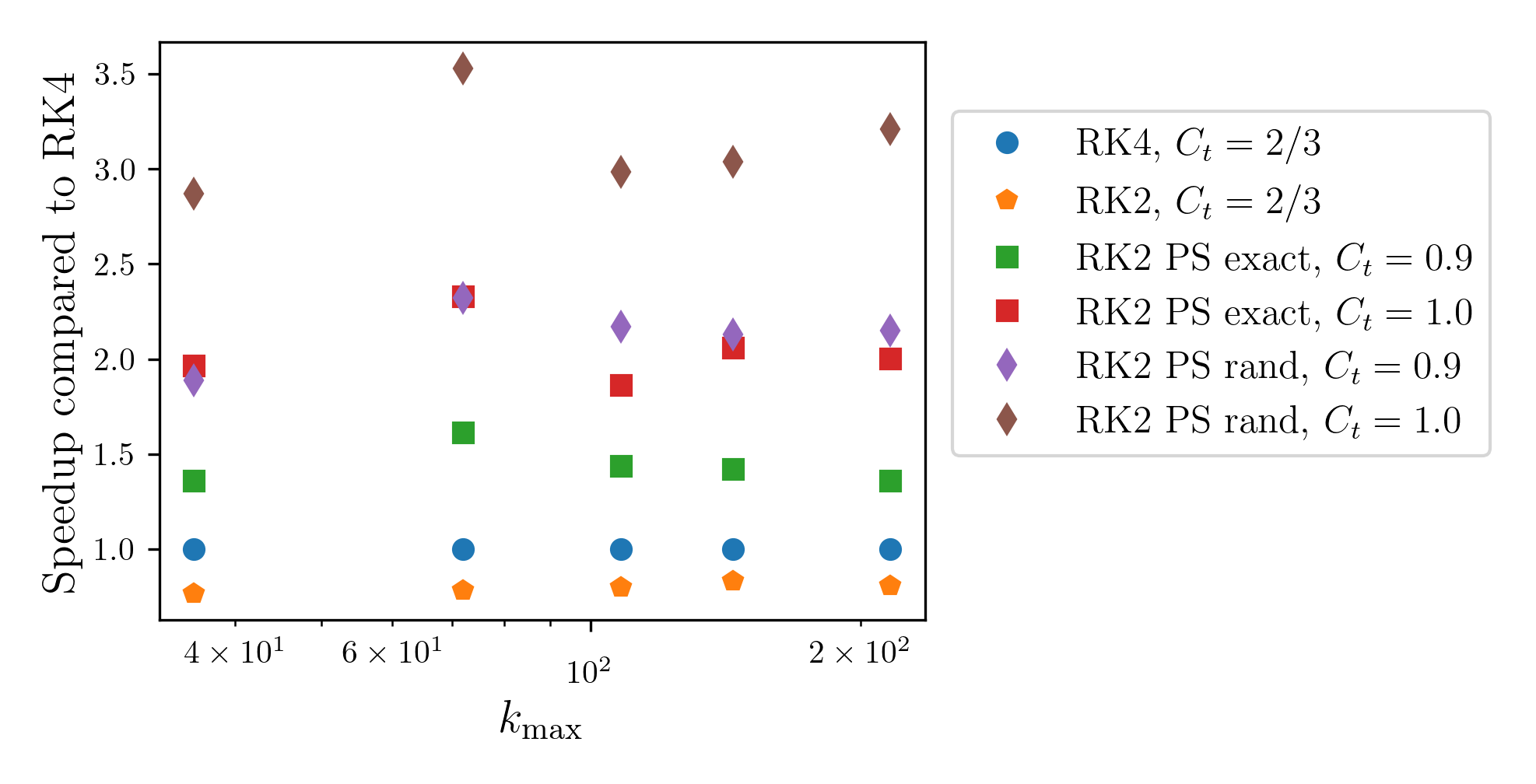}}
\caption{Speedup (elapsed time of the RK4 simulation at $C_t = 2/3$ divided by the
elapsed time of each simulation, for the same $\kmax$) as a function of $\kmax$, for
simulations with different dealiasing methods. For RK4 ($C_t = 2/3$), the values of
$\kmax$ correspond to $N = 108$, $216$, $324$, $432$, and $648$. \label{fig:speedup}}
\end{figure}

Fig.~\ref{fig:speedup} shows the speedup relative to the RK4 simulation, at $C_t = 2/3$
with the same $\kmax$, at various resolutions (various $\kmax$ values). By definition,
all RK4 points (blue) have a speedup of $1$. The plain RK2 scheme (orange) is
consistently slower than RK4, for the reasons explained above. All phase-shifting
methods are faster; for example, the elapsed time for ``RK2 phase-shift random'' is
approximately three times smaller than for RK4. No clear trend in speedup as a function
of $\kmax$ is observed.

\subsubsection{Effect of time scheme and Reynolds number}

We now present three series of four simulations each, listed in
Table~\ref{tab:all_series}. The three series correspond to different $(Re,\,
\kmax\etamin)$ combinations: $(1600,\, 1)$, $(2800,\, 0.6)$, and $(1600,\, 1.6)$. Each
series consists of four simulations: a reference run with the RK4 scheme and $C_t =
2/3$; a second RK4 run with $C_t = 1$; and two runs at $C_t = 1$ using the ``RK2
phase-shift exact'' and ``RK2 phase-shift random'' schemes.

\begin{table}[]
\centering \begin{tabular}{| c | c | c | c | c | c | c | c | c |}
\toprule
$N$ & $Re$ & scheme & $C_t$ & $k_{\max} \tilde\eta$ & number of proc. & CPU.h & CPU.h speedup & error (\%) \\
\midrule
256 & 1600 & RK4 & 2/3 & 1.0 & 4 & 13.4 & 1.0 & 0.0 \\
168 & 1600 & RK4 & 1 & 0.99 & 4 & 1.6 & 8.3 & 17.63 \\
168 & 1600 & RK2 PS exact & 1 & 0.99 & 4 & 3.6 & 3.7 & 6.31 \\
168 & 1600 & RK2 PS rand & 1 & 0.99 & 4 & 2.3 & 5.9 & 6.27 \\ \hline
256 & 2800 & RK4 & 2/3 & 0.63 & 4 & 14.9 & 1.0 & 0.0 \\
168 & 2800 & RK4 & 1 & 0.62 & 4 & 1.7 & 8.8 & 18.99 \\
168 & 2800 & RK2 PS exact & 1 & 0.62 & 4 & 3.5 & 4.3 & 8.19 \\
168 & 2800 & RK2 PS rand & 1 & 0.62 & 4 & 3.3 & 4.5 & 7.53 \\ \hline
400 & 1600 & RK4 & 2/3 & 1.56 & 10 & 82.0 & 1.0 & 0.0 \\
280 & 1600 & RK4 & 1 & 1.64 & 10 & 13.5 & 6.0 & 10.07 \\
280 & 1600 & RK2 PS exact & 1 & 1.64 & 10 & 38.4 & 2.1 & 4.37 \\
280 & 1600 & RK2 PS rand & 1 & 1.64 & 10 & 20.1 & 4.1 & 4.33 \\
\bottomrule
\end{tabular}
 \caption{Parameters for the
three series of decaying flow simulations with various time schemes. Errors are
expressed as a percentage relative to the reference simulation: the RK4 simulation at
$N = 256$ with $C_t = 2/3$ (first and fifth rows for the first two series) and at $N =
400$ (tenth row) for the third series.}\label{tab:all_series}
\end{table}

\begin{figure}[]
\centerline{
  \includegraphics[width=\figwidth]{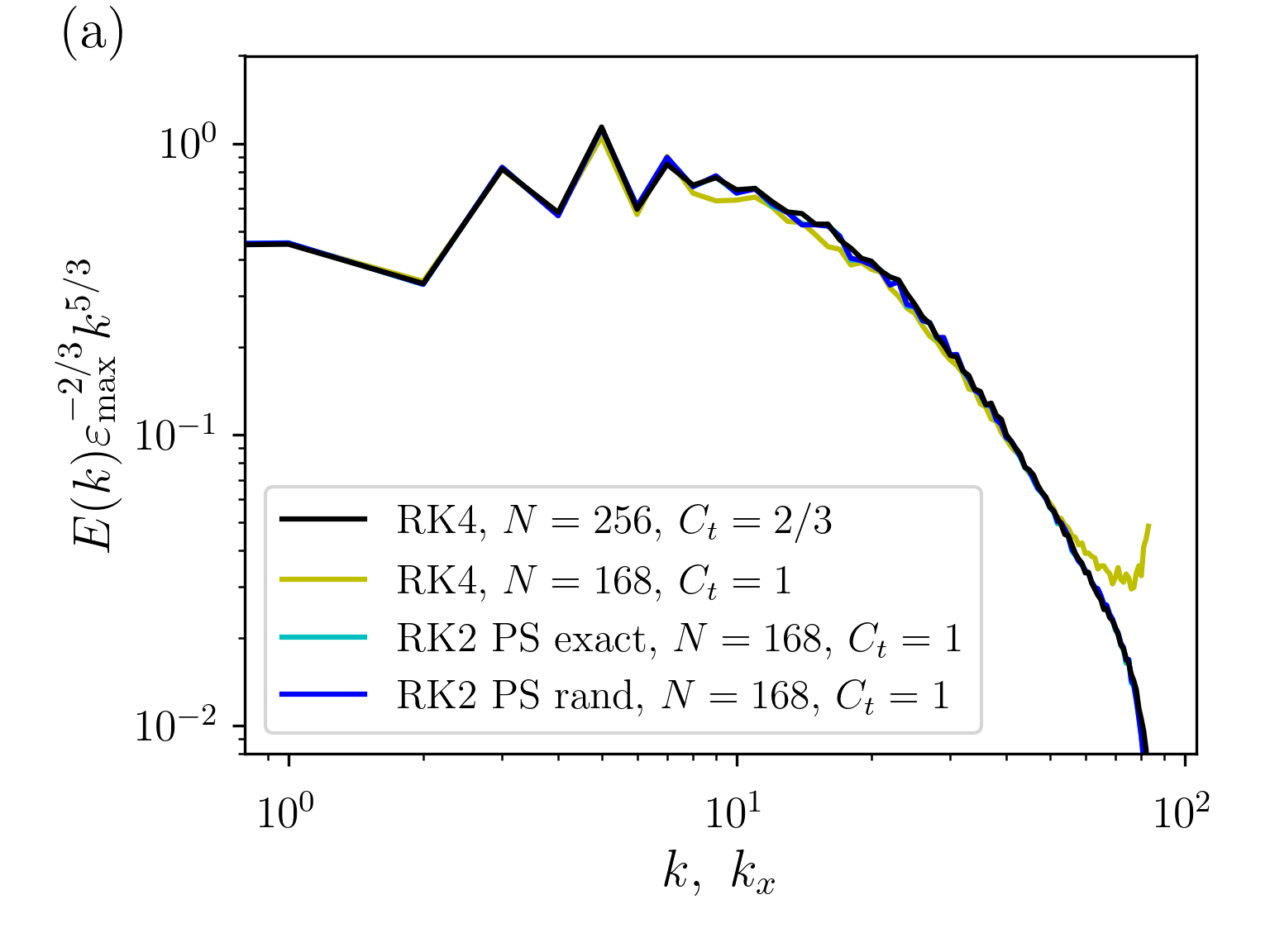}
  \includegraphics[width=\figwidth]{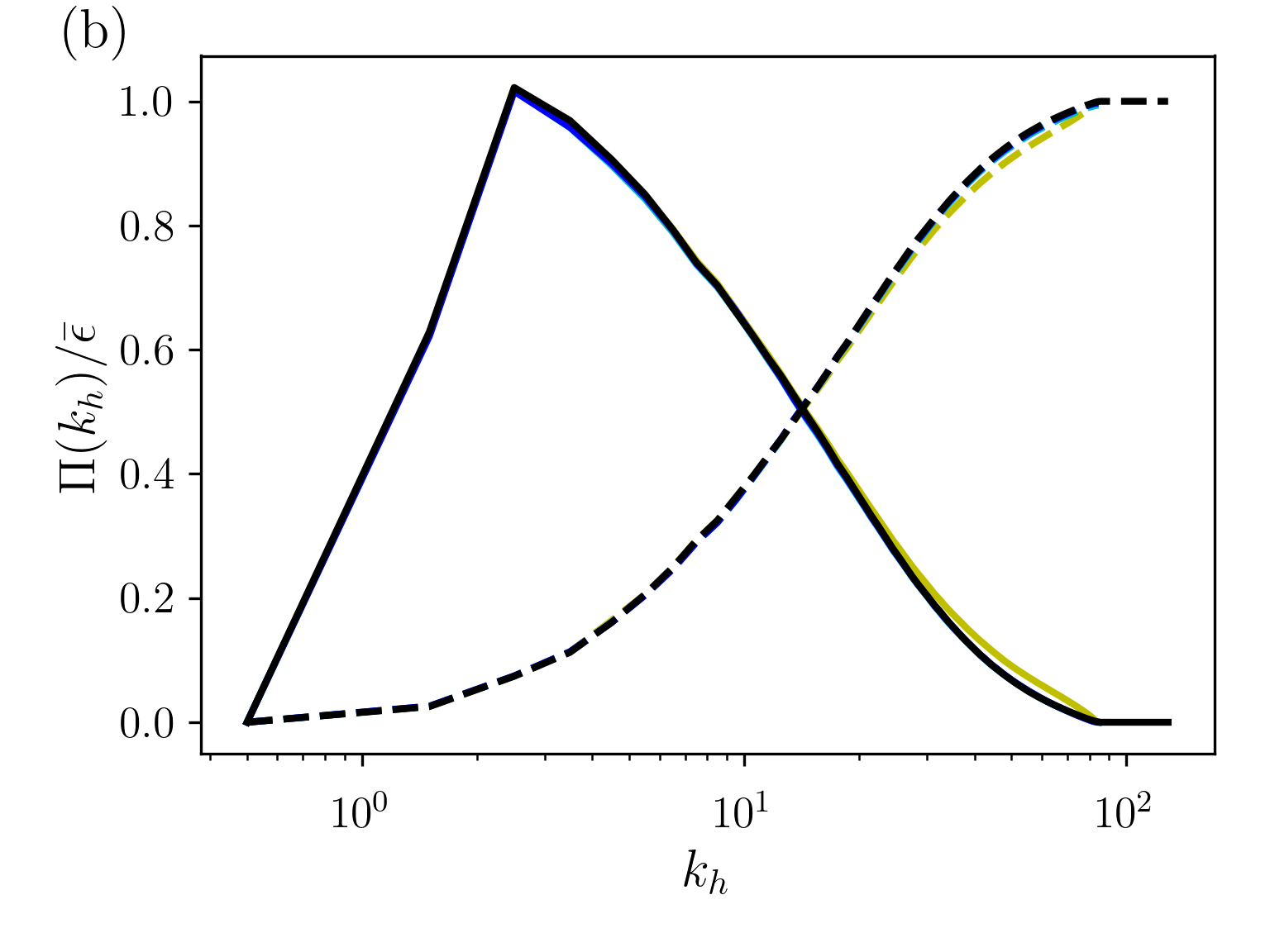}
}
\caption{Same as Fig.~\ref{fig:spectra:192} but for $Re = 1600$, $\kmax\etamin \simeq
1$, and various time schemes (first series of Table~\ref{tab:all_series}).
\label{fig:spectra:256}}
\end{figure}

\begin{figure}[]
\centerline{
  \includegraphics[width=\figwidth]{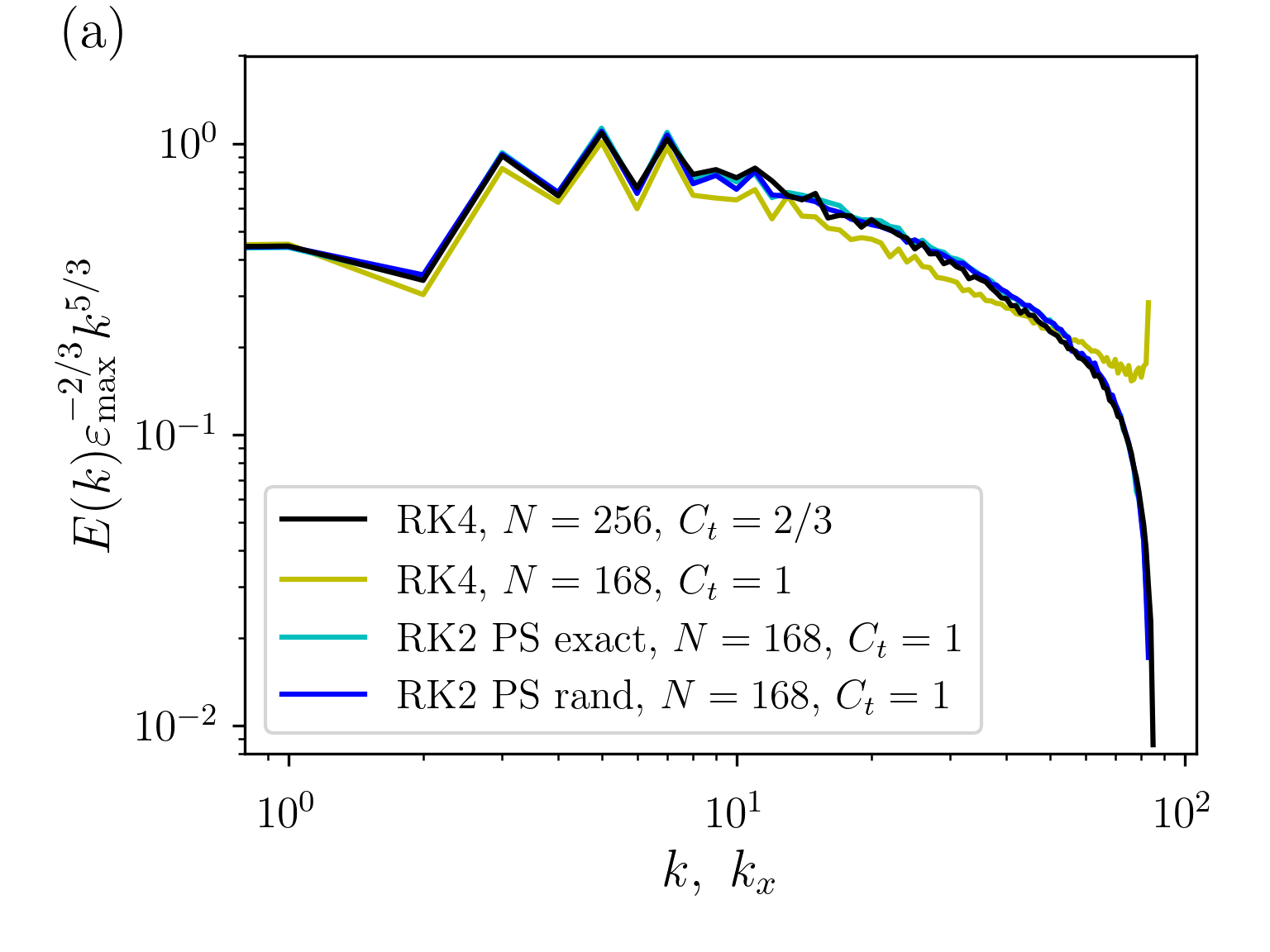}
  \includegraphics[width=\figwidth]{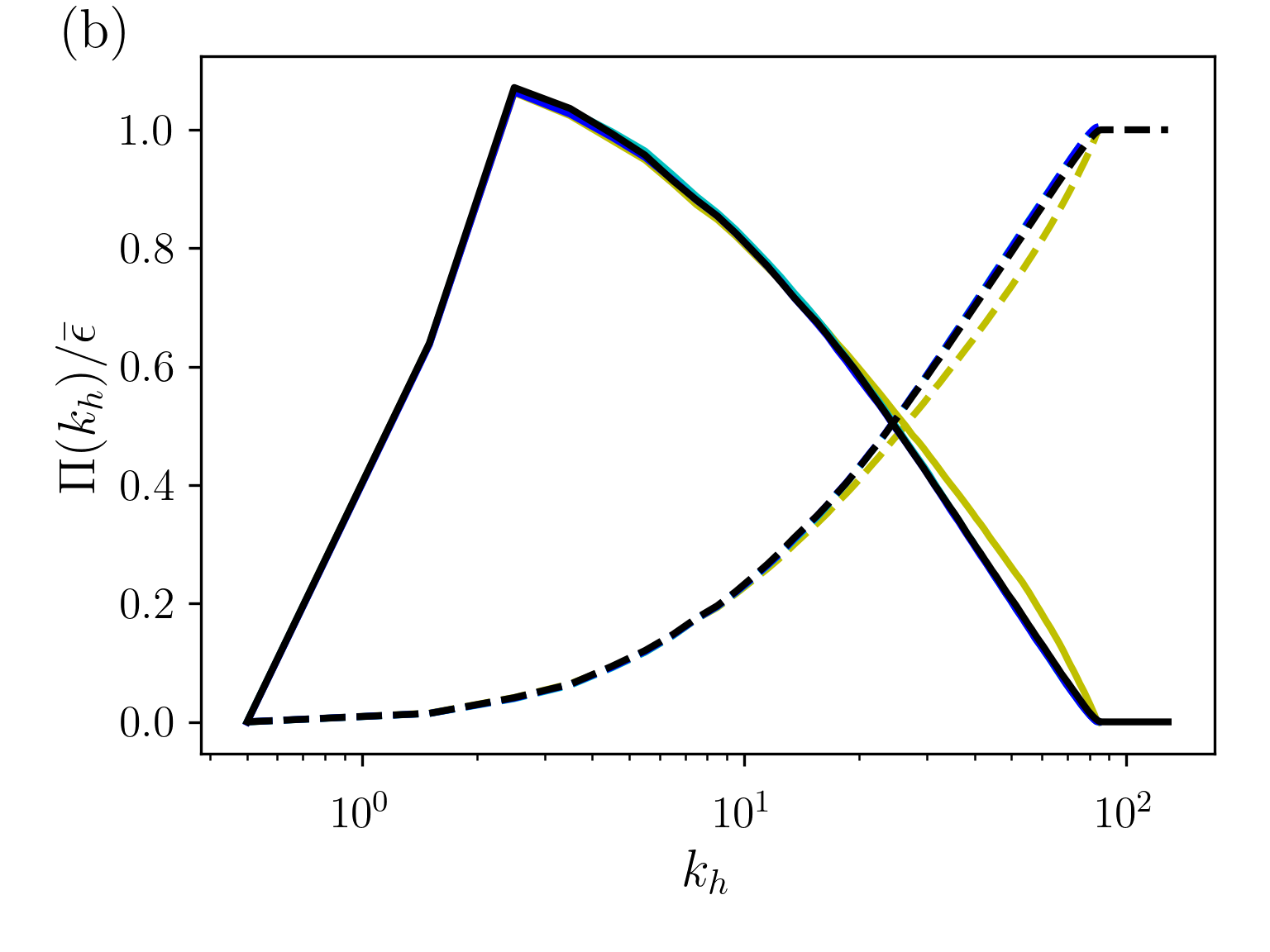}
}
\caption{Same as Fig.~\ref{fig:spectra:192} but for $Re = 2800$, $\kmax\etamin \simeq
0.6$, and various time schemes (second series of Table~\ref{tab:all_series}).
\label{fig:spectra:256:Re2800}}
\end{figure}

Figs.~\ref{fig:spectra:256} and \ref{fig:spectra:256:Re2800} show the spectra for both
cases sharing the same grid size but differing in Reynolds number. Spectra for the
third series are provided in the supplementary material \cite{supp}. Increasing $Re$
from $1600$ to $2800$ produces a more turbulent flow with an inertial range that more
closely follows the $-5/3$ law (Fig.~\ref{fig:spectra:256:Re2800}(a)), as opposed to
the lower-$Re$ case (Fig.~\ref{fig:spectra:256}). However, at fixed grid size,
increasing $Re$ reduces $\kmax\etamin$, leading to larger differences between
phase-shifting and truncated simulations in Fig.~\ref{fig:spectra:256:Re2800}. At a
given resolution, the phase-shifting schemes produce very similar spectra to one
another, while the RK4 scheme with $C_t = 1$ shows noticeably larger deviations from
the reference.

\begin{figure}[]
\centerline{
  \includegraphics[width=1.1\figwidth]{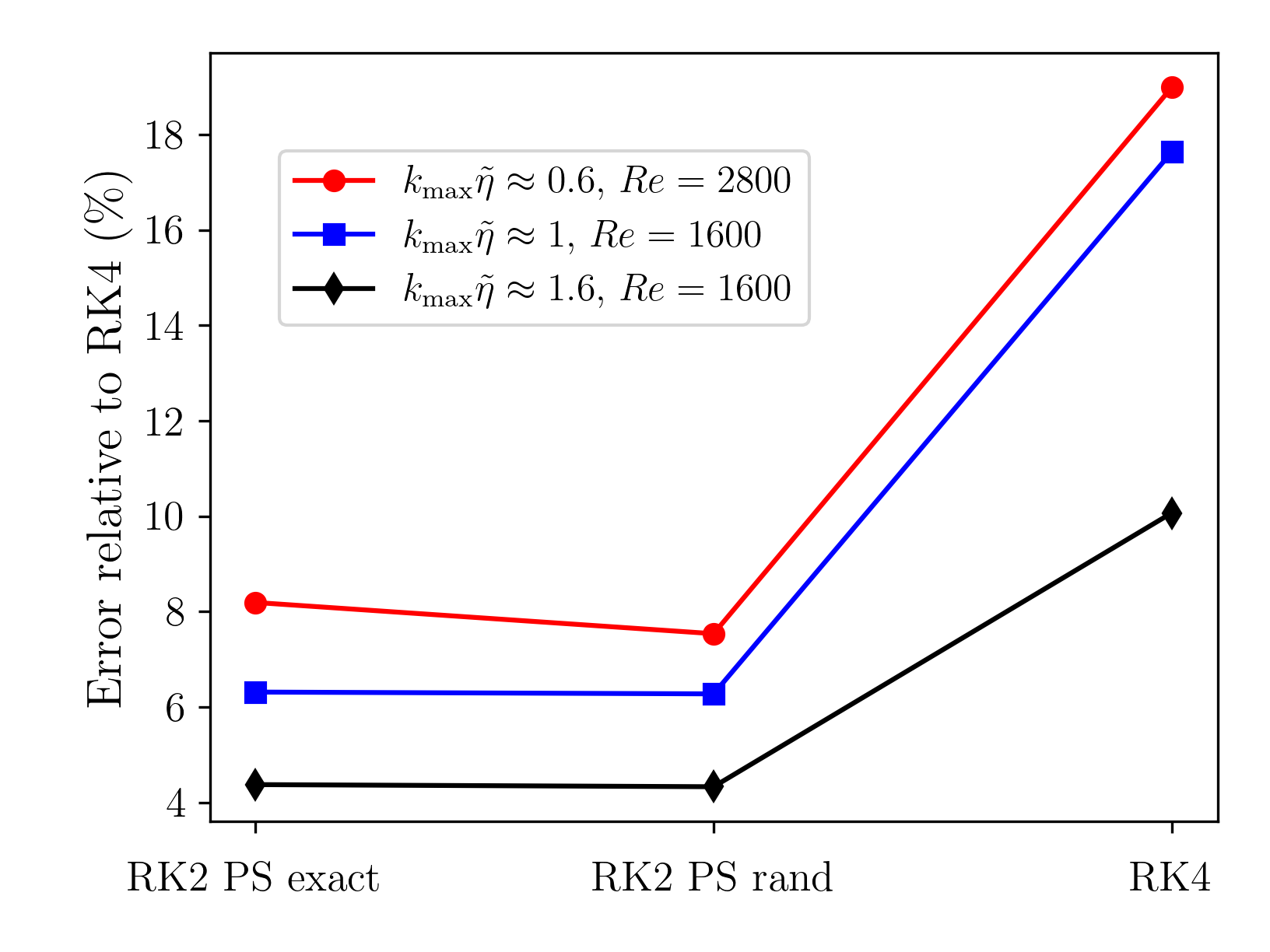}}
\caption{Spectral error in \% relative to the reference simulation (RK4 with $C_t =
2/3$, giving $0\%$ error, see Eq.~\eqref{eq:quality}) versus time scheme at various
$\kmax \etamin$. Simulations correspond to
Table~\ref{tab:all_series}.\label{fig:error-all}}
\end{figure}

Fig.~\ref{fig:error-all} shows the errors for all simulations listed in
Table~\ref{tab:all_series}. The phase-shifting methods yield errors of comparable
magnitude, all significantly lower than the fully aliased case. As resolution
decreases, the error increases, consistently with the spectral observations above and
the result from Fig.~\ref{fig:error-sizes} which can thus be generalized to
phaseshifting-based time schemes. Suprisingly, the errors between the random and the
exact time schemes are the same order, with a higher speedup, the former appears
therefore as a good candidate to replace the 2/3-truncation method.

\subsubsection{Quality and speedup}

\begin{figure}[]
\centerline{
  \includegraphics[width=1.4125\figwidth]{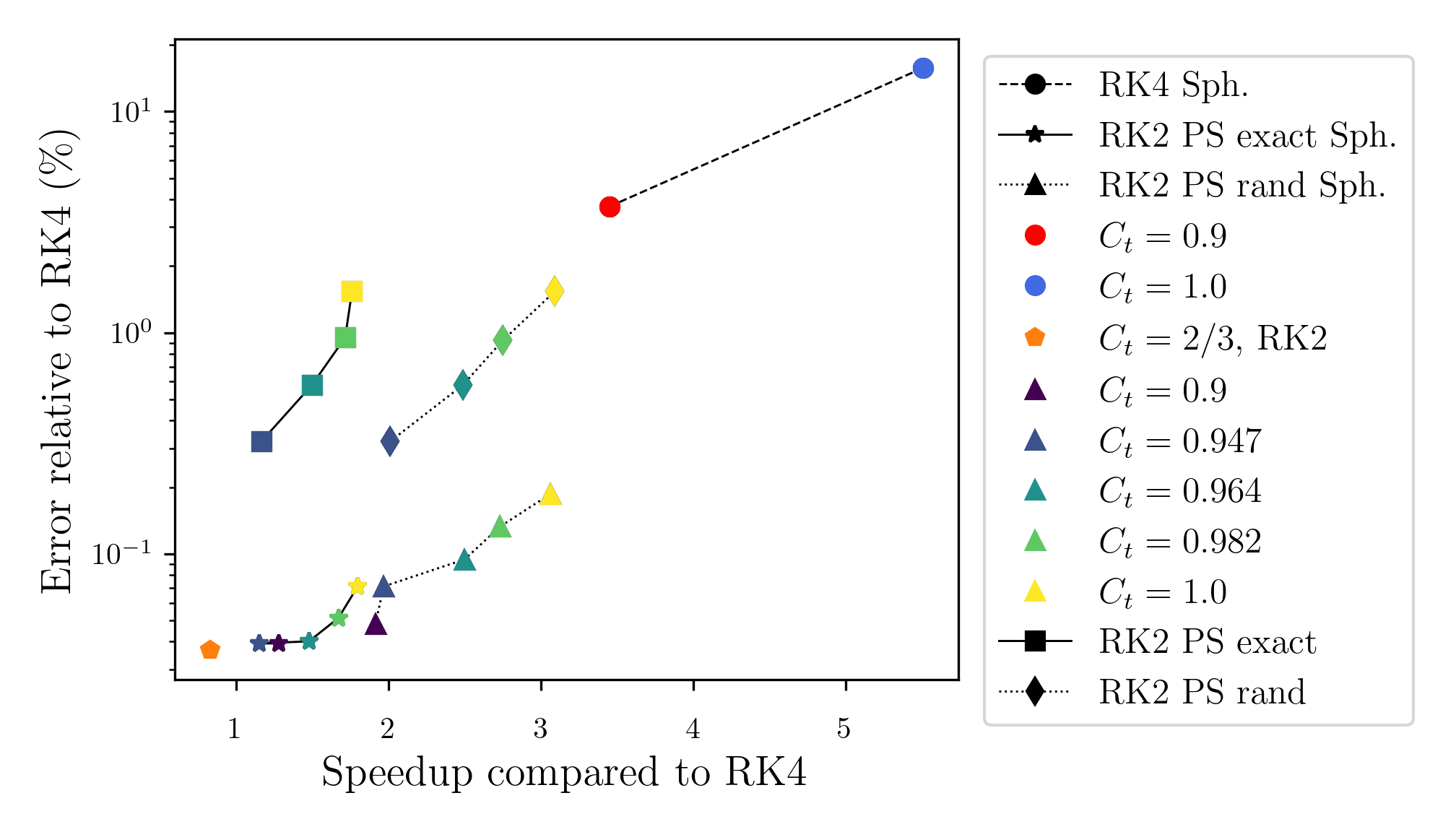}}
\caption{Spectral error in \% relative to the reference simulation (RK4 with $C_t =
2/3$, see Eq.~\eqref{eq:quality}) as a function of speedup for $\kmax = 108$
($\kmax\etamin \approx 1.27$), for simulations with different dealiasing methods and
computed in parallel on $6$ cores. The truncation coefficients $C_t = 2/3$, $0.9$,
$216/228 \approx 0.947$, $216/224 \approx 0.964$, $216/220 \approx 0.982$ and $1$
correspond respectively to grid sizes $N = 324$, $240$, $228$, $224$, $220$ and $216$.
``Sph.'' means that the truncation is purely spherical (double aliases remains), when
it is not specified it means that double aliases are also truncated. For each points of
simulations with phase-shifted time schemes, the speedup is averaged over $10$
profiles. \label{fig:quality}}
\end{figure}

To generalize the conclusions of the previous section, we explore whether intermediate
combinations of truncation and phase shifting can improve the compromise between
speedup and accuracy. Because the speedup trends appear to be independent of $\kmax$,
we fix $\kmax = 108$ ($\kmax\etamin \approx 1.27$) and compare all combinations of time
scheme and truncation coefficient against the reference simulation (RK4, $C_t = 2/3$)
with purely spherical truncation and combinations of phase-shifting methods and
truncation coefficient above $0.94$ with double aliases removal. For $C_t > 0.94$ and
double aliases removal, the truncation is no longer exactly isotropic, so we can
compare the departure from isotropy to the error due to double aliases remaining in
purely spherical truncation shape.

Fig.~\ref{fig:quality} shows the spectral error versus speedup for each combination.
The results follow the overall trend already visible in Fig.~\ref{fig:error-all} and
Table~\ref{tab:all_series}: low error and low speedup in the lower-left, high error and
high speedup in the upper-right and the random scheme always faster than the exact.
Three groups emerges clearly at three ranges of error: (1) high-quality simulations at
low error ($<0.2\%$) with purely spherical truncation, and (2) lower-quality
simulations at intermediate error (between $0.3\%$ to $2\%$) with double aliases
removed and (3) bad quality simulations at large error ($>4\%$, red and blue circles).
The latter group allows aliased simulations to be used as a reference for bad quality
simulations. Quality gaps of at least one order of magnitude separate the two first
groups, and gaps of up to almost two orders of magnitude separate phase-shifted from
non-phase-shifted simulations at the same $C_t$. Suprisingly, double alias removal
increases error instead of reducing it relatively to purely spherical simulations.
However, since double alias removal with phase-shifting exactly removes all aliases,
the gap between both cases is not due to aliasing error. This error is solely due to
the anisotropy in the shape of the truncation. In spectral space, this shape differs
from the isotropic purely spherical shape by the intersection between the 3D shape from
Fig.~\ref{fig:truncation:double}(a) and a sphere of radius $C_t\kmax\delta k$. Note
that since phaseshifted simulations from Table~\ref{tab:all_series} and
Fig.~\ref{fig:error-all} are performed with double aliases removed, their high errors
are probably due to the truncation shape, and an additional departure ($4.33\%$ and
$1.54\%$ error for the random scheme simulations at $N = 280$ in
Fig.~\ref{fig:error-all} and $N = 216$ in Fig.~\ref{fig:quality} respectively) due to
an adaptative time step unlike simulations from Fig.~\ref{fig:quality}. Also, even
though small non-systematic departure is observed, the speedup at a given $C_t$ is
about the same between both groups and only depends on the resolution.

Within the two first groups, notable variations in speedup are observed. The general
trend from Fig.~\ref{fig:speedup}—faster simulations for the random method than the
exact one, and increasing speedup with $C_t$—is broadly confirmed, but with two
exceptions. For the exact phase-shifting method, the simulations at $C_t = 0.947$ ($N =
228$) are slower than at $C_t = 0.9$ ($N = 240$). These anomalies can be traced to the
dependence of FFT performance on the prime factorization of the grid size $N$. Detailed
profiling shows that for the exact method, the FFT accounts for $86\%$ of the elapsed
time at $N = 228$ ($81\,\text{s}$) but only $76\%$ at $N = 240$ ($51\,\text{s}$). The
FFT is significantly faster for grid sizes with large powers of 2 in their
factorization: $240 = 2^4 \times 3 \times 5$ outperforms $228 = 2^2 \times 3 \times
19$. For the random method, the FFT accounts for only $68\%$ of the elapsed time at $N
= 240$ ($27\,\text{s}$), meaning the speedup gain from reducing $N$ is dominant over
the prime factorization effect, and the speedup trend with $C_t$ is regular.

Overall, these results demonstrate that removing double aliases lowers simulations
quality and that intermediate combinations of phase shifting and truncation can
usefully optimize the speedup–accuracy trade-off even for truncation coefficients over
the classical $2\sqrt{2}/3$. The best compromise identified here is for purely
spherical truncation with the ``RK2 phase-shift random'' scheme and a truncation
corfficient in between $C_t = 0.964$ and $C_t = 1$.

\section{Homogeneous isotropic forced turbulence}


\label{section-forced}

In this section, we describe the time-stepping scheme adapted to allow phase-shifting
methods for forced flows, then present results from simulations of homogeneous
isotropic turbulence (HIT) using phase shifting and truncation. \citet{ESWARAN1988}
previously combined isotropic forcing with a phase-shifting scheme adapted from
\citet{Rogallo1981}, but do not discuss the scheme adaptation itself, and their study
focuses on flow statistics rather than on the quantification of aliasing errors.

\subsection{Description of the phase-shifting method with forcing}

We consider an extended form of Eq.~\eqref{eq:base-eq} \cite{ESWARAN1988}:

\begin{equation}
\partial_t S = \sigma S + \mathcal{N}(S) + O(S),
\label{eq:forcing-eq}
\end{equation}
where $O$ represents forcing and any other nonlinear terms that are not sources of
aliasing (see below). The scheme used here is identical to ``RK2 phase-shift random''
except for the addition of forcing terms. The first stage uses $\tau = dt$, $S_a =
S_0$, and $\mathcal{N}_a = \widetilde{\mathcal{N}_{0\alpha}} + O_0$; the second stage
uses $\tau = dt$, $S_a = S_0$, and:

\begin{equation}
\left. \frac{\mathcal{N}}{S}\right\vert_a =
  \frac{\tfrac{1}{2}\!\left(\widetilde{\mathcal{N}_{0\alpha}} + \widetilde{\mathcal{N}_{1\beta}} + O_0 + O_1\right)}{S_0 e^{\sigma dt /2}}.
\end{equation}

This formulation is valid provided that $O$ does not generate aliasing, so that it does
not need to be dealiased by phase shifting, as in \cite{ESWARAN1988}. This condition is
satisfied here: the flow is forced at large scales (small $k$ relative to $\kmax$) and
the forcing involves no nonlinear products between modes. In the following, this scheme
is referred to as ``RK2 phase-shift random split'' and has been implemented in the
\fluidsimsolver{ns3d} solver.

\subsection{Performance and accuracy for isotropically forced flows}

We perform simulations for various grid sizes and dealiasing methods, exploring
combinations of purely spherical truncation and the random phase-shifting method for
HIT flows. For all simulations, the Reynolds number is fixed at $Re = 850$ (Taylor
Reynolds number $Re_\lambda \approx 200$), which allows reaching $\kmax\etamin \approx
2.07$ with moderate grid sizes (up to $N = 960$ for the RK4 reference simulation). The
numerical setup follows Section~\ref{section-num-methods}, and all simulations are
listed in Table~\ref{tab:960}.

\begin{table}[]
\begin{tabular}{| c | c | c | c | c | c | c | c |}
\toprule
$N$ & scheme & $C_t$ & nb of proc. & CPU.h speedup & $\mathrm{Err}_{\mathrm{early}}$ & $\mathrm{Err}_{\mathrm{late}}$ & $\mathrm{Err}(\overline{E})$ \\
\midrule
960 & RK4 & 2/3 & 60 & 1.0 & 0.0 & 0.0 & 0.0 \\
896 & RK2 PS rand split & 0.714 & 64 & 0.7 & 0.6 & 5.42 & 3.37 \\
768 & RK2 PS rand split & 0.833 & 64 & 1.4 & 0.73 & 4.41 & 2.67 \\
696 & RK2 PS rand split & 0.920 & 58 & 1.7 & 0.73 & 4.22 & 1.6 \\
672 & RK2 PS rand split & 0.952 & 56 & 2.4 & 0.81 & 3.78 & 1.83 \\
640 & RK2 PS rand split & 1 & 64 & 2.9 & 0.77 & 4.83 & 2.5 \\
\bottomrule
\end{tabular}

\caption{Forced HIT simulations carried out in Section~\ref{section-forced}. Errors are
defined by Eq.~\eqref{eq:quality} as percentages relative to the reference simulation
at $N = 960$ and $C_t = 2/3$. In all simulations $Re = 850$ and $\kmax\etamin =
2.07$.}\label{tab:960}
\end{table}

\begin{figure}[]
\centerline{\includegraphics[width=\figwidth]{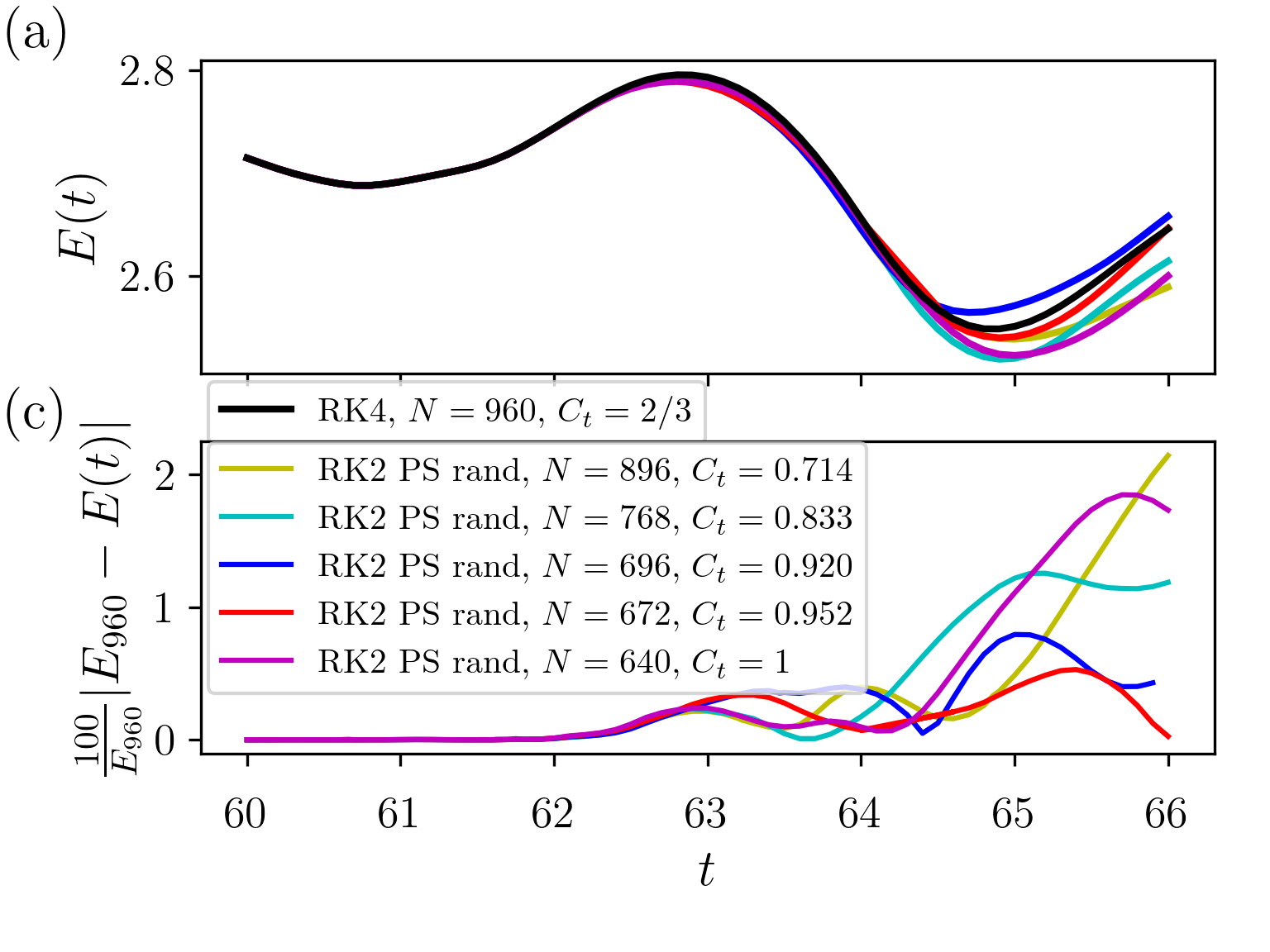}
  \includegraphics[width=\figwidth]{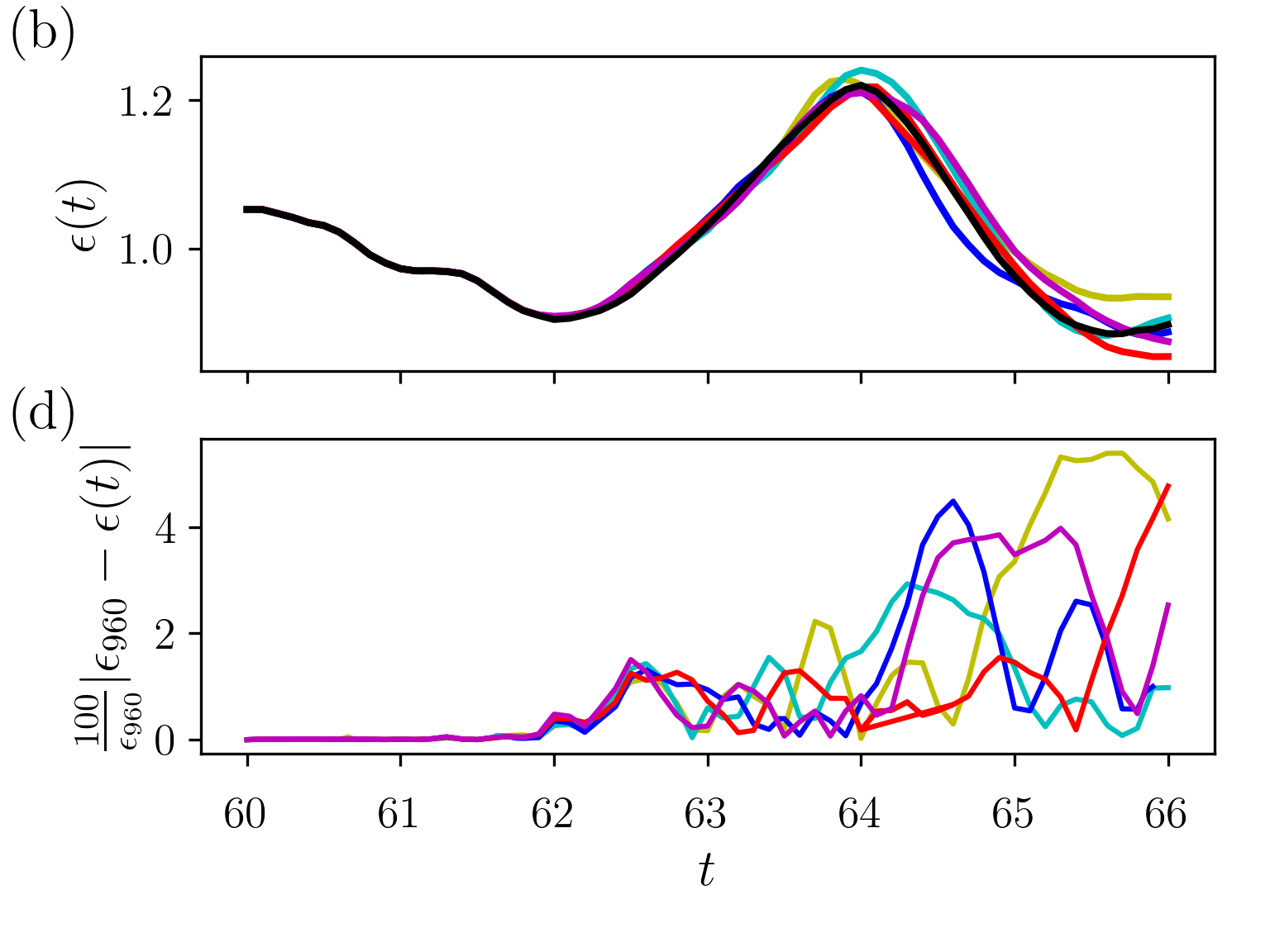} }
\caption{Same as Fig.~\ref{fig:spatialmeans:192} for HIT flows with $Re = 850$,
$\kmax\etamin = 2.07$, and various combinations of truncation and phase shifting.
Simulations correspond to Table~\ref{tab:960}. Note that curves ovelap in (a) and (b)
\label{fig:spatialmeans:960}}
\end{figure}

The temporal evolution of the total kinetic energy and energy dissipation rate is shown
in Fig.~\ref{fig:spatialmeans:960}(a) and (b). All simulations are initialized from a
state that has already reached statistical stationarity, which is why the time axis
starts at $t_0 = 60$ overturn times. The slow oscillation of both quantities around a
constant mean is characteristic of a statistically stationary forced flow, in contrast
to the decaying Taylor-Green flows studied previously.

Figs.~\ref{fig:spatialmeans:960}(c) and (d) show the corresponding errors relative to
the $N = 960$ reference simulation. In the first time interval ($60 < t < 63$), all
simulations show small errors. In the second interval ($63 < t < 66$), the errors grow
and begin to oscillate for all resolutions. This growth is expected: because the
forcing depends on the instantaneous flow state, small differences in the numerical
parameters (grid size, time scheme, truncation coefficient) are amplified in the same
way as physical perturbations in a turbulent flow. Specifically, the perturbation
$\delta S$ between two trajectories satisfies a linearized equation of the form
$\partial_t \delta S \sim \lambda \delta S$, where $\lambda$ is related to the Lyapunov
exponent of the turbulent flow, leading to approximately exponential growth of the
error at short times. The growth eventually saturates because both the energy and the
dissipation rate are statistically stationary. The errors will therefore fluctuate
around a finite long-time mean, rather than growing indefinitely. Since no systematic
trend across methods is apparent in the time traces, the errors are quantified more
precisely using time-averaged energy spectra in the following.

\begin{figure}[]
\centerline{
  \includegraphics[width=\figwidth]{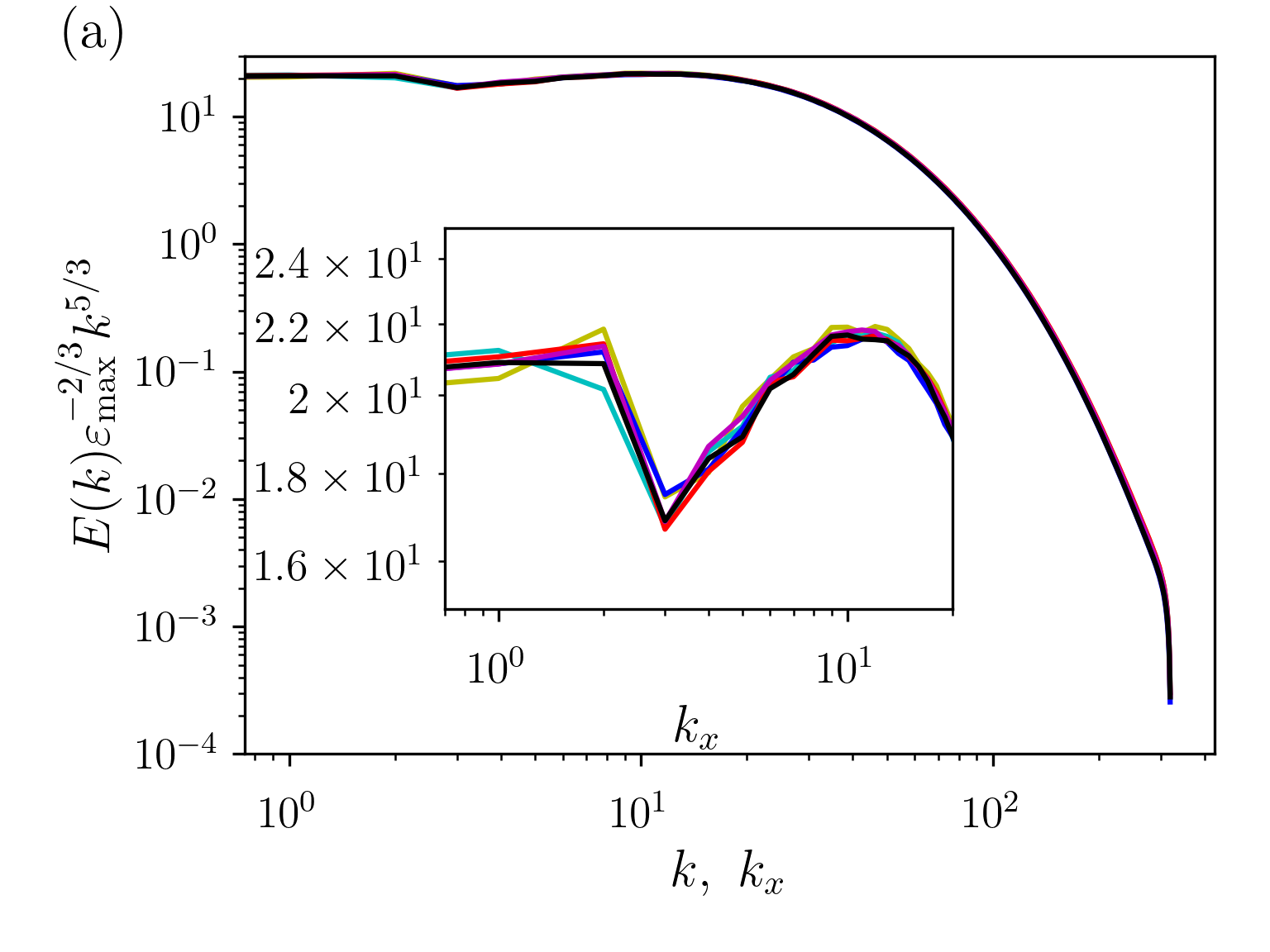}
  \includegraphics[width=\figwidth]{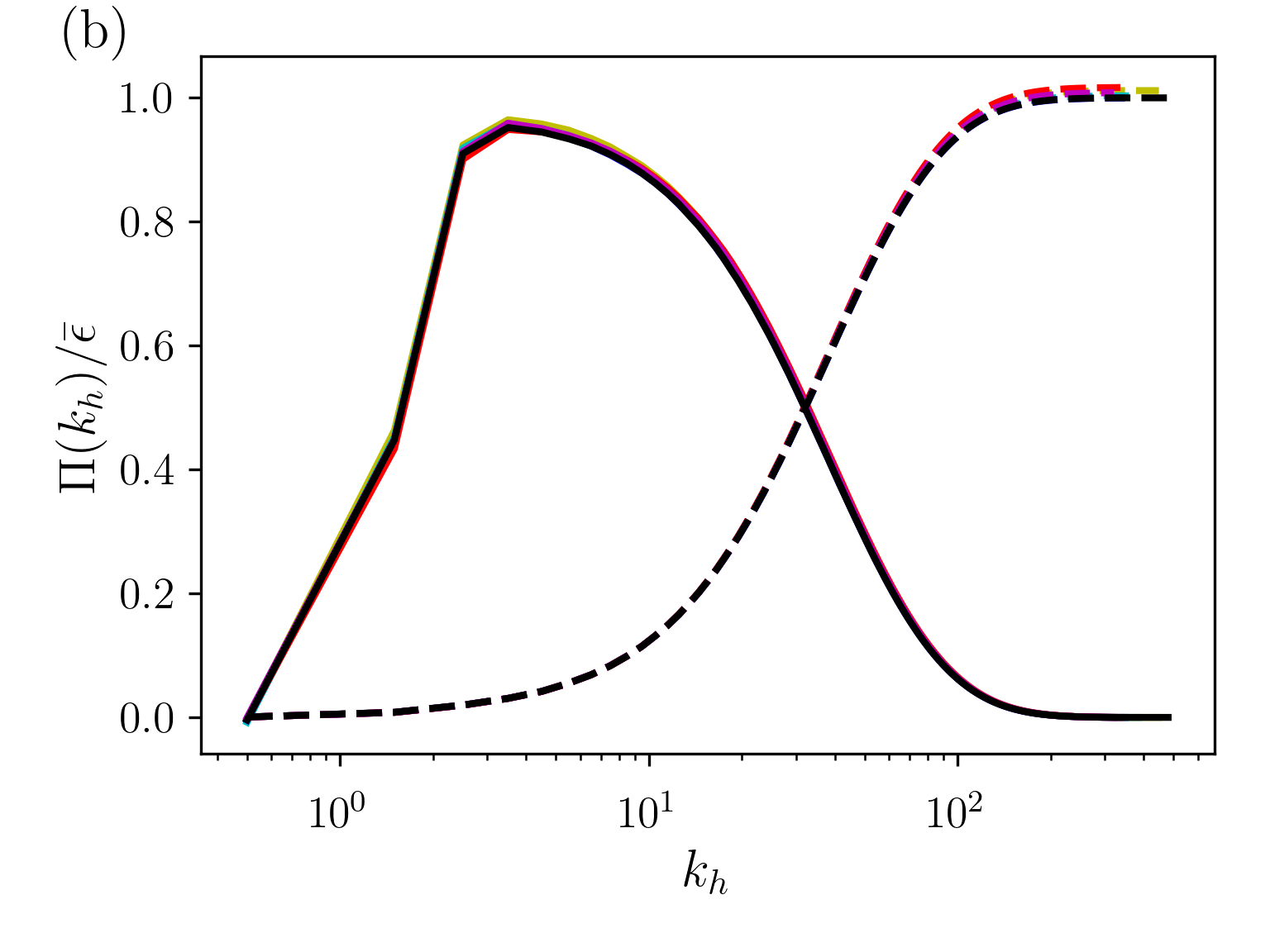}
}
\caption{Same as Fig.~\ref{fig:spectra:192}, averaged in time from $63$ to $66$, for
HIT flows with $Re = 850$, $\kmax\etamin = 2.07$, and various combinations of
truncation and phase shifting. Simulations correspond to Table~\ref{tab:960}. The
legend is the same as in Fig.~\ref{fig:spatialmeans:960}(a). Note that curves overlap
in most of the spectral domain. The inset in (a) is a zoom of the energy spectra at
large scales. \label{fig:spectra:960}}
\end{figure}

Fig.~\ref{fig:spectra:960} shows (a) the energy spectra and (b) the energy fluxes and
cumulated dissipation, averaged in time from $63$ to $66$. The results are remarkably
similar across all simulations and the reference (black lines). Despite small
variations at large scales (see the inset of Fig.~\ref{fig:spectra:960}(a)), all
simulations share essentially the same spectral properties.

These spectral differences are quantified in Table~\ref{tab:960}, together with speedup
values. Three error metrics are reported. The first, $\mathrm{Err}_{\mathrm{early}}$,
is the spectral error computed exactly as in Eq.~\eqref{eq:quality} over the first half
of the simulation time, from $t = 60$ to $t = 63$ overturn times. This interval
corresponds to the regime of small, slowly growing errors identified above, and
provides the cleanest measure of the intrinsic numerical error. The second,
$\mathrm{Err}_{\mathrm{late}}$, is the same quantity computed over the second half,
from $t = 63$ to $t = 66$ overturn times, and reflects the regime in which chaotic
divergence has amplified the differences between trajectories. The third metric,
$\mathrm{Err}(\overline{E})$, is computed from the time-averaged spectra rather than
instantaneous ones: it evaluates $100|\overline{E}(k_x) -
\overline{E}_{\mathrm{ref}}(k_x)|/\overline{E}_{\mathrm{ref}}(k_x)$ after averaging in
time from $63$ to $66$, and is therefore insensitive to trajectory divergence. It
measures how well each simulation reproduces the statistically stationary spectral
distribution.

The results show that $\mathrm{Err}_{\mathrm{early}}$ is generally low at high
resolutions, even for forced flows. Consistently with previous observations, it
decreases as the truncation coefficient is reduced (i.e., as more modes are removed),
and so does the speedup, with one exception: the simulation at $N = 896$ is slower than
the reference. This occurs because when the grid size is similar for phase-shifted and
purely truncated simulations, the computational gain from the smaller grid is
outweighed by the cost of the smaller time step required by the RK2 scheme relative to
RK4. This effect was also visible in Fig.~\ref{fig:speedup}.

In contrast, $\mathrm{Err}_{\mathrm{late}}$ and $\mathrm{Err}(\overline{E})$ show no
clear trend with resolution or truncation. For instance, $\mathrm{Err}_{\mathrm{late}}$
at $N = 696$ is lower than at $N = 768$, even though $\mathrm{Err}(\overline{E})$ is
higher for the former. The simulation at $N = 896$ has the highest
$\mathrm{Err}(\overline{E})$ and elevated $\mathrm{Err}_{\mathrm{late}}$, while its
$\mathrm{Err}_{\mathrm{early}}$ is among the lowest. These apparently inconsistent
rankings confirm that the late-time and time-averaged errors at these resolutions are
dominated by the chaotic trajectory divergence discussed above, rather than by
systematic numerical biases. The apparent randomness of the error fluctuations in
Figs.~\ref{fig:spatialmeans:960}(c) and (d) is consistent with this interpretation.

As a consequence, no particular advantage is gained from combining phase shifting and
truncation with relatively small $C_t$ for well-resolved HIT flows. The optimal choice
is to use phase-shifted time schemes with purely spherical truncation at $C_t = 1$,
which yields simulations up to $2.9$ times faster than the reference (pure truncation
with RK4), with a global error $\mathrm{Err}(\overline{E})$ of $2.5\%$.

\section{Conclusions}


This study provides a comprehensive treatment of phase-shifting dealiasing methods for
pseudo-spectral simulations of the incompressible Navier-Stokes equations, addressing a
long-standing gap in the literature. We have described the physical origin of aliasing
errors arising from quadratic nonlinearities in discrete Fourier space and explained
how phase-shifting methods cancel these errors exactly or approximately depending on
the time-stepping scheme. Our work includes detailed derivations of the aliasing
mechanism, systematic comparisons of different phase-shifting algorithms, and thorough
evaluations of their performance and accuracy on realistic turbulent flows.

The algorithms evaluated include the exact and approximate RK2 phase-shifting schemes
of \citet{PattersonOrszag1971} and \citet{Rogallo1981}, as well as an extension to
forced flows. All methods have been implemented in the open-source framework Fluidsim,
providing the first publicly available implementation of phase-shifting dealiasing for
pseudo-spectral Navier-Stokes solvers, and bridging the gap between textbook
descriptions and practical use in open-source research codes.

Our evaluations on decaying Taylor-Green vortex flows and forced homogeneous isotropic
turbulence demonstrate significant computational efficiency gains. For decaying flows
with $\kmax\etamin \approx 1$, the ``RK2 phase-shift random'' scheme with purely
spherical truncation at $C_t \approx 0.964$ achieves speedups of up to a factor of $3$
relative to the reference RK4 scheme with 2/3 truncation, while maintaining errors
below $0.1\%$. For forced flows at $Re_\lambda \approx 200$, the same method with
minimal truncation ($C_t = 1$) reproduces the statistically stationary spectral
properties to within $2.5\%$ while running $2.9$ times faster than the reference
simulation. These results confirm that phase-shifting methods can reduce the
computational cost of dealiasing from approximately $80\%$ of total simulation time to
a much smaller fraction.

Based on these findings, we offer the following practical guidelines. Neither
double-alias removal nor $C_t = 2\sqrt{2}/3$ are necessary to obtain qualitative
turbulent flow simulations. For most applications, the ``RK2 phase-shift random''
scheme with purely spherical truncation offers the best compromise between
computational efficiency and simulation quality. For decaying flows, a truncation
coefficient $C_t$ slightly below $1$ provides a good balance between accuracy and
efficiency. For forced flows, minimal truncation at $C_t = 1$ is sufficient, because
the errors at late times are dominated by chaotic trajectory divergence rather than by
systematic numerical bias, and the time-averaged spectral properties are accurately
reproduced regardless of the truncation. More generally, the choice of numerical
parameters should account for the desired $\kmax\etamin$, the specific flow
characteristics, and the required precision.

While this study focuses on the incompressible Navier-Stokes equations, the methods are
directly applicable to other systems with quadratic nonlinearities. Our evaluations
were conducted at moderate Reynolds numbers, but the fundamental aliasing mechanisms
and dealiasing strategies are expected to extend to higher Reynolds numbers. The
current implementation uses CPU-based computation, leaving opportunities for further
performance gains through GPU acceleration.

Finally, it is worth placing these efficiency gains in a broader perspective. The
numerical cost of a DNS scales as $N^4$, combining $N^3$ grid operations and a
CFL-limited time step proportional to $1/N$. A simulation at $N = 32768$
\cite{YEUNG2025} is therefore at least $10^6$ times more expensive than one at $N =
1000$ with the same code. Methods such as phase shifting can reduce this cost by a
factor of roughly $3$, and modern GPUs offer a similar factor over CPUs for
double-precision arithmetic. These gains are welcome, but they are negligible compared
to the six orders of magnitude separating moderate- and extreme-resolution simulations.
The $N^4$ scaling implies that even a factor-of-$3$ reduction in $N$ yields a $3^4
\approx 80$-fold decrease in cost—a much more effective lever than algorithmic
improvements alone. This underscores the importance of carefully choosing the minimum
resolution required to answer a given scientific question, rather than defaulting to
the largest affordable grid. Without breakthroughs in energy or computing technology,
routinely running simulations at $N \gtrsim 10^4$ is difficult to reconcile with the
imperative to limit greenhouse gas emissions from scientific research. We hope that the
efficiency improvements demonstrated here, together with their open-source
implementation, will help the turbulence community pursue ambitious science within more
sustainable computational budgets.

\acknowledgments

Part of this work was performed using resources provided by CINES (under GENCI
allocation number A0180107567) and the GRICAD infrastructure
(https://gricad.univ-grenoble-alpes.fr), which is supported by Grenoble research
communities. This project received initial financial support from the European Research
Council (ERC) under the European Union's Horizon 2020 research and innovation programme
(Grant Agreement No. 647018 – WATU). It is supported by the Simons Foundation through
the Simons Collaboration on Wave Turbulence. We gratefully acknowledge the open-source
movement, especially the Python community, whose tools and collaborative spirit were
essential to this work. A CC-BY public copyright license has been applied by the
authors to the present document and will be applied to all subsequent versions up to
the Author Accepted Manuscript arising from this submission, in accordance with the
grant’s open access conditions. The authors would like to sincerly acknowledge Nicolas
Mordant for his insightful discussions and for reviewing this manuscript.

\appendix


\section{Discrete error formula}
\label{anx:annex_0}

The discrete form of Eq.~\eqref{eq:quality} is expressed as follows, with the spectral
step $\Delta k$ vanishing in the ratio:

\begin{subequations} \label{eq:quality-i-disc}
\begin{alignat}{2}
\mathrm{Error}_x^{disc} = \frac{1}{\left(N_t\sum\limits_{n=0}^{\kmax}k_{x,n}^{-1}\right)} \sum\limits_{j=t_{start}}^{t_{end}} \sum\limits_{n=0}^{\kmax}k_{x,n}^{-1}\frac{100|E_{ref}(k_{x,n}, j)-E(k_{x,n}, j))|}{E_{ref}(k_{x,n}, j)},
\end{alignat}
\end{subequations}

where $N_t$ is the number of steps between $t_{start}$ and $t_{end}$.

\bibliography{references.bib}

\end{document}


\title{Supplementary material: Aliasing and phase shifting in pseudo-spectral
simulations of the incompressible Navier-Stokes equations}

\author{Clovis Lambert}
\author{Jason Reneuve}
\author{Pierre Augier}

\email[]{pierre.augier@univ-grenoble-alpes.fr}

\affiliation{Laboratoire des Ecoulements G\'eophysiques et Industriels, Universit\'e
Grenoble Alpes, CNRS, Grenoble-INP,  F-38000 Grenoble, France}

\maketitle

\section{Mathematical explaination of aliasing}
\label{anx:annex_1}

In this section, the mathematical basis enabling to understand how aliasing error is
generated in pseudo-spectral simulations is detailed. The mathematical reason behind
the presence of alias for a given field is first detailed, then the 3D generalization
of alias creation in bi-linear product is presented.

\subsection{Error in the discretization of Fourier Transform}

This part follow the work of \citet{Canuto1988,Canuto2007}. We consider a
$2\pi$-periodic function $u(x)$ whose Fourier series is given by

\begin{equation}
S u = \sum_{k=-\infty}^{+\infty} \check{u}_k e^{ikx},
\label{eq:fourier_series}
\end{equation}
where $ \check{u}_k $ are the exact (continuous) Fourier coefficients.

In pseudo-spectral methods, $ u(x) $ is represented at a finite number of equally
spaced grid points:

\begin{equation}
x_j = \frac{2\pi j}{N},
\qquad j = 0, 1, \dots, N-1.
\label{eq:grid_points}
\end{equation}
The discrete Fourier coefficients are then defined by the Discrete Fourier Transform
(DFT):

\begin{equation}
\hat{u}_k = \frac{1}{N} \sum_{j=0}^{N-1} u(x_j) e^{-ikx_j},
\qquad k = -\frac{N}{2}, \dots, \frac{N}{2}-1.
\label{eq:dft_def}
\end{equation}
which are then computed by the Fast Fourier Transform (FFT), see \cite{CooleyTukey} and
\cite{fluidfft}.

If the Fourier series $Su$ converges pointwise to $ u(x) $, then we can replace $
u(x_j) $ by its exact Fourier expansion Eq.~\eqref{eq:fourier_series}
\cite{Canuto1988,Canuto2007}:

\begin{equation}
u(x_j) = \sum_{p=-\infty}^{+\infty} \check{u}_p e^{ipx_j}.
\label{eq:fourier_on_grid}
\end{equation}
Substituting this into \eqref{eq:dft_def} yields:

\begin{equation}
\hat{u}_k = \frac{1}{N} \sum_{j=0}^{N-1}
\left( \sum_{p=-\infty}^{+\infty} \check{u}_p e^{ipx_j} \right)
e^{-ikx_j}
= \sum_{p=-\infty}^{+\infty} \check{u}_p
\left( \frac{1}{N} \sum_{j=0}^{N-1} e^{i(p-k)x_j} \right).
\label{eq:insert_sum}
\end{equation}
The inner sum is a periodic geometric series:

\begin{equation}
\frac{1}{N} \sum_{j=0}^{N-1} e^{i(p-k)x_j}
= \frac{1}{N} \sum_{j=0}^{N-1} e^{i(p-k)\frac{2\pi j}{N}}.
\label{eq:geom_sum}
\end{equation}
Using the discrete orthogonality property of the Fourier basis, see
\cite{Canuto1988,Canuto2007}, we have:

\begin{equation}
\frac{1}{N} \sum_{j=0}^{N-1} e^{i(p-k)\frac{2\pi j}{N}} =
\begin{cases}
1, & \text{if } p - k = q2k_N, \quad q \in \mathbb{Z},\\[4pt]
0, & \text{otherwise.}
\end{cases}
\label{eq:orthogonality_2}
\end{equation}
Which can also be written using the Kronecker symbol:

\begin{equation}
\frac{1}{N} \sum_{j=0}^{N-1} e^{i(p-k)\frac{2\pi j}{N}} = \delta[p - k \bmod 2k_N]
\label{eq:Kronecker_2}
\end{equation}
where $k_N = (2\pi / L) N/2$. This means that frequencies differing by integer
multiples of $ 2k_N $ are indistinguishable when sampled at $ N $ points.

From \eqref{eq:insert_sum} and \eqref{eq:orthogonality_2}, only terms with $ p = k +
q2k_N $ contribute:

\begin{equation}
\hat{u}_k = \sum_{q=-\infty}^{+\infty} \check{u}_{k+q2k_N}.
\label{eq:aliasing_sum}
\end{equation}
Separating the primary mode ($ q = 0 $) gives the canonical aliasing relation
\cite{Canuto1988,Canuto2007}:

\begin{equation}
\hat{u}_k = \check{u}_k + \sum_{q=-\infty,\, q \neq 0}^{+\infty} \check{u}_{k+q2k_N},
\qquad
k = 1-2k_N, \dots, 2k_N.
\label{eq:aliasing_relation}
\end{equation}
Eq.~\eqref{eq:aliasing_relation} shows that each discrete coefficient $ \hat{u}_k $
contains not only the true Fourier coefficient $ \check{u}_k $, but also the
contributions from all modes whose wavenumbers differ by multiples of $2k_N $. This is
the mathematical source of spectral aliasing: sampling at $ N $ points causes the
spectrum to repeat periodically in Fourier space with period $ 2k_N $, folding energy
modes outside the span of wavenumber into the spectral band.

Note that the relation \eqref{eq:aliasing_relation} holds exactly if the convergence
pointwise of the Fourier series is true. According to the Carleson's Theorem, $Su$
converges almost everywhere to $u$ if $u$ is an $L^2$-function \cite{Carleson1966}.
However, in the case of Navier-Stokes equations, although convergence pointwise is not
proven yet, convergence in $L^2$ norms was proven depending on the resolution
smoothness \cite{HALD1981} and the presence of alias error found in many studies
developing spectral methods to resolve Navier-Stokes equations is a good start to
assume that the solutions developped to suppress alias error by
\citet{PattersonOrszag1971} and \citet{Rogallo1981} and explained in this article have
coherent mathematical basis.

\subsection{Aliasing error generated by nonlinear products in 3D}

Here, we present how the bilinear product of two fields can generate aliasing in three
dimensions. All of the steps described in the aliasing description section of the
article can be used replacing wavenumbers and positions by vectors as $\boldsymbol{k}$,
$\boldsymbol{x}$

\begin{equation}
\hat{w}(\boldsymbol{k}) = \frac{1}{L^3} \int_{0}^{L}\int_{0}^{L}\int_{0}^{L} w(\boldsymbol{x}) e^{-i \boldsymbol{k} \cdot \boldsymbol{x}}dx^3
\label{eq:discrete_transform}
\end{equation}
as explained in the core of the paper,this equation is equivalent to:

\begin{equation}
\hat{w}(\boldsymbol{k}) = \sum_{\boldsymbol{k_n} \in \boldsymbol{\mathcal{K}}} \sum_{\boldsymbol{k_m} \in \boldsymbol{\mathcal{K}}} \hat{u}(\boldsymbol{k_n}) \hat{v}(\boldsymbol{k_m})  \frac{1}{L^3} \int_{0}^{L}\int_{0}^{L}\int_{0}^{L} e^{i(\boldsymbol{k_n} + \boldsymbol{k_m} - \boldsymbol{k}) \cdot \boldsymbol{x}}dx^3.
\label{eq:w_product_developed}
\end{equation}
Which is the same as:

\begin{equation}
\hat{w}(\boldsymbol{k})
= \sum_{\boldsymbol{k_n} \in \boldsymbol{\mathcal{K}}} \sum_{\boldsymbol{k_m} \in \boldsymbol{\mathcal{K}}}
  \hat{u}(\boldsymbol{k_n}) \hat{v}(\boldsymbol{k_m}) \,
  \delta[(\boldsymbol{k_n}+\boldsymbol{k_m}) - \boldsymbol{k} \bmod 2\boldsymbol{k_N}].
\label{eq:rogallo_double_sum}
\end{equation}
In three dimensional space, it is interesting to follow \citet{PattersonOrszag1971}
work and decompose Eq.~\eqref{eq:w_product_developed} or
Eq.~\eqref{eq:rogallo_double_sum} into two parts: the physical and the aliased parts.
To do so, it is important to understand that in three dimensions, we have the internal
sum corresponding to:

\begin{equation}
\frac{1}{L^3} \int_{0}^{L}\int_{0}^{L}\int_{0}^{L} dx^3 e^{\,i(\boldsymbol{k_n}+\boldsymbol{k_m}-\boldsymbol{k})\cdot\boldsymbol{x}}
=
\begin{cases}
1, & \text{if } \boldsymbol{k_n}+\boldsymbol{k_m}-\boldsymbol{k} = 2\boldsymbol{e}\odot\boldsymbol{k_N},\quad e_\alpha\in\{-1,0,1\},\\[4pt]
0, & \text{otherwise}.
\end{cases}
\label{eq:orthogonality_3D}
\end{equation}
With $\odot$ the Hadamard or Schur product which is a component by component product:
\begin{equation}
\boldsymbol{e}\odot\boldsymbol{k_N}
=
\begin{pmatrix}
  e_1 k_{N_1} \\
  e_2 k_{N_2} \\
  e_3 k_{N_3} \\
\end{pmatrix}
=
k_N
\begin{pmatrix}
  e_1 \\
  e_2 \\
  e_3 \\
\end{pmatrix}
= \boldsymbol{e} k_N
\end{equation}
since $k_{N_1} = k_{N_2} = k_{N_3} = k_N$ here. Eq.~\eqref{eq:orthogonality_3D} is
equivalent to select the modes:

$$
\boldsymbol{k_n}+\boldsymbol{k_m} = \boldsymbol{k} \qquad\text{or}\qquad
\boldsymbol{k_n}+\boldsymbol{k_m} = \boldsymbol{k} + \boldsymbol{e} 2k_N,
$$
as $\boldsymbol{k_n},\boldsymbol{k_m}\in\boldsymbol{\mathcal{K}}$ each component
$k_{n_\alpha}+k_{m_\alpha}$ is in $[2-2k_N, 2k_N]$. Then , there exist integers
$e_\alpha\in\{-1,0,1\}$ such that $k_{n_\alpha}+k_{m_\alpha} - k_\alpha = e_\alpha
2k_N$. These integers are the components of the alias vector $\boldsymbol{e}$ that
takes the values $\boldsymbol{e} \in \boldsymbol{\mathcal{E}} : e_\alpha \in
\{-1,0,1\}$. Yielding to the exact decomposition:
\begin{equation}
\hat{w}(\boldsymbol{k})
=
\sum_{\boldsymbol{e} \in \boldsymbol{\mathcal{E}}}
\;\sum_{\substack{
\boldsymbol{k_n},\boldsymbol{k_m}\in\boldsymbol{\mathcal{K}} \\
\boldsymbol{k_n}+\boldsymbol{k_m} - \boldsymbol{k}
=\boldsymbol{e} 2k_N
}}
\hat{u}(\boldsymbol{k_n})\,\hat{v}(\boldsymbol{k_m})
\label{eq:alias_full}
\end{equation}

From here, we can seperate the Fourier coefficient of the nonlinear product into clean
and aliased terms:
\begin{equation}
\hat{w}(\boldsymbol{k})
=
\hat{w}_{\text{clean}}(\boldsymbol{k})
+
\hat{w}_{\text{alias}}(\boldsymbol{k}).
\end{equation}
The clean part corresponds to $\boldsymbol{e}=(0,0,0)$ :

\begin{equation}
\hat{w}_{\text{clean}}(\boldsymbol{k})
=
\sum_{\substack{\boldsymbol{k_n}+\boldsymbol{k_m}=\boldsymbol{k}}}
\hat{u}(\boldsymbol{k_n})\,\hat{v}(\boldsymbol{k_m}).
\end{equation}
The aliased part corresponds to $\boldsymbol{e}\neq (0,0,0)$ :

\begin{equation}
\hat{w}_{\text{aliased}}(\boldsymbol{k})
=
\sum_{\boldsymbol{e}\neq (0,0,0)}
\sum_{\substack{\boldsymbol{k_n}+\boldsymbol{k_m}=\boldsymbol{k}+\boldsymbol{e} 2k_N}}
\hat{u}(\boldsymbol{k_n})\,\hat{v}(\boldsymbol{k_m}).
\end{equation}

This latter quantity contains three types of alias. \emph{Simple aliases}, for all
combinations of $\boldsymbol{e}$ containing two zero value components. \emph{Double
aliases}, for all combinations of $\boldsymbol{e}$ containing exactly one zero value
component and finally, \emph{triple aliases} for all combinations of $\boldsymbol{e}$
containing no zero value component.

\section{Error versus time step for 1D example}

\begin{figure}[h]
\includegraphics[width=\figwidth]{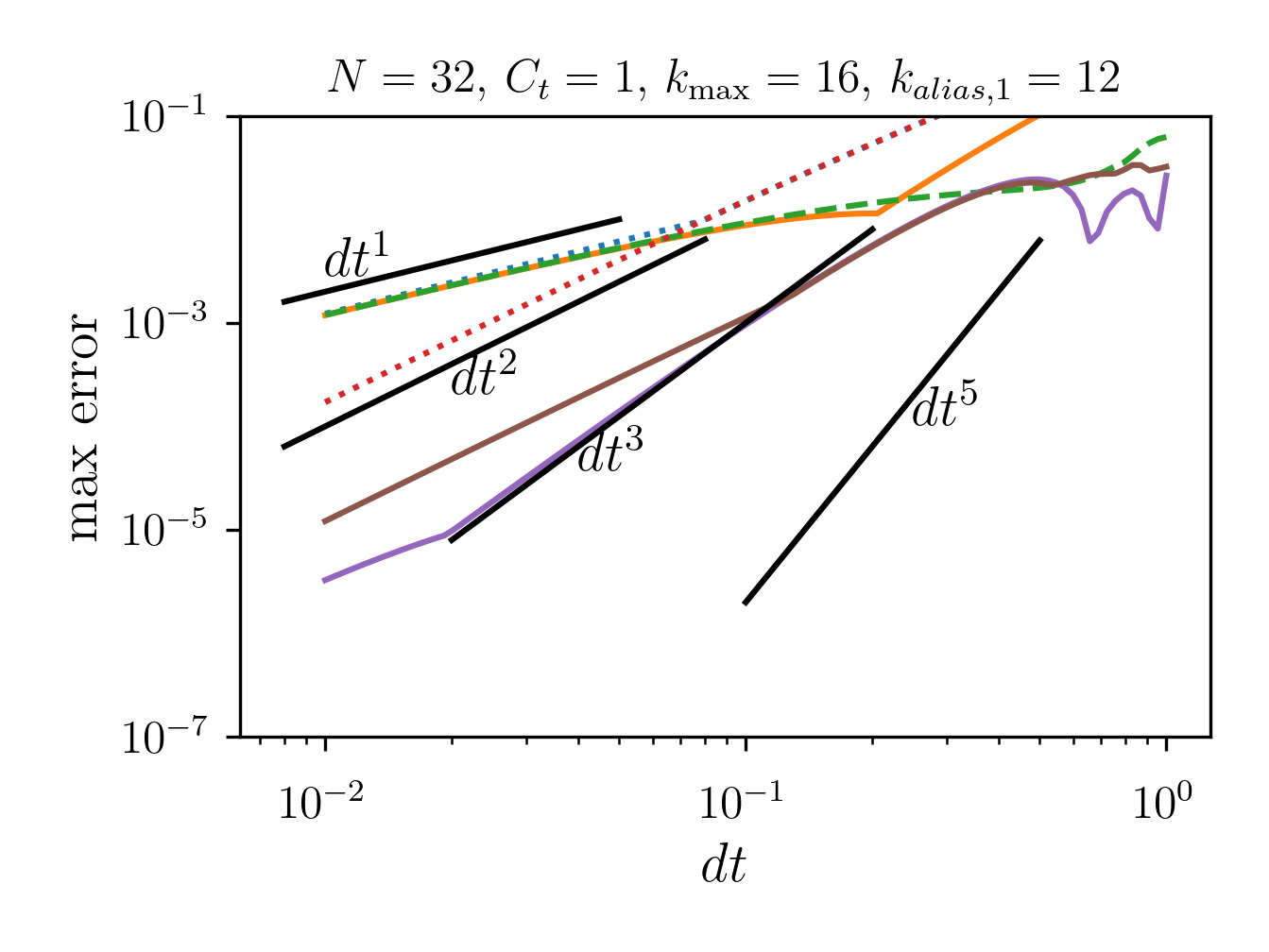}
\caption{Maximum error ($max|S(k, dt) - S_\text{exact}(k, dt)|$) for one time step
versus time increment $dt$ for all time schemes implemented at $N = 32$ and $C_t = 1$.
With $k_0 = 10$, one time step creates one nonlinear mode $k_{nl,1} = 20$ for all
schemes and a second one at $k_{nl,2} = 30$ for multi-steps schemes.
\label{fig:error1d-32}}
\end{figure}

Fig.~\ref{fig:error1d-32} presents the scalings of various time schemes with the time
step for size $N = 32$ and $C_t = 1$. In this example, any scheme that does not contain
phase-shifting containing aliasing errors as in the first example of the quadratic
model presented in the core of the article.

\section{Profiles comparison at various resolutions}
\label{anx:annex_0_2}

\begin{figure}[]
\centerline{\includegraphics[width=\figwidth]{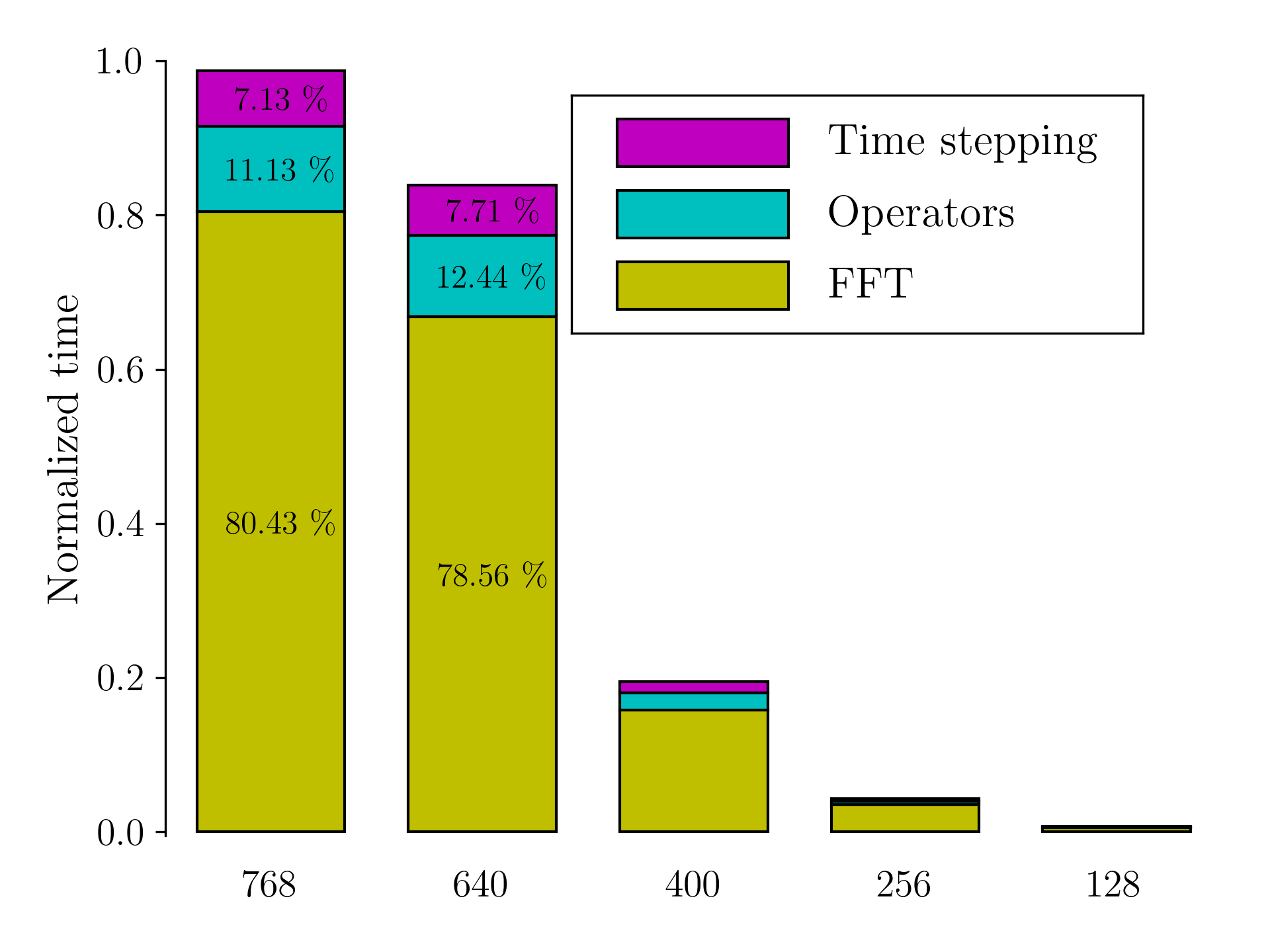}}
\caption{Distribution of times spent in different tasks (Fourier transforms, operators,
time stepping) for simulations run with different resolutions. The simulations have
been run up to $0.01$ overturn times in sequential. The heights of the rectangles are
proportional to the times normalized by the elapsed time for the biggest simulation at
$N = 768$. By contrast, the percentages are computed with the elapsed time for each
size. \label{fig:profiling_resol}}
\end{figure}

Fig.~\ref{fig:profiling_resol} presents the evolution of the computation time of
simulations with various sizes, normalized by the computation time of the longest, i.e.
biggest, simulation. It also shows the distribution of times spent in different tasks
classified into three categories. This result needs to be read with the table from
Appendix C.

\section{Spectra}

\begin{figure}[H]
\centerline{
    \includegraphics[width=\figwidth]{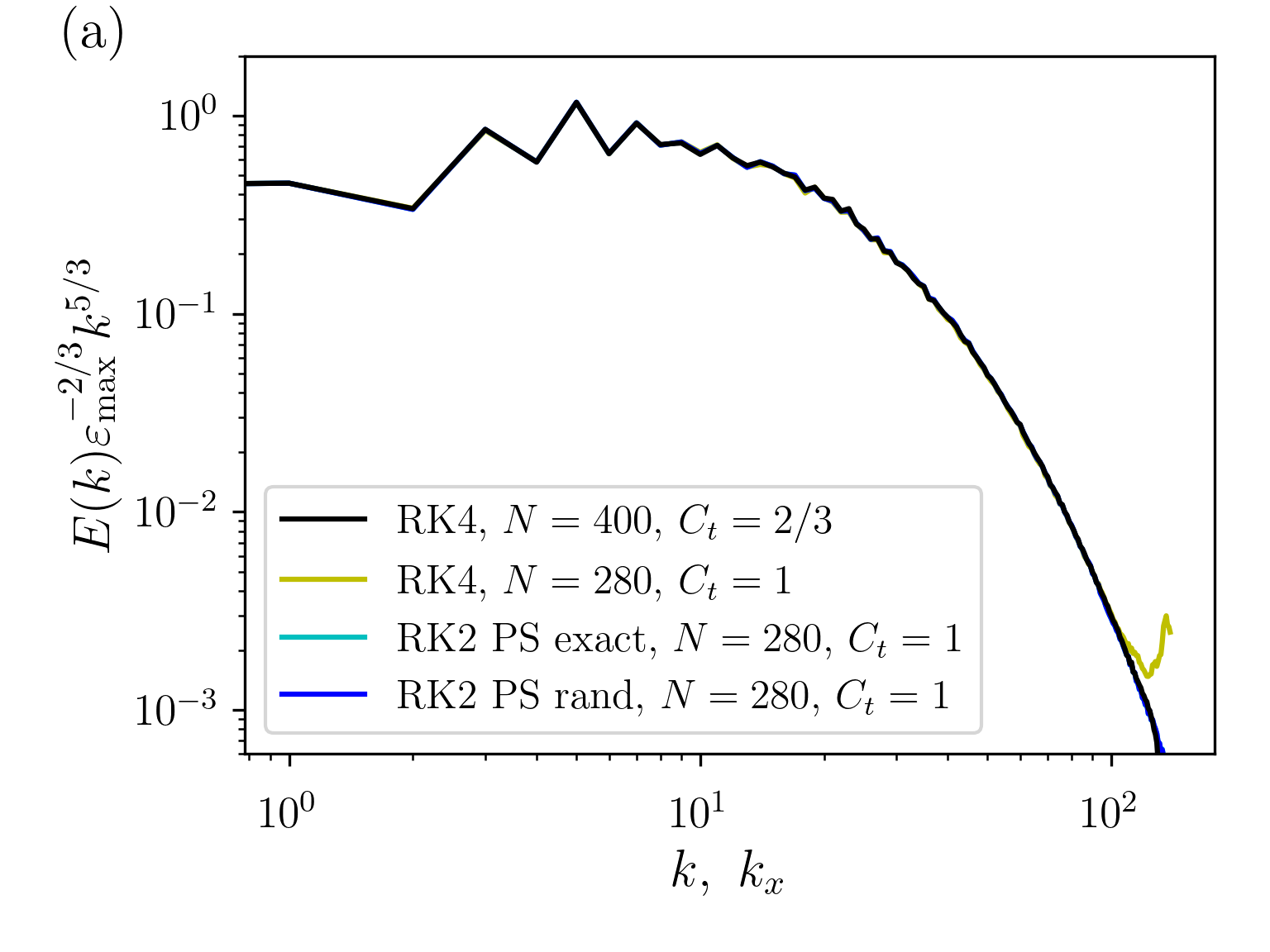}
    \includegraphics[width=\figwidth]{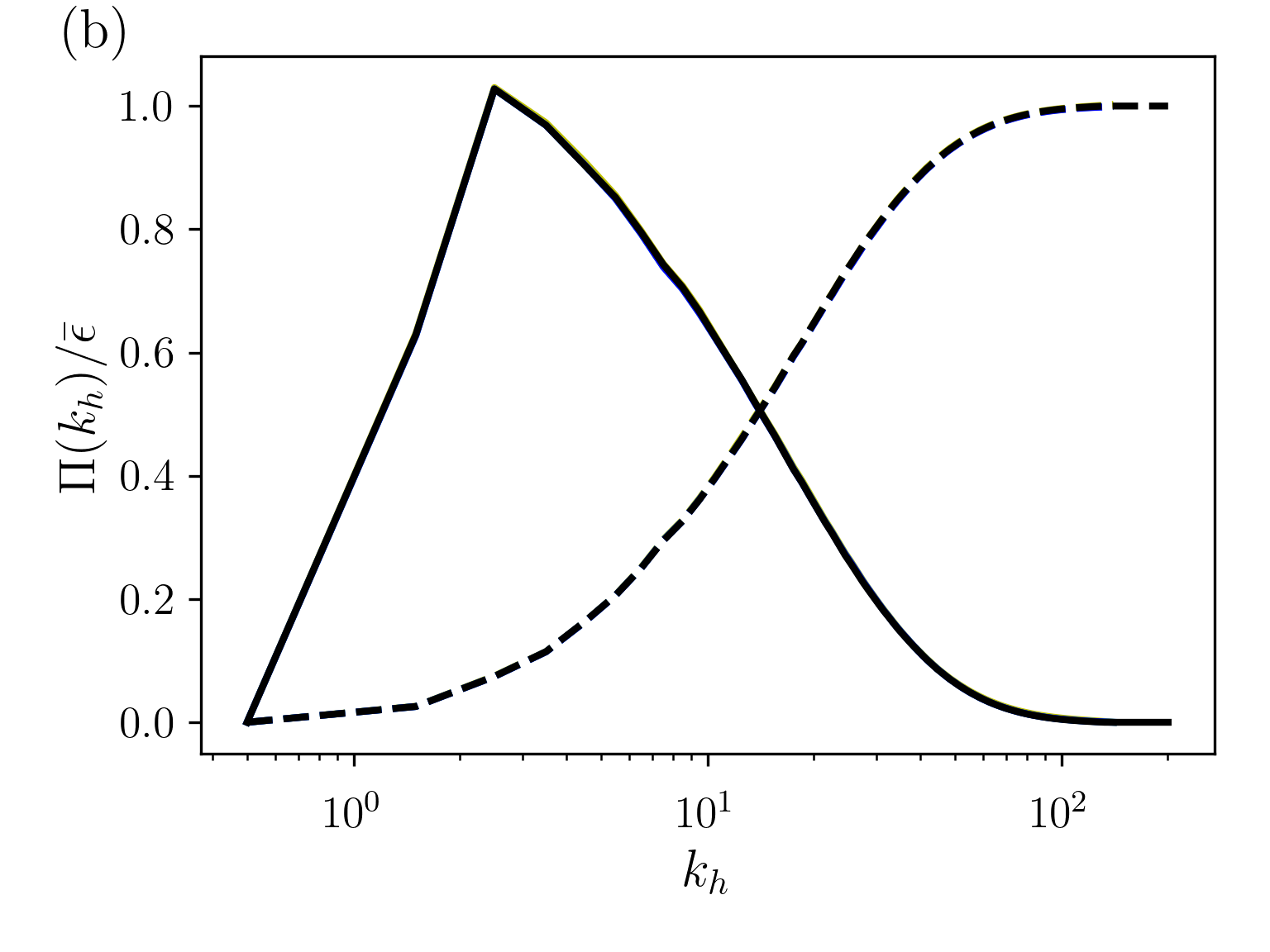} }
\caption{(a) Compensated energy spectra (3D spectra in dotted lines) versus wavenumber
modulus $k$ and one-dimensional (along $x$) energy spectra (solid lines) versus
$x$-direction wavenumber $k_x$. (b) Nonlinear energy flux (solid lines) and cumulated
dissipation energy (dashed lines), averaged in time between $t_{start} = 9$ and
$t_{end} = 14$ overturn times, for $Re = 1600$ and $\kmax \etamin \simeq 1.6$ and
various time schemes. Simulations corresponding to the third serie of Table IV from the
article. \label{fig:spectra:400} }
\end{figure}

Fig.~\ref{fig:spectra:400} presents the energy, energy flux, and cumulated dissipation
energy spectra for simulations at various time schemes and maximum size $N = 400$.

\section{Temporal evolutions}

\begin{figure}[H]
\centerline{\includegraphics[width=\figwidth]{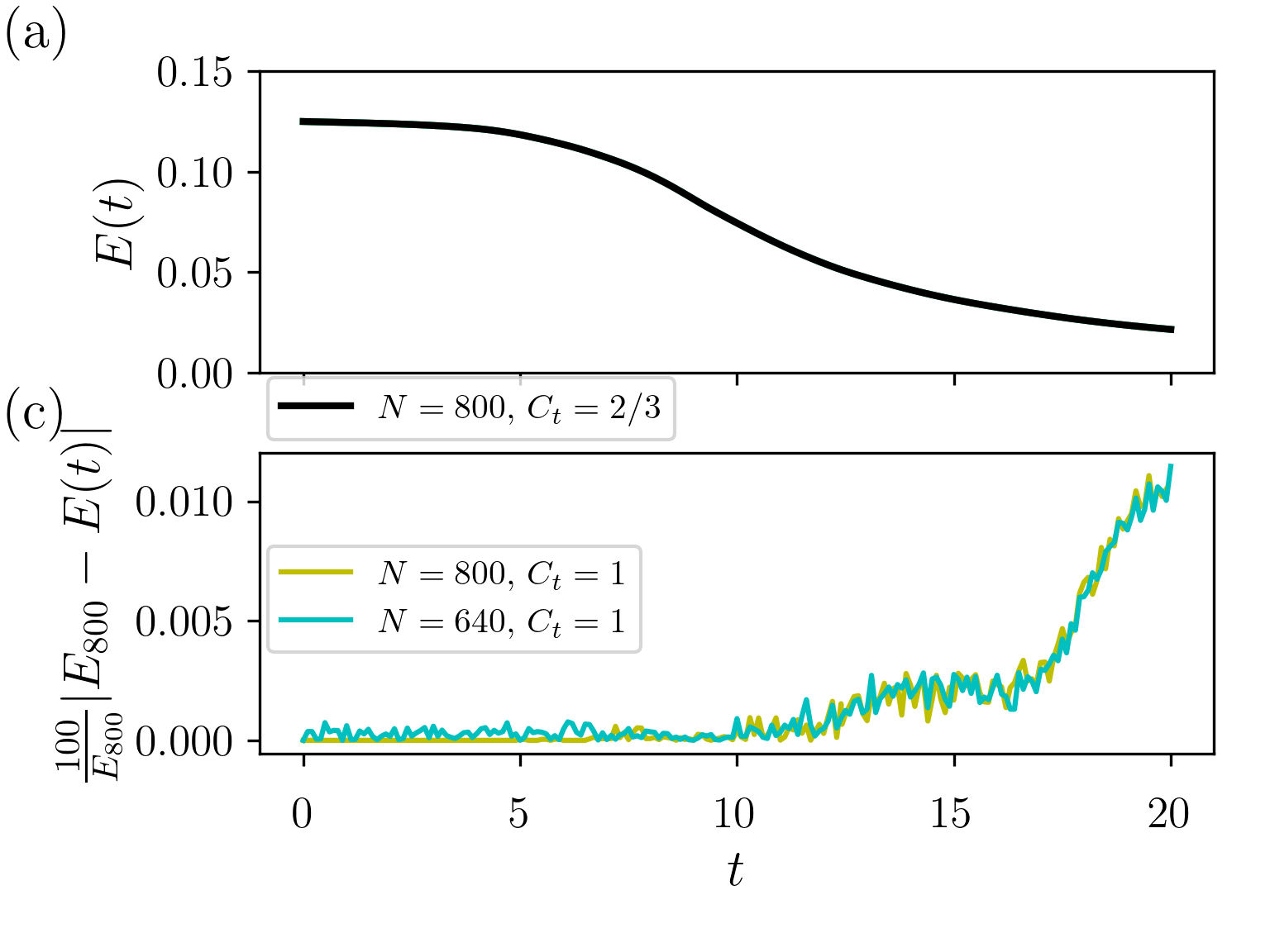}
  \includegraphics[width=\figwidth]{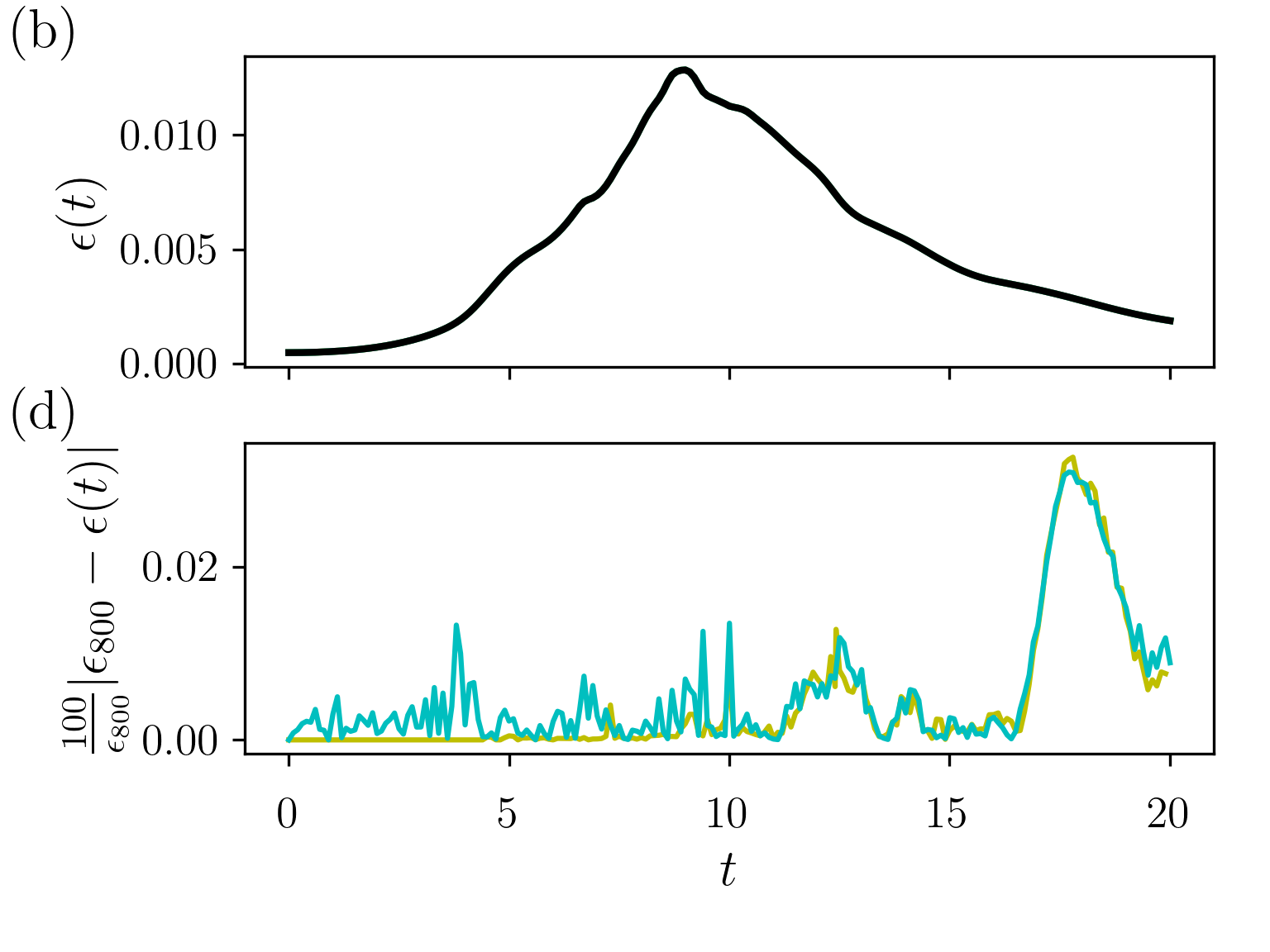} }
\caption{Temporal evolution of (a) total kinetic energy $E(t)$ (b) energy dissipation,
(c) energy relative error, and (d) energy dissipation relative error for simulations
computed with RK4 at various $\kmax \etamin > 3$ (equal to $3.13$, $3.75$ and  $4.69$
from lowest to biggest resolution respectively). Simulations corresponding to Table III
in the article.\label{fig:spatialmeans:RK4} }\end{figure}

\begin{figure}[H]
\centerline{\includegraphics[width=\figwidth]{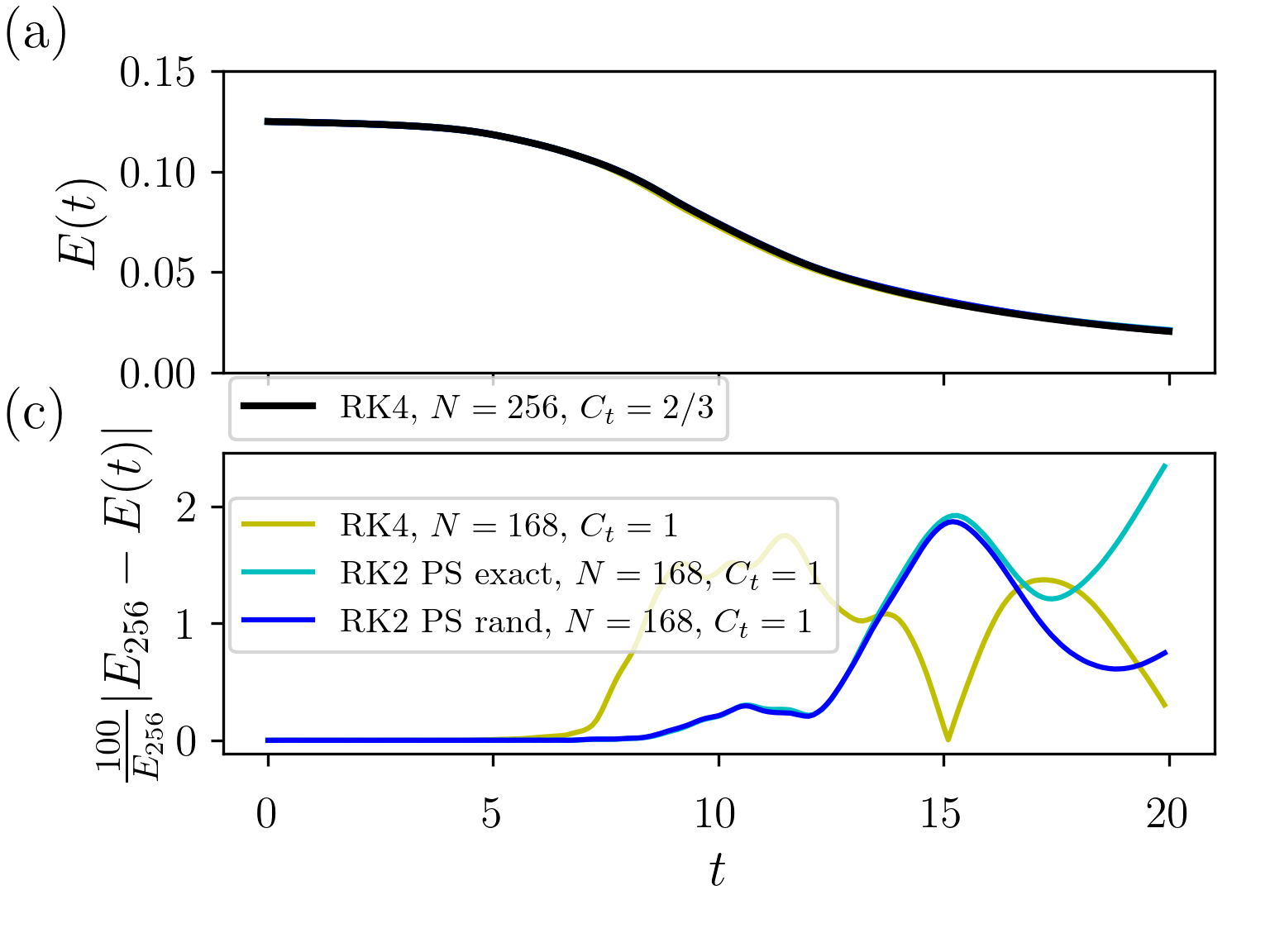}
  \includegraphics[width=\figwidth]{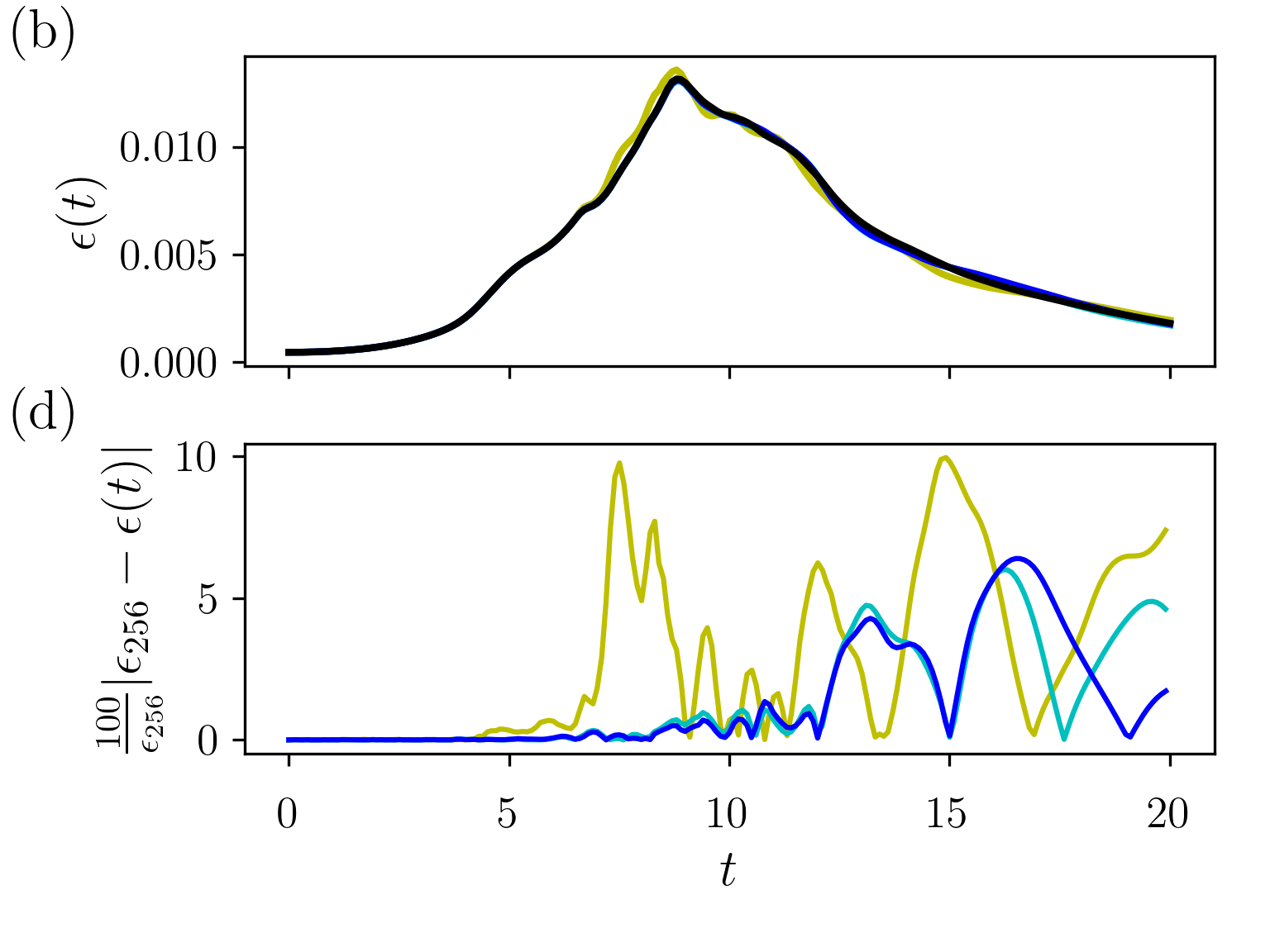} }
\caption{Same as Fig.~\ref{fig:spatialmeans:RK4} for $Re = 1600$ and $\kmax \etamin
\simeq 1$ and various time schemes. Simulations corresponding to the first serie of
Table IV in the article. \label{fig:spatialmeans:256} }\end{figure}

\begin{figure}[H]
\centerline{\includegraphics[width=\figwidth]{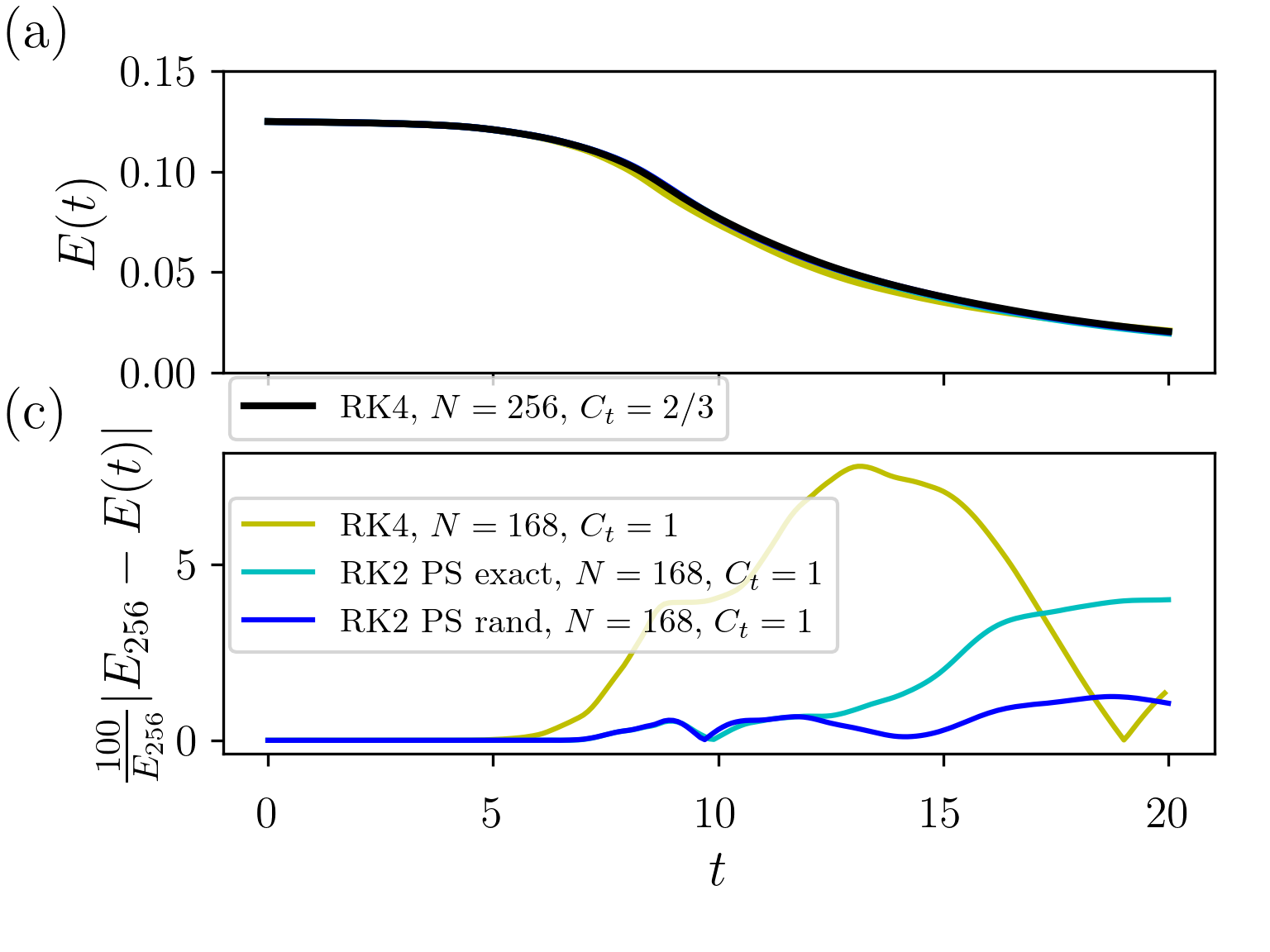}
  \includegraphics[width=\figwidth]{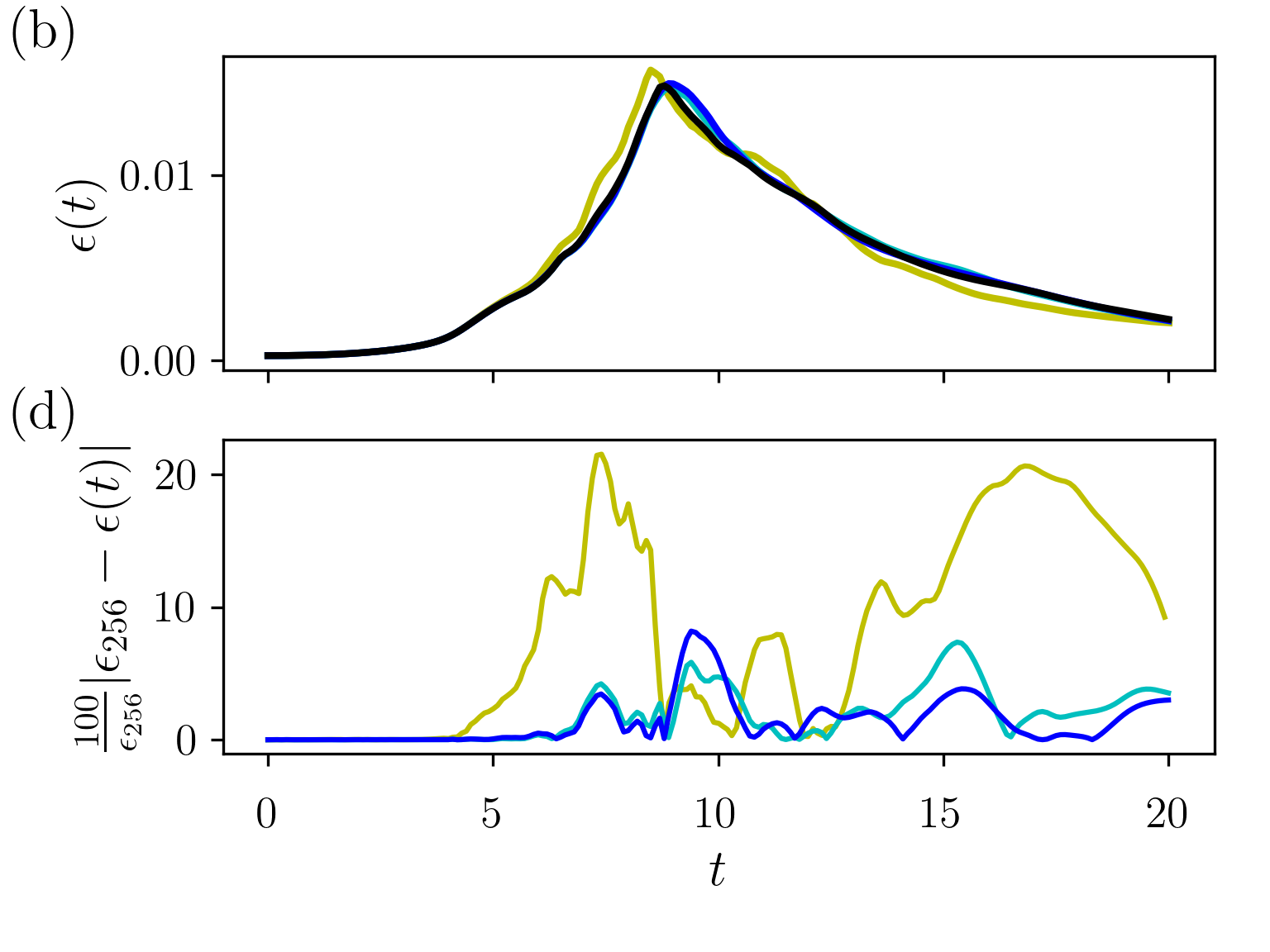} }
\caption{Same as Fig.~\ref{fig:spatialmeans:RK4} for $Re = 2800$ and $\kmax \etamin
\simeq 0.6$ and various time schemes. Simulations corresponding to the second serie of
Table IV in the article. \label{fig:spatialmeans:256:Re2800}}
\end{figure}

\begin{figure}[H]
\centerline{\includegraphics[width=\figwidth]{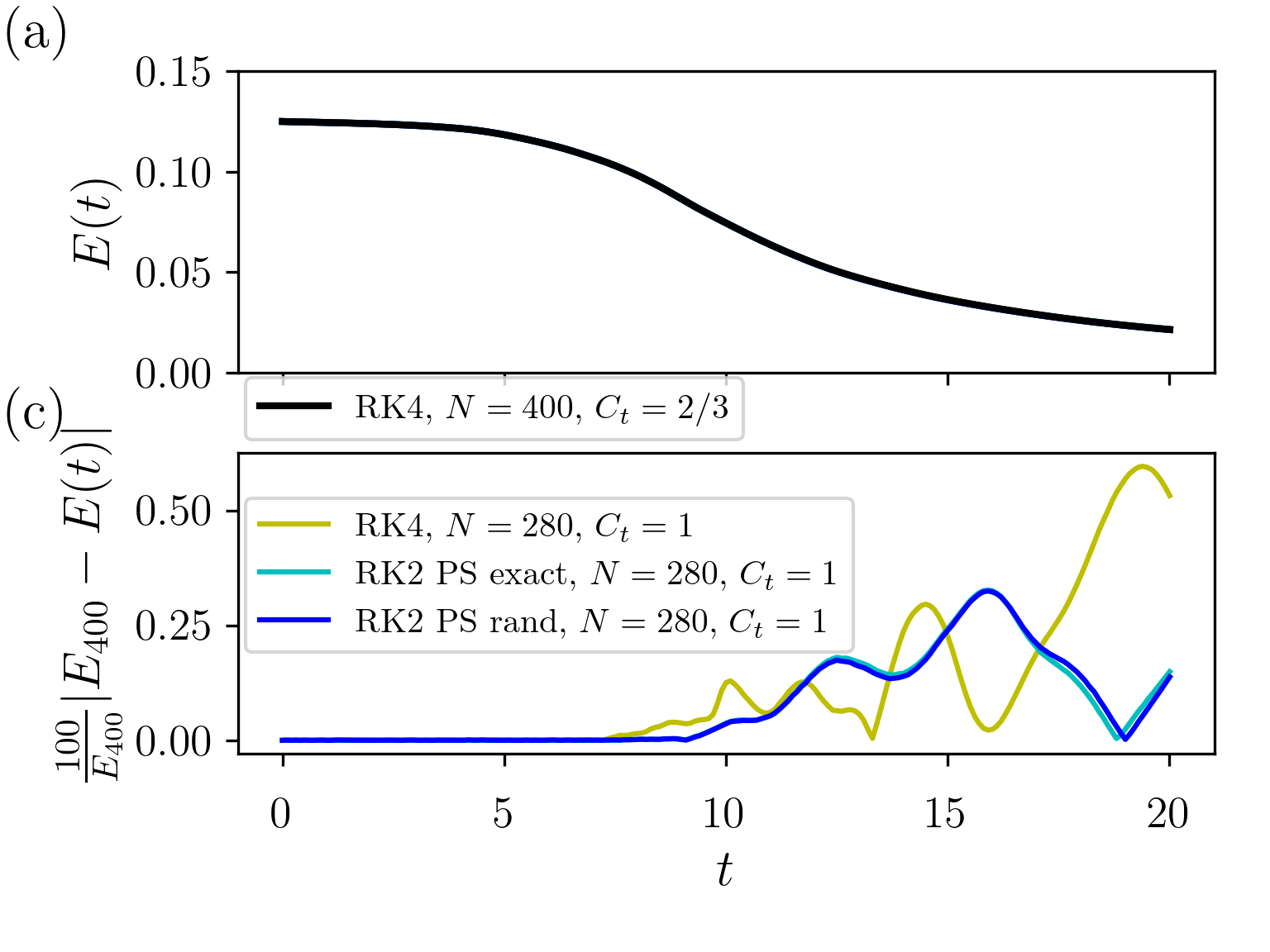}
  \includegraphics[width=\figwidth]{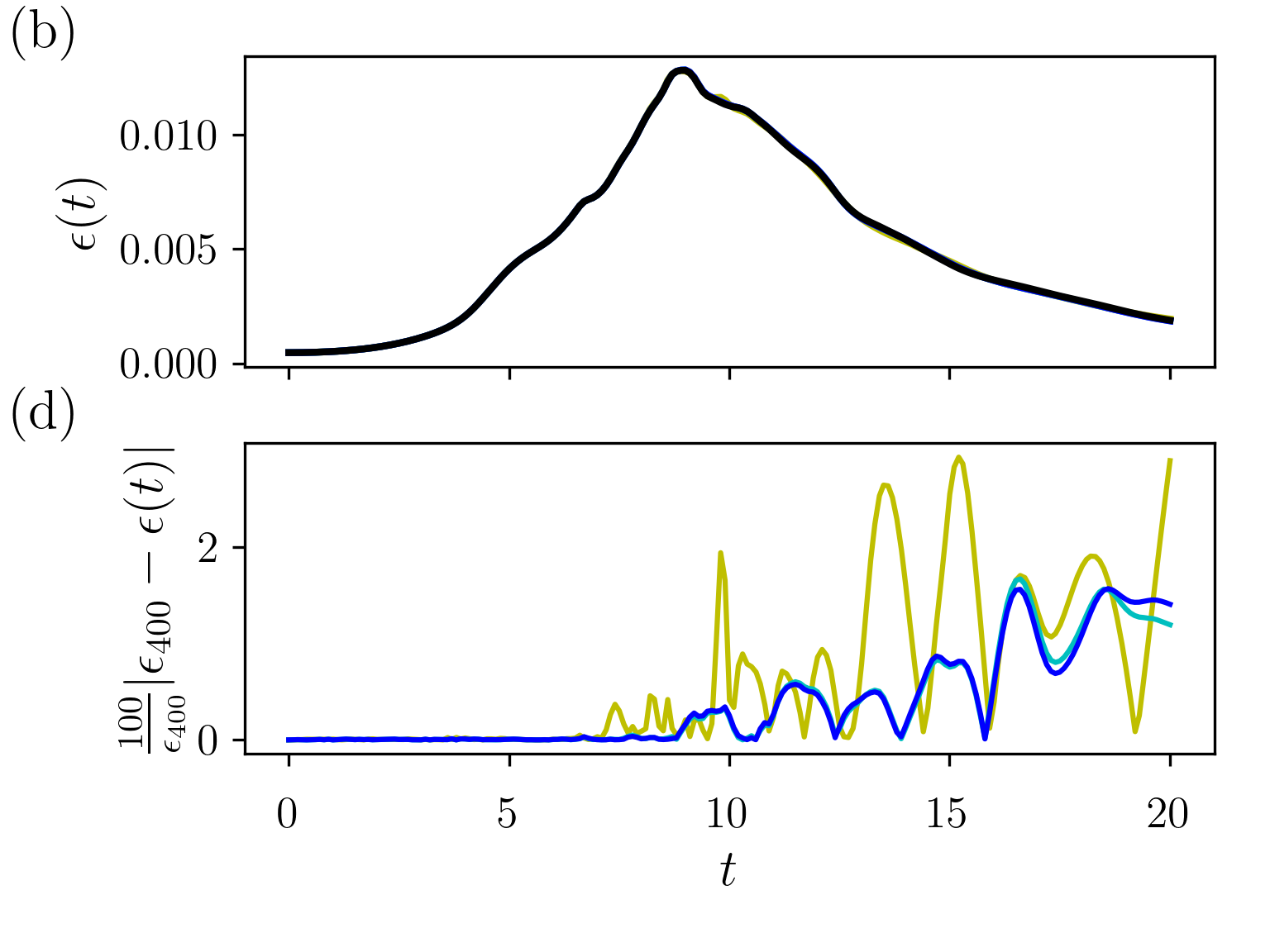} }
\caption{Same as Fig.~\ref{fig:spatialmeans:RK4} for $Re = 1600$ and $\kmax \etamin
\simeq 1.6$ and various time schemes. Simulations corresponding to the third serie of
Table IV in the article. \label{fig:spatialmeans:400}}\end{figure}

Figs.~\ref{fig:spatialmeans:RK4}, ~\ref{fig:spatialmeans:256},
~\ref{fig:spatialmeans:400} and ~\ref{fig:spatialmeans:256:Re2800} presents the
temporal evolutions of energy dissipation energy and their corresponding errors for
Taylor-Green flow simulations presented in the article.

\section{Table of speedups and errors}

Tables~\ref{tab:quality} presents the speedup and errors to Taylor-Green flow
simulations presented in the article.


\begin{table}[H]
\centering \begin{tabular}{| c | c | c | c | c | c | c | c | c |}
\toprule
$N$ & scheme & $C_t$ & $k_{\max} \tilde\eta$ & Trunc. shape & speedup & error (\%) \\
\midrule
324 & RK2 & 2/3 & 1.27 & Spherical & 0.8 & 0.04 \\
324 & RK4 & 2/3 & 1.27 & Spherical & 1.0 & 0.0 \\
240 & RK2 PS exact & 0.9 & 1.27 & Spherical & 1.3 & 0.04 \\
240 & RK4 & 0.9 & 1.27 & Spherical & 3.4 & 3.72 \\
240 & RK2 PS rand & 0.9 & 1.27 & Spherical & 1.9 & 0.05 \\
228 & RK2 PS exact & 0.947 & 1.27 & Spherical & 1.2 & 0.04 \\
228 & RK2 PS rand & 0.947 & 1.27 & Spherical & 2.0 & 0.07 \\
224 & RK2 PS rand & 0.964 & 1.27 & Spherical & 2.5 & 0.09 \\
224 & RK2 PS exact & 0.964 & 1.27 & Spherical & 1.5 & 0.04 \\
220 & RK2 PS exact & 0.982 & 1.27 & Spherical & 1.7 & 0.05 \\
220 & RK2 PS rand & 0.982 & 1.27 & Spherical & 2.7 & 0.13 \\
216 & RK2 PS exact & 1 & 1.27 & Spherical & 1.8 & 0.07 \\
216 & RK2 PS rand & 1 & 1.27 & Spherical & 3.1 & 0.19 \\
216 & RK4 & 1 & 1.27 & Spherical & 5.5 & 15.72 \\
228 & RK2 PS exact & 0.947 & 1.27 & No double aliases & 1.2 & 0.32 \\
228 & RK2 PS rand & 0.947 & 1.27 & No double aliases & 2.0 & 0.32 \\
224 & RK2 PS rand & 0.964 & 1.27 & No double aliases & 2.5 & 0.58 \\
224 & RK2 PS exact & 0.964 & 1.27 & No double aliases & 1.5 & 0.58 \\
220 & RK2 PS rand & 0.982 & 1.27 & No double aliases & 2.7 & 0.92 \\
220 & RK2 PS exact & 0.982 & 1.27 & No double aliases & 1.7 & 0.95 \\
216 & RK2 PS exact & 1 & 1.27 & No double aliases & 1.8 & 1.54 \\
216 & RK2 PS rand & 1 & 1.27 & No double aliases & 3.1 & 1.54 \\
\bottomrule
\end{tabular}
 \caption{Speedups and
errors (see Eq. (15) in the article) for Taylor-Green flows simulations at various
combinations of time schemes and truncation. Errors are in \% of the reference
simulation: RK4 simulation at $N = 324$ with $C_t = 2/3$.}\label{tab:quality}
\end{table}

\bibliography{references.bib}